\documentclass{aa}

\usepackage{graphicx}
\usepackage{txfonts}
\usepackage{natbib}
\usepackage{longtable}
\usepackage{lscape}
\usepackage{upgreek}
\usepackage[colorlinks,citecolor=blue,urlcolor=blue,filecolor=blue,linkcolor=blue]{hyperref}
\usepackage{amsmath}
\usepackage{amssymb}
\usepackage{xcolor}

\DeclareMathOperator{\sinc}{sinc}

\begin{document} 

\title{The LOFAR Tied-Array All-Sky Survey (LOTAAS): Survey overview
  and initial pulsar discoveries}

\titlerunning{An overview of the LOFAR Tied-Array All-Sky Survey}

\authorrunning{S.\ Sanidas et al.}

\author{S.\ Sanidas\inst{\ref{uva}, \ref{jb}}
  \and
  S.\ Cooper\inst{\ref{jb}}
  \and
  C.\ G.\ Bassa\inst{\ref{astron}}
  \and
  J.\ W.\ T.\ Hessels\inst{\ref{astron}, \ref{uva}}
  \and
  V.\ I.\ Kondratiev\inst{\ref{astron},\ref{asc}}
  \and
  D.\ Michilli\inst{\ref{astron}, \ref{uva}, \ref{mcgill}, \ref{msi}}
  \and
  B.\ W.\ Stappers\inst{\ref{jb}}
  \and
  C.\ M.\ Tan\inst{\ref{jb}}
  \and
  J.\ van Leeuwen\inst{\ref{astron}, \ref{uva}}
  \and
  L.\ Cerrigone\inst{\ref{alma}}
  \and
  R.\ A.\ Fallows\inst{\ref{astron}}
  \and
  M.\ Iacobelli\inst{\ref{astron}}
  \and
  E.\ Orr\'u\inst{\ref{astron}}
  \and
  R.\ F.\ Pizzo\inst{\ref{astron}}
  \and
  A.\ Shulevski\inst{\ref{uva}}
  \and
  M.\ C.\ Toribio\inst{\ref{leiden}}
  \and
  S.\ ter Veen\inst{\ref{astron}}
  \and
  P.\ Zucca\inst{\ref{astron}}
  \and
  L.\ Bondonneau\inst{\ref{lpc2e}}
  \and
  J.-M.\ Grie{\ss}meier\inst{\ref{lpc2e}, \ref{nancay}}
  \and
  A.\ Karastergiou\inst{\ref{oxford}, \ref{western_cape}, \ref{rhodes}}
  \and
  M.\ Kramer\inst{\ref{mpifr}, \ref{jb}}
  \and
  C.\ Sobey\inst{\ref{csiro}}
}

\institute{
  Anton Pannekoek Institute for Astronomy, University of Amsterdam, Postbus 94249, 1090 GE Amsterdam, The Netherlands\label{uva}
  \and
  Jodrell Bank Centre for Astrophysics, School of Physics and Astronomy, University of Manchester, Manchester M13 9PL, UK\label{jb}
  \and
  ASTRON, the Netherlands Institute for Radio Astronomy, Oude Hoogeveensedijk 4, 7991 PD Dwingeloo, The Netherlands\\\email{\url{bassa@astron.nl}}\label{astron}
  \and
  Astro Space Centre, Lebedev Physical Institute, Russian Academy of Sciences, Profsoyuznaya Str. 84/32, Moscow 117997, Russia\label{asc}
  \and
  Department of Physics, McGill University, 3600 rue University, Montr\'eal, QC H3A 2T8, Canada\label{mcgill}
  \and
  McGill Space Institute, McGill University, 3550 rue University, Montr\'eal, QC H3A 2A7, Canada\label{msi}
  \and
  Joint ALMA Observatory, Alonso de Cordova 3107, Vitacura, Chile\label{alma}
  \and
  Leiden Observatory, Niels Bohrweg 2, 2333 CA Leiden, The Netherlands\label{leiden}
  \and
  LPC2E - Universit\'{e} d'Orl\'{e}ans / CNRS, 45071 Orl\'{e}ans cedex 2, France\label{lpc2e}
  \and
  Station de Radioastronomie de Nan\c{c}ay, Observatoire de Paris, PSL Research University, CNRS, Univ. Orl\'{e}ans, OSUC, 18330 Nan\c{c}ay, France\label{nancay}
  \and
  Oxford Astrophysics, Denys Wilkinson Building, Keble Road, Oxford OX1 3RH, UK\label{oxford}
  \and
  Department of Physics \& Astronomy, University of the Western Cape, Private Bag X17, Bellville 7535, South Africa\label{western_cape}
  \and
  Department of Physics and Electronics, Rhodes University, PO Box 94, Grahamstown 6140, South Africa\label{rhodes}
  \and
  Max-Planck-Institut f\"ur Radioastronomie, Auf dem H\"ugel 69, 53121 Bonn, Germany\label{mpifr}
  \and
  CSIRO Astronomy and Space Science, PO Box 1130 Bentley, WA 6102, Australia\label{csiro}
}

\date{Received April\,3, 2019; accepted May\,7, 2019}

\abstract{We present an overview of the LOFAR Tied-Array All-Sky
  Survey (LOTAAS) for radio pulsars and fast transients. The survey
  uses the high-band antennas of the LOFAR Superterp, the dense inner
  part of the LOFAR core, to survey the northern sky ($\delta>0\degr$)
  at a central observing frequency of 135\,MHz. A total of 219
  tied-array beams (coherent summation of station signals, covering 12
  square degrees), as well as three incoherent beams (covering
  67\,square degrees) are formed in each survey pointing. For each of
  the 222 beams, total intensity is recorded at 491.52\,$\upmu$s time
  resolution. Each observation integrates for 1\,hr and covers 2592
  channels from 119 to 151\,MHz. This instrumental setup allows LOTAAS
  to reach a detection threshold of 1 to 5\,mJy for periodic
  emission. Thus far, the LOTAAS survey has resulted in the discovery
  of 73 radio pulsars. Among these are two mildly recycled binary
  millisecond pulsars ($P=13$ and 33\,ms), as well as the
  slowest-spinning radio pulsar currently known ($P=23.5$\,s). The
  survey has thus far detected 311 known pulsars, with spin periods
  ranging from 4\,ms to 5.0\,s and dispersion measures from 3.0 to
  217\,pc\,cm$^{-3}$. Known pulsars are detected at flux densities
  consistent with literature values. We find that the LOTAAS pulsar
  discoveries have, on average, longer spin periods than the known
  pulsar population. This may reflect different selection biases
  between LOTAAS and previous surveys, though it is also possible that
  slower-spinning pulsars preferentially have steeper radio
  spectra. LOTAAS is the deepest all-sky pulsar survey using a digital
  aperture array; we discuss some of the lessons learned that can
  inform the approach for similar surveys using future radio
  telescopes such as the Square Kilometre Array.}

\keywords{pulsars: general -- pulsars: individual -- methods: data
  analysis -- methods: observational}

\maketitle
\section{Introduction}
\label{sec:introduction}

To date, there are over 2200 slow rotation-powered radio pulsars
($P_\mathrm{spin} \gtrsim 0.1$\,s; $B_\mathrm{surf} \sim 10^{12}$\,G)
and about 360 rotation-powered radio millisecond pulsars (MSPs;
$P_\mathrm{spin} \lesssim 30$\,ms; $B_\mathrm{surf} \sim 10^{8}$\,G)
known \citep[see ATNF catalogue\footnote{Catalogue version 1.59,
    \url{http://www.atnf.csiro.au/people/pulsar/psrcat/}};][]{mht+05}.
However, these represent only a small fraction ($\lesssim 10$\%) of
the total expected Galactic population \citep[e.g.][and references
  therein]{kbk+15}, and there remains a strong scientific drive to
find more pulsars.

Radio pulsar searches are motivated by understanding the total
Galactic population of neutron stars: for example, the various classes
of neutron stars \citep{tkb+15}, their spatial distribution
\citep{lor11}, evolution \citep{jk17}, and birth rates
\citep{fk06}. Population synthesis models are based on the yields of
previous pulsar surveys, and a variety of targeted, wide-field, low
and high-frequency surveys are needed to constrain these models, while
also averaging out the observational biases inherent to different
survey approaches \citep[e.g.][]{lfl+06,slm+14}.

Furthermore, the discovery of individual pulsar systems has continued
to provide important new insights into gravitational theories
\citep[e.g.][]{afw+13,agh+18}, the dense matter equation of state
\citep[e.g.][]{dpr+10}, exotic stellar evolution \citep{rsa+14},
accretion physics \citep[e.g.][]{asr+09} and the interstellar medium
\citep{smc+01}. A massive international effort is also underway with
the goal of using a set of MSPs to directly detect gravitational waves
\citep{det79,haa+10,vlh+16}. Pulsar discoveries thus also have a
wide-reaching and significant impact outside the field of pulsar
astrophysics itself; in essence, they are nature's clocks and can be
used in many applications: for example, a pulsar based-timescale
\citep{hcm+12} and space navigation \citep{rww17}.

Pulsar radio spectra are typically characterised as a power-law, where
the flux density at a particular frequency, $S_\nu$, is proportional
to the observing frequency $\nu$ to some power, i.e.\ $S_\nu \propto
\nu^{\alpha}$. This is not always a good characterization of the
spectrum \citep{lrkm15,jsk+18}, however, and a low-frequency turnover
around 100 to 200\,MHz appears to be present for some sources
\cite[][and references therein]{bkk+16}. Recently, \citet{blv13} found
that the average spectral index is $\alpha = -1.4$, with a $1\sigma$
dispersion of $\sim1$\footnote{This is a somewhat shallower scaling
  compared to the $\alpha = -1.8$ found by \citet{mkkw00}.}. Indeed,
pulsars have been observed with a wide range of spectral indices, $-4
\lesssim \alpha \lesssim 0$. This variation in spectral index is not
well understood. In principle, it could be intrinsic, a function of
viewing geometry, and/or related to the interstellar medium
\citep{lrkm15,rla16}. In general, however, pulsars are steep-spectrum
radio sources, meaning that they become significantly brighter towards
lower radio frequencies.

While the steep radio spectra of pulsars may at first seem to suggest
that low radio frequencies (here defined as $\nu \lesssim 400$\,MHz)
are the best option for searches, there are a number of chromatic
effects that create challenges for low-frequency pulsar
surveys. Propagation through the ionised and magnetised interstellar
medium (ISM) influences the observed pulsar signal via i)
scintillation, a frequency and time-dependent modulation/variability
of the signal strength \citep{ric70}; ii) scattering, multi-path
propagation \citep{ric77}; iii) dispersion, a frequency-dependent
light-travel time through the intervening medium \citep[e.g.][and
  references therein]{hsh+12}; and iv) Faraday rotation, a rotation of
the angle of linearly polarised emission \citep[e.g.][]{man72, bb05,
  sbg+19}. A review of these propagation effects can be found in
\citet{ric90}. Finally, the synchrotron sky background temperature,
$T_\mathrm{sky}$, which increases rapidly towards lower radio
observing frequencies: $T_\mathrm{sky} \propto \nu^{-2.55}$
\citep{hks+81,hssw82,lmop87}, further reduces sensitivity at low
Galactic latitudes.

Most relevant to low-frequency pulsar searches are dispersion and
scattering. Dispersion introduces a time delay $\Delta t \propto
\mathrm{DM}\ \nu^{-2}$ (where DM is the dispersion measure, the
integrated column density of free electrons along the
line-of-sight). This effect can be compensated for using incoherent
dedispersion, but requires high frequency resolution. However, at low
observing frequencies, dispersion within a frequency channel strongly
limits the effective time resolution that can be achieved, especially
for relatively high DM ($>50-100$\,pc\,cm$^{-3}$). Therefore,
searching for fast-spinning millisecond pulsars at the lowest radio
frequencies requires coherent dedispersion \citep{bph17a,bph+17b}.

Scattering causes a pulse broadening $\tau_\mathrm{scat}
\propto\nu^{-4.4}$ (under the assumption of a Kolmogorov turbulence
spectrum; the exact frequency scaling can deviate from $\nu^{-4.4}$)
that cannot be corrected for in pulsar surveys (in practice). The
magnitude of scattering depends on the distribution of the material
along the line-of-sight, and is loosely correlated with DM
\citep{bcc+04, gkv+17}. Scattering becomes a significant limitation
for low-frequency pulsar searches for $\mathrm{DM} >
50-100$\,pc\,cm$^{-3}$.  DM is a proxy for distance, given a model for
the free electron distribution in the Galaxy \citep{cl02,ymw17}.
Because of dispersive smearing and scattering, low-frequency searches
are limited to detecting pulsars with $\mathrm{DM} \lesssim
100$\,pc\,cm$^{-3}$.  This is not a major limitation for the search
volume at high Galactic latitudes, but it precludes finding pulsars at
large distances ($\gtrsim 3$\,kpc) within the Galactic plane.

Despite these challenges, pulsars were originally discovered at
81.5\,MHz \citep{hbp+68}, and low-frequency searches continue to be
fruitful.  In the last decade, low-frequency pulsar surveys have
discovered $\sim 300$ pulsars, and have mapped the nearby population
both towards the Galactic plane and at higher Galactic latitudes. A
350-MHz survey of the northern Galactic plane with the Green Bank
Telescope (GBT) discovered 33 pulsars \citep{hrk+08}.  Using a very
similar observing strategy, the GBT Driftscan survey found 31 pulsars
\citep{blr+13,lbr+13}, including the pulsar stellar triple system
PSR~J0337+1715 \citep{rsa+14} and the transitional millisecond pulsar
PSR~J1023+0038 \citep{asr+09}.  An ongoing, full-sky 350-MHz survey
with the GBT, the GBNCC, has found 160 pulsars to date
\citep{slr+14,klk+18,lsk+18}.  An ongoing driftscan survey with
Arecibo at 327\,MHz has found 82
pulsars\footnote{\url{http://www.naic.edu/~deneva/drift-search}}
(AO327; \citealt{dsm+13}), while the GMRT High Resolution Southern Sky
(GHRSS; \citealt{bcm+16}) survey discovered 10\,pulsars at 322\,MHz in
the Southern sky ($-54\degr<\delta<-40\degr$). Targeted low-frequency
searches of unidentified {\it Fermi} gamma-ray sources have also
discovered 62 millisecond pulsars
\citep[e.g.][]{hrm+11,rap+12,cck+16,bph+17b,bph+18,pbh+17}.

In \citet{clh+14} we presented a pilot survey for pulsars and fast
transients using the Low-Frequency Array (LOFAR; \citealt{hwg+13}) and
its high-time-resolution `beam-formed' modes \citep{sha+11}.  We have
subsequently built on those observations and started the LOFAR
Tied-Array All-Sky Survey (LOTAAS).  Compared with other modern,
wide-field pulsar surveys, LOTAAS is novel because of its very low
observing frequency ($119-151$\,MHz) and long dwell time per pointing
(1\,hr).  These characteristics were instrumental in enabling the
LOTAAS discovery of 7 rotating radio transients \citep{mhl+18} and a
23.5-second pulsar, which is by far the slowest-spinning radio pulsar
known \citep{tbc+18}.

More broadly, the discoveries of the rotating radio transients
\citep[RRATs;][]{mll+06}, radio-emitting magnetars \citep{crh+06},
intermittent pulsars \citep{klo+06}, and the fast radio bursts
\citep[FRBs;][]{lbm+07,tsb+13,sch+14} have shown that radio-emitting
neutron stars have a host of emission properties and can sometimes be
very sporadic in their detectability.  These insights strongly
motivate pulsar and fast transient surveys that achieve a large
`on-sky' time, $\Sigma = N_\mathrm{obs} \Omega t_\mathrm{obs}$, here
defined as the product of the total number of survey observations,
$N_{\rm obs}$, field-of-view per pointing, $\Omega$, and dwell time
per pointing, $t_\mathrm{obs}$.  This makes LOTAAS complementary to
other ongoing, low-frequency surveys, like the GBNCC and AO327 drift
\citep{slr+14,dsm+13}, which have higher instantaneous sensitivity but
$\sim 30-60\times$ lower dwell time and instantaneous field-of-view:
$\Sigma_\mathrm{LOTAAS} \simeq 23400$\,h\,deg$^2$ at 135\,MHz,
$\Sigma_\mathrm{GBNCC} \simeq1430$\,h\,deg$^2$ at 350\,MHz above
$\delta>-40\degr$, and $\Sigma_\mathrm{AO327} \simeq 132$\,h\,deg$^2$
at 327\,MHz.

Here we present an overview of the LOTAAS survey and its first
discoveries.  In \S\ref{sec:survey_description} we describe the novel
observational setup of the survey, along with parameters such as
time/frequency resolution and sensitivity.  In \S\ref{sec:analysis}
and \S\ref{sec:results} we describe the search pipeline and the
results of the processing to date---in terms of pulsar discoveries and
redetections. These results are discussed in
\S\ref{sec:discussion}. Lastly, we conclude in
\S\ref{sec:conclusions}.

\begin{figure*}
  \includegraphics[width=\textwidth]{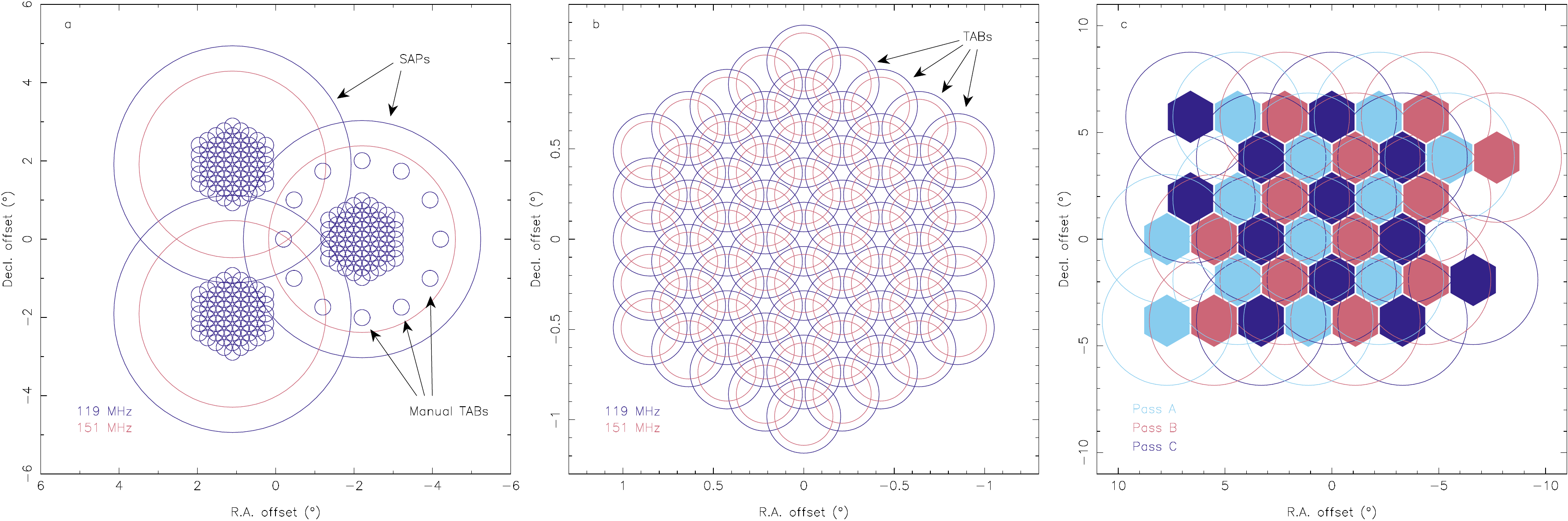}
  \caption{The beam setup of a LOTAAS pointing. \textit{(a)} A LOTAAS
    pointing consists of three sub-array pointings (SAPs) separated by
    $3\fdg82$ on the vertices of an equilateral triangle. Within each
    SAP, the LOFAR correlator and beamformer forms 61 coherently
    summed tied-array beams (TABs; small circles) and an incoherent
    beam (large circle). The TABs are hexagonally tiled to cover the
    centre of each SAP. Within each SAP an additional 12 TABs are
    formed, and can be pointed towards known pulsars (shown here only
    for one SAP). The SAP full-width at half maximum (FWHM) varies
    from 6\fdg1 to 4\fdg7 over the LOTAAS band (119\,MHz to
    151\,MHz). \textit{(b)} The hexagonal tiling of the 61 TABs within
    a single SAP. The TAB FWHM varies from 0\fdg41 to 0\fdg32 over the
    LOTAAS band (119\,MHz to 151\,MHz). Adjacent TABs overlap by 28\%
    at 119\,MHz and 14\% at 151\,MHz. \textit{(c)} LOTAAS pointings
    are tessellated in three passes, as indicated. Three passes are
    required to cover the sky with TABs, while each single pass covers
    the sky with incoherent beams from the SAPs.}
  \label{fig:pointing}
\end{figure*}

\section{Survey description}
\label{sec:survey_description}
\subsection{Array configuration and beamforming}
\label{ssec:array_conf}

The LOFAR Tied-Array All-Sky Survey (LOTAAS) is an ongoing survey of
the Northern sky for pulsars and transients with LOFAR. The survey
uses the high-band antennas (HBAs) on the Superterp, the dense central
part of the LOFAR core, since these provide the highest filling factor
of LOFAR stations in the array and hence the best balance of
field-of-view and raw sensitivity \citep{hwg+13,sha+11}. This compact
configuration also removes the need to compensate for differential
ionospheric phase delays between stations. The HBAs on the Superterp
are spread over 6 stations (CS002-007; \citealt{hwg+13}), each of
which has two sub-stations of 24 HBA tiles. Each HBA tile consists of
16 dual-polarisation antenna elements (dipoles) arranged in a
$4\times4$ pattern. Hence, the total number of dual-polarisation
dipoles on the Superterp is 4608.

LOTAAS uses LOFAR's three stages of beamforming to create tied-array
beams for the HBA antennas \citep{hwg+13,sha+11}. First, the analogue
beamformer of each HBA tile forms a \textit{tile beam} out of the
signals from the 16 antenna elements within a tile. After digitization
and coarse channelization of the signals, these tile-beams are
digitally beamformed into \textit{station beams}. The final stage of
beamforming is performed by the central LOFAR correlator and
beamformer, which combines the signals of 12 HBA sub-stations into
\textit{tied-array beams} (see \citealt{sha+11} for details). At the
beginning of the LOTAAS survey in December\,2012 a CPU-based IBM Blue
Gene/P was used as the central LOFAR correlator and beamformer
\citep{mr11}. In May\,2014 its functionality was transferred to
\textsc{Cobalt}, a GPU-based correlator and beamformer \citep{bmn+18}.

For the LOTAAS band of 119\,MHz to 151\,MHz (see
\S\,\ref{ssec:bandwidth_and_sampling}), the Superterp HBA station
beams vary in full-width at half maximum (FWHM) at zenith from 6\fdg1
at 119\,MHz to 4\fdg7 at 151\,MHz. Tied-array beams from the 12 HBA
sub-stations of the Superterp yield beams for which the FWHM varies
with frequency between 0\fdg41 (119\,MHz) to 0\fdg32 (151\,MHz) at the
zenith. Both the station and tied-array beams will become elongated in
zenith angle when pointing away from the zenith, as the baselines
between tiles in a station and the baselines between stations
foreshorten.

\subsection{Beam setup and tessellation}
\label{ssec:beam_setup_and_tessellation}

To maximise the number of tied-array beams as well as their
sensitivity within station beams (sensitivity is higher towards the
centre), LOTAAS forms three station beams (hereafter called sub-array
pointings, or SAPs), each of which is tessellated by 61 tied-array
beams (TABs). The TABs fill the central region of each SAP with a
central TAB surrounded by four hexagonally filled TAB rings. The TAB
rings are spaced at $0\fdg245$ ($14\farcm7$) separation, where the
FWHM of one TAB overlaps for 28\% and 14\% with adjacent TABs at
119\,MHz and 151\,MHz, respectively. This overlap provides
pseudo-Nyquist sampling of the sky. The same is true for the SAPs,
which are separated by $3\fdg82$ at the vertices of an equilateral
triangle. Figure\,\ref{fig:pointing} shows the layout of the SAPs and
TABs.

Furthermore, an additional 12 TABs are formed within each SAP. These
are pointed towards known pulsars that happen to coincide with the SAP
field-of-view, or at predetermined positions in the absence of known
sources. Finally, the signals from the 12 HBA sub-stations per SAP are
summed incoherently \citep{sha+11,bmn+18} to form an incoherent beam
with a field-of-view equivalent to that of the station beam. Hence, a
total of 222 beams are formed for each LOTAAS pointing. At zenith, the
hexagonally tiled TABs of the three SAPs of a single pointing cover
approximately 12 square degrees, while the incoherent beams cover 67
square degrees.

The setup of the tied-array and incoherent LOTAAS beams allows
pointings to be efficiently tessellated to cover the entire sky. To
fill in the gaps between the tied-array beams, three interleaved
pointings are required to fully sample the sky, as depicted in
Fig.\,\ref{fig:pointing} and \ref{fig:coverage}. The large
field-of-view of the incoherent beams has the advantage that the three
pointings can be performed in three sequential passes, such that for
each pass the incoherent beams completely cover the sky. As a result,
each sky position is observed once with a coherently summed TAB and
three times with an incoherently summed station beam. A total of 1953
individual pointings, 651 per pass, are required to cover the sky
above a declination of $\delta>0\degr$. At the time of writing,
January 2019, all pointings of Pass A and B have been obtained, and 43
pointings remain to be observed for Pass C. Initial processing of Pass
A and B has been completed using a `Version 1.0' search pipeline, and
the analysis of the pass\,C pointings is ongoing.

\subsection{Bandwidth and sampling}
\label{ssec:bandwidth_and_sampling}

LOFAR stations are capable of digitizing dual-polarisation signals at
8-bit resolution for up to 488 sub-bands. Station beams can be formed
for individual or groups of sub-bands. By choosing three station
beams, LOTAAS can record data at a maximum of 162 sub-bands for each
station beam. To optimise the LOTAAS sensitivity and TAB field-of-view
we use the bottom part of the HBA band where LOFAR is most sensitive
\citep{hwg+13}. The 162 sub-bands of 195.3125\,kHz bandwidth are
centred at 135.25\,MHz with a total bandwidth of 31.64\,MHz. The
central LOFAR correlator and beamformer uses a polyphase filter to
channelise each sub-band to 16 channels, providing a total of 2592
channels over the LOTAAS band. The signals from the two linear
polarisations are summed in quadrature to form total intensity (Stokes
I). To comply with the network limitations between the
correlator/beamformer and the central processing cluster, the data are
decimated by a factor of 6 in time, to a time resolution of
491.52\,$\upmu$s. Finally, time/frequency samples, represented by
32-bit floating point values, are streamed to the central processing
cluster at a data rate of 37.46\,Gb\,s$^{-1}$.

The large instantaneous field-of-view of a single LOTAAS pointing
allows for long dwell times. Each LOTAAS pointing has a 1-hr
integration time, compared to the few-minute integration times used by
other ongoing wide-field surveys. The long integration times increase
the probability of discovering RRATs, intermittent pulsars and
transient signals. With 1\,hr integration times, the data volume of a
single LOTAAS pointing is 16.9\,TB (with 32-bit samples).

\subsection{Survey sensitivity}
\label{ssec:survey_sensitivity}

\begin{figure*}[!ht]
  \includegraphics[width=\columnwidth]{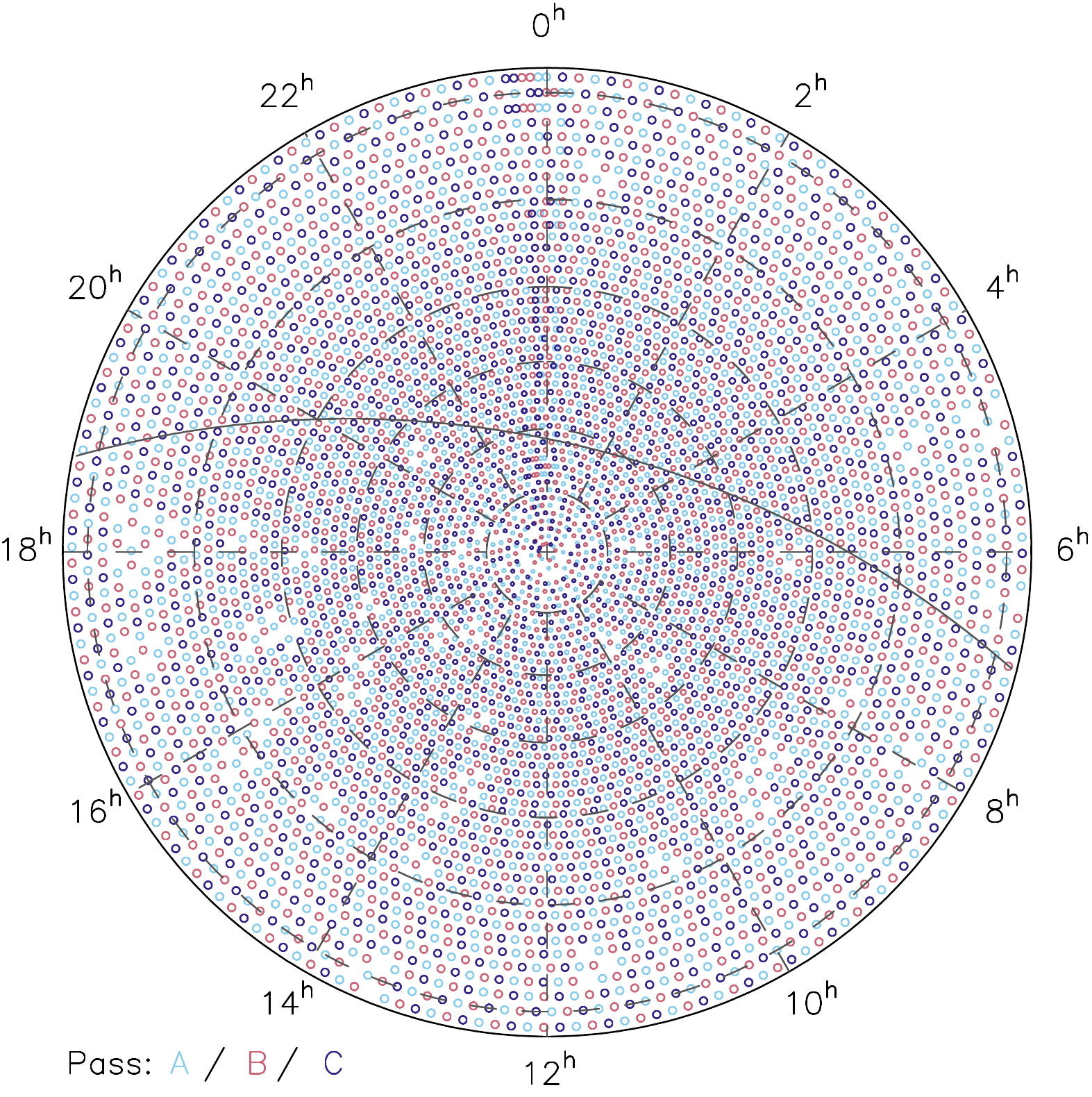}
  \includegraphics[width=\columnwidth]{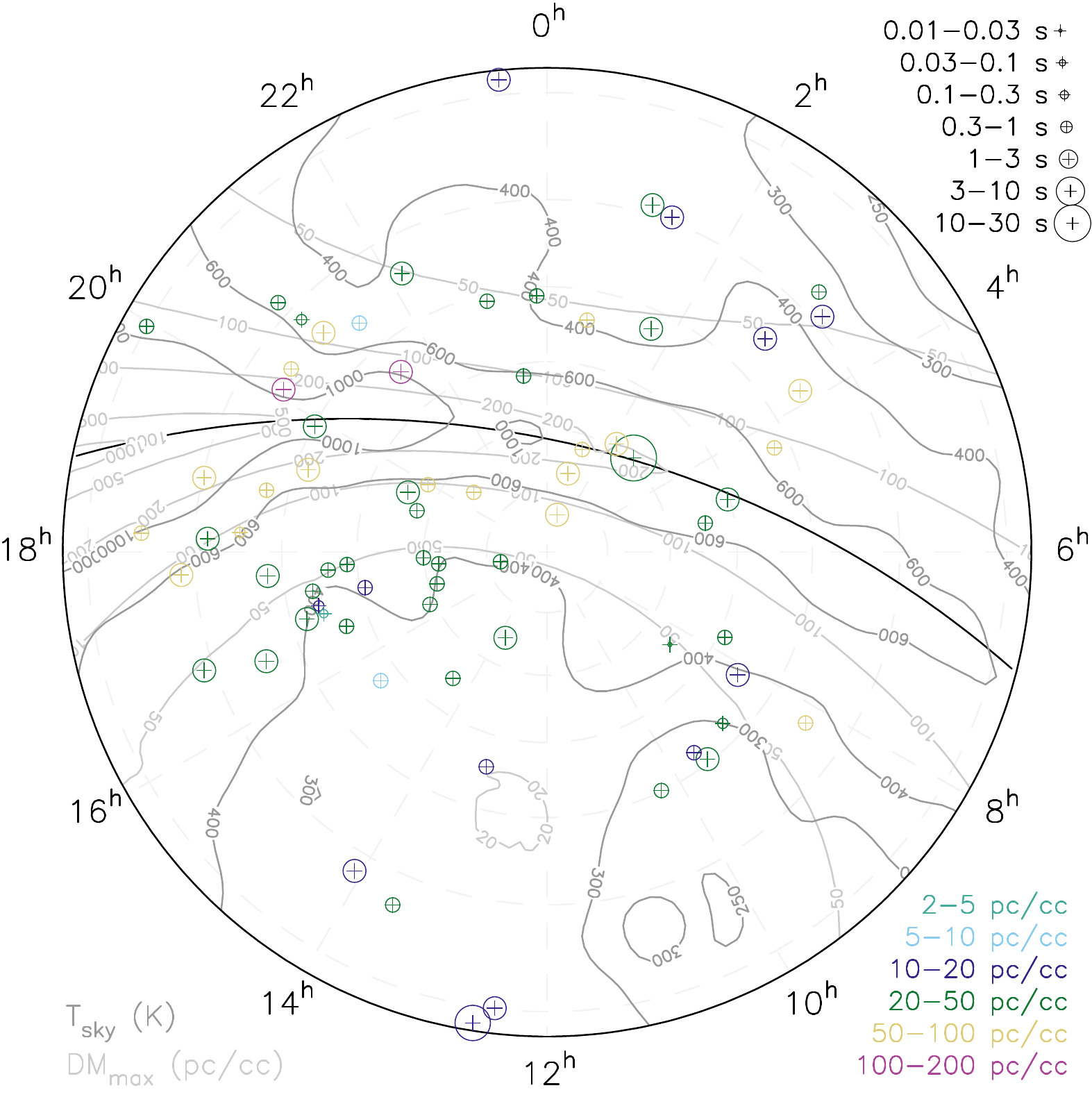}
  \caption{Stereographic projections of the Northern hemisphere in
    equatorial coordinates ($\delta>-3\degr$). Lines of equal right
    ascension and declination are indicated with dashed gray lines
    ($2^\mathrm{h}$ steps in R.A.\ and $15\degr$ in decl., from
    $0\degr$ to $75\degr$). The Galactic plane is shown as the black
    curve.  \textit{(left)} Observed LOTAAS pointings as of January
    2019, following the LOTAAS tessellation scheme. The sky coverage
    of the incoherent beams is shown for the three passes with
    different colours. Note that beamsizes have been scaled down for
    readability of the plot. \textit{(right)} The sky location of the
    first 73 LOTAAS pulsar discoveries. The period $P$ and dispersion
    measure $\mathrm{DM}$ of the pulsars are indicated with different
    symbols and colours. Contours show the sky temperature
    $T_\mathrm{sky}$ at 135\,MHz, as extrapolated from the
    \citet{hssw82} 408\,MHz map, as well as the maximum predicted
    $\mathrm{DM}$ within our Galaxy by the \citet{cl02} NE2001 model.}
  \label{fig:coverage}
\end{figure*}
Following \citet{dtws85}, the sensitivity of the LOTAAS survey for a
pulsar with period $P$ and effective pulse width $W_\mathrm{eff}$ down
to a minimum signal-to-noise $\mathrm{S}/\mathrm{N}_\mathrm{min}$,
relates to the gain $G$ and system temperature $T_\mathrm{sys}$ of the
telescope, and the integration time $t_\mathrm{obs}$, bandwidth
$\Delta \nu$ and number of summed polarisations $n_\mathrm{p}$ of the
observation through
\begin{equation}
S_\mathrm{min} = \frac{\mathrm{S}/\mathrm{N}_\mathrm{min} T_\mathrm{sys}}{G\sqrt{n_\mathrm{p} t_\mathrm{obs} \Delta \nu}} \sqrt{\frac{W_\mathrm{eff}}{P-W_\mathrm{eff}}}.\nonumber
\end{equation}

The gain $G$ depends on the effective area $A_\mathrm{eff}$ of the
telescope ($G=\frac{1}{2} A_\mathrm{eff}/k_\mathrm{b}$, with
$k_\mathrm{b}$ the Boltzmann constant). For LOFAR, the number of
stations and active dipoles defines the effective area, as well as the
coherency of the tied-array beamforming. Following \citet{hwg+13}, the
effective area for a HBA sub-station of 16 dipoles per tile and 24
tiles per sub-station at the central LOTAAS frequency of 135\,MHz
($\lambda=2.2$\,m) is $A_\mathrm{eff}=600$\,m$^2$. \citet{kvh+16}
finds that the coherent summation of $N$ stations scales as
$N^{0.85}$, and typically, 5\% of dipoles are not in operation. Hence,
the effective area of the LOFAR Superterp is
$(1-0.05)12^{0.85}600\,\mathrm{m}^2=4712$\,m$^2$. This yields a gain
of $G=1.7$\,K\,Jy$^{-1}$ at zenith.

The system temperature $T_\mathrm{sys}$ is the sum of the sky
temperature $T_\mathrm{sky}$ and the receiver or antenna temperature
$T_\mathrm{ant}$. The latter varies over the LOTAAS observing band,
with a minimum value of 330\,K at 125\,MHz to 390\,K at 151\,MHz; we
use the mean $T_\mathrm{ant}$ of 360\,K \citep{kvh+16}. At the
observing frequencies of LOTAAS, the system temperature
$T_\mathrm{sys}$ can be dominated by the sky temperature
$T_\mathrm{sky}$. We use the reference sky temperature map at 408\,MHz
from \citet{hssw82}, and the $\lambda^{2.55}$ scaling law by
\citet{lmop87} to estimate the sky temperature at 135\,MHz. The
all-sky averaged sky temperature at 135\,MHz is
$T_\mathrm{sky}=510$\,K, but can be as low as 200\,K at high Galactic
latitudes to $\ga1000$\,K in regions at low Galactic latitude towards
the Galactic centre (see Fig.\,\ref{fig:coverage}). For determining
the sensitivity limits we use the best-case system temperature of
$T_\mathrm{sys}=T_\mathrm{ant}+T_\mathrm{sky}=560$\,K, as well as the
all-sky averaged value of $T_\mathrm{sys}=870$\,K.

The intrinsic pulse width is broadened by effects due to the
propagation of the pulse through the interstellar medium, depending on
the way these are corrected for in the analysis. As defined in
\citet{lk12}, the total temporal smearing $\tau_\mathrm{tot}$ is the
quadratic sum of the finite sampling time $\tau_\mathrm{samp}$, the
dispersive smearing within a single channel $\tau_\mathrm{chan}$, the
dispersive smearing across the full bandwidth due to the finite steps
in trial $\mathrm{DM}$ $\tau_\mathrm{BW}$, and the dispersive smearing
due to the piece-wise linear approximation of the quadratic dispersion
law in the sub-band dedispersion algorithm \citep{mla+96}
$\tau_\mathrm{sub}$.

\begin{figure}
  \includegraphics[width=\columnwidth]{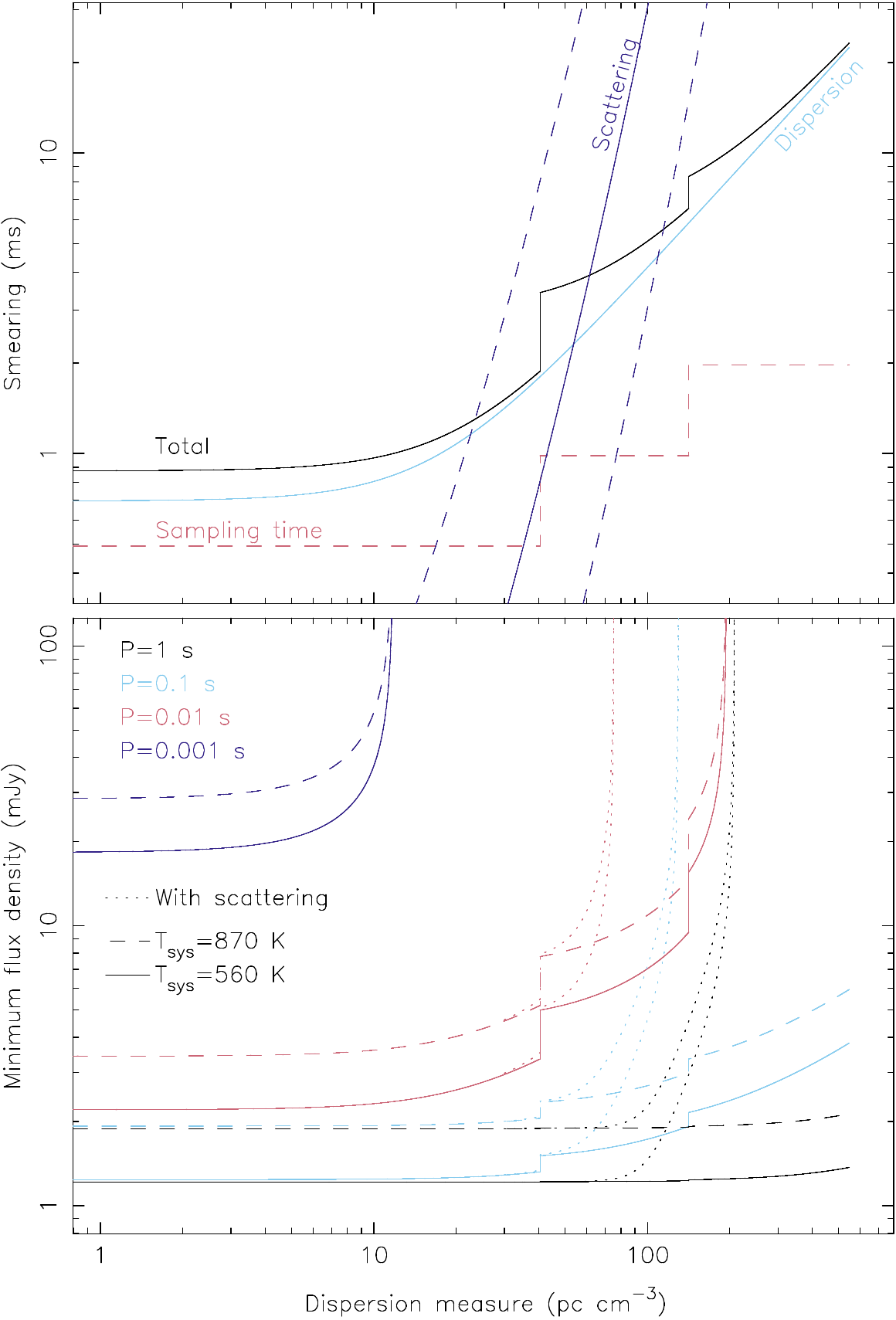}
  \caption{\textit{(top)} Pulse broadening due to the finite sampling
    time, dispersive smearing due to the incoherent dedispersion
    algorithm used, and the effects of scattering using the scattering
    relation of \citet{bcc+04}. The diagonal dashed lines denote a one
    order-of-magnitude larger or smaller range in predicted
    scattering. \textit{(bottom)} The minimum detectable flux density
    for the LOTAAS survey as a function of $\mathrm{DM}$. Sensitivity
    limits are plotted for different pulse periods $P$, and different
    system temperatures $T_\mathrm{sys}$. The effect of pulse
    broadening due to interstellar scattering is shown by the dotted
    lines.}
  \label{fig:sensitivity}
\end{figure}

The top panel of Fig.\,\ref{fig:sensitivity} shows the dispersive
smearing as a function of $\mathrm{DM}$. Using the dedispersion plan
described in \S\ref{ss:dedispersion}, the dispersive smearing at low
DM in LOTAAS is 0.9\,ms, but rises to 20\,ms at the highest-searched DM of
550\,pc\,cm$^{-3}$. Besides dispersive smearing,
the pulsar pulses will also be smeared due to scattering in the
interstellar medium. Whereas the effects of dispersion can largely be
mitigated by dedispersion, scattering can not be corrected for in a
blind search. Figure\,\ref{fig:sensitivity} also shows the smearing
due to scattering as a function of DM, as predicted by the empirical
relation from \citet{bcc+04}. Observationally, the smearing due to
scattering for pulsars at the same DM but different location shows
variations up to 2 or 3 orders of magnitude, and hence we plot a range
of possible scattering variations. Based on this range, we expect
smearing due to interstellar scattering to become dominant over
dispersive smearing for DMs in the range of $20-100$\,pc\,cm$^{-3}$.

The sensitivity limit of the LOTAAS survey is shown in the bottom
panel of Fig.\,\ref{fig:sensitivity} as a function of dispersion
measure for different pulse periods $P$ and system temperature
$T_\mathrm{sys}$ in the presence (or absence) of interstellar
scattering. All these curves assume an intrinsic pulse width that is
3\% of the pulse period, and a minimum detection significance of
$\mathrm{S}/\mathrm{N}_\mathrm{min}=10$. These curves indicate that
LOTAAS has a best sensitivity of about 1.2\,mJy for nearby slow
pulsars at high Galactic latitude ($\mathrm{DM}<50$\,pc\,cm$^{-3}$,
$P>0.1$\,s and $T_\mathrm{sys}=560$\,K). Sensitivity to millisecond
pulsars with $P<0.01$\,s is limited by the coarse sampling of
491.52\,$\upmu$s, though theoretically they could be detected if they
are at low $\mathrm{DM}$ and relatively bright. In the presence of
interstellar scattering following the \citet{bcc+04} predictions,
LOTAAS would not be able to detect pulsars with $\mathrm{DM}$s above
200\,pc\,cm$^{-3}$, regardless of their spin period.

\subsection{Confirmation and follow-up observations}
\label{ssec:confirmation_observations}

\begin{figure}
  \includegraphics[width=\columnwidth]{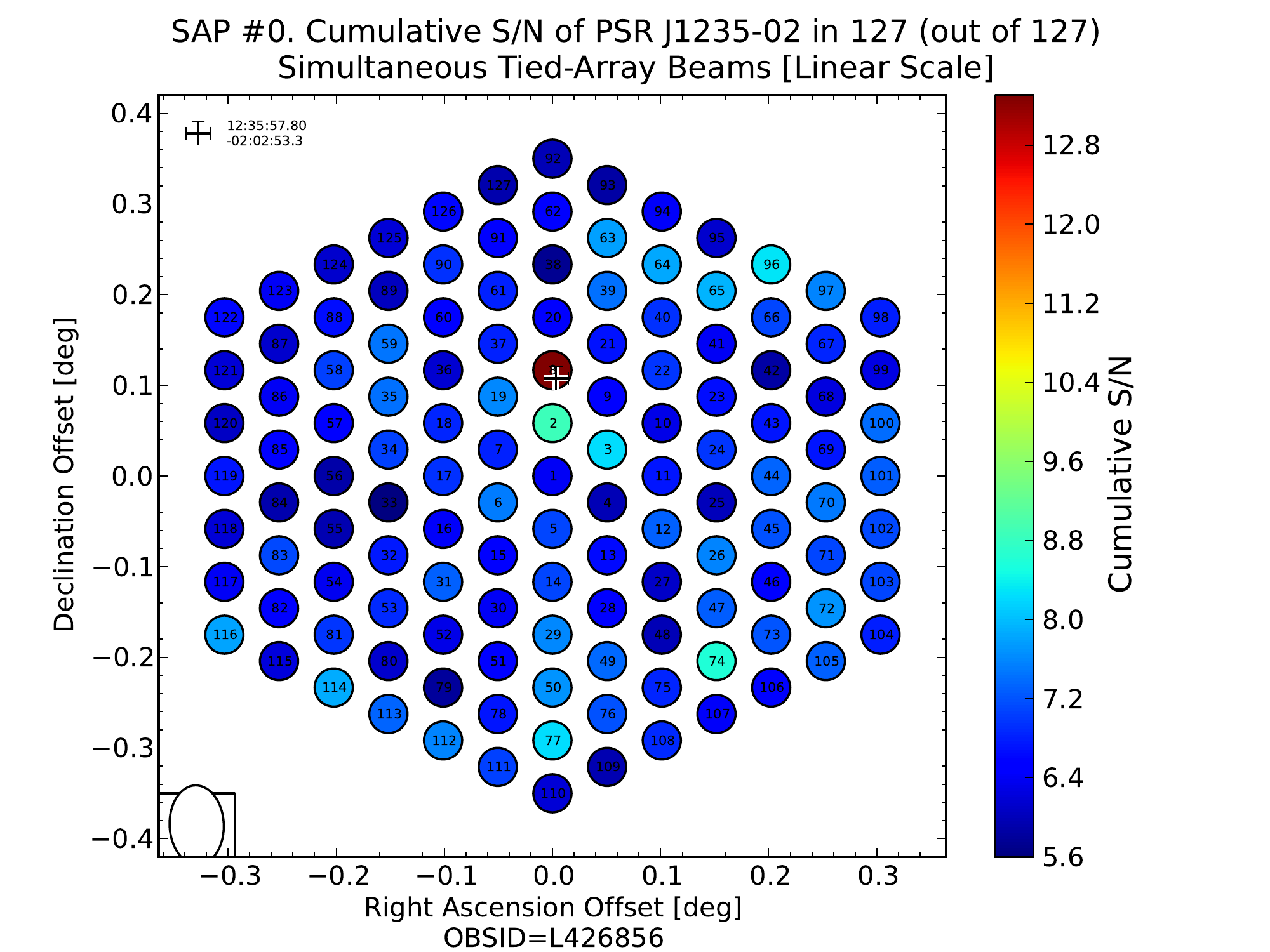}
  \includegraphics[width=\columnwidth]{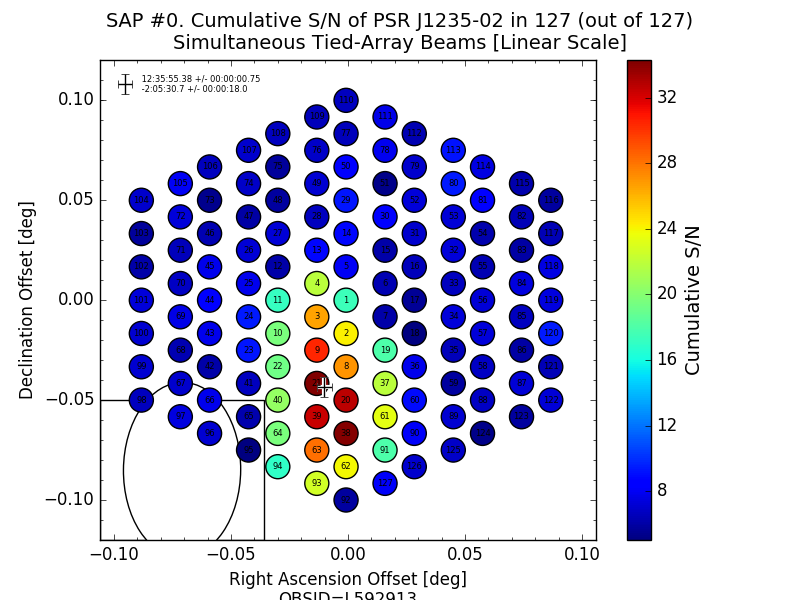}
  \caption{Tied-array beam (TAB) locations for confirmation
    observations \textit{(top)} and localization observations
    \textit{(bottom)}. In these observations a single sub-array
    pointing is tiled with 127 TABs, where the colour indicates the
    cumulative signal-to-noise of the pulse profile in each beam. Note
    the change of scale between the two plots, denoted by the
    approximate TAB beamsize depicted in the lower left of each
    panel. In the localization observations the TABs overlap
    significantly to obtain a signal-to-noise weighted localization of
    the pulsar.}
  \label{fig:localization}
\end{figure}

The different observing modes and station locations of LOFAR allow for
flexibility when confirming pulsar candidates and performing follow-up
observations. For confirming pulsar candidates, LOTAAS uses the HBA
dipoles of the 24 LOFAR core stations with 15-min exposures, yielding
a factor of approximately 2 increase in sensitivity over the 6
stations of the Superterp using 1-hr observations. The maximum
baseline between core stations is 3.5\,km, compared to 300\,m for the
stations on the Superterp, and hence the TABs using the core stations
have a FWHM of $3\farcm5$, a factor 7 smaller than Superterp TABs. To
tile out the Superterp discovery beam, we form a single sub-array
pointing and use 127 TABs in a hexagonal
pattern. Depending on the accuracy of the discovery localization, the
spacing between the TABs in confirmation observations can
be decreased from Nyquist sampling to improve the localization by
comparing the pulsar profile signal-to-noise between different
beams. This approach allows the pulsar to be localised to an accuracy
of about $3\arcmin$. Examples of the tied-array beam tiling in
confirmation and localization observations are shown in
Fig.\,\ref{fig:localization}.

Besides the larger number of stations and shorter integration time,
confirmation observations use the same observational setup as the
search observations. Hence, 162 sub-bands cover 31.64\,MHz of bandwidth
over the same frequency range, channelised to 2592 channels, with
polarisations summed to form Stokes\,I and downsampled to a sampling
time of 491.52\,$\upmu$s.

Once a newly discovered pulsar is confirmed and localised, the timing
programme is started. The follow-up timing observations again use all
the HBA dipoles of the 24 LOFAR core stations, but now only a single
tied-array beam is formed. For this tied-array beam, dual-polarisation
complex voltages, sampled at the 5.12\,$\upmu$s Nyquist rate of a
195.3125\,kHz sub-band are recorded for 400 sub-bands. Hence,
78.125\,MHz of bandwidth is recorded between frequencies of 110\,MHz
to 188\,MHz. The complex voltages allow phase-coherent dedispersion
removal with \textsc{dspsr} \citep{sb10}. Subsequent timing analysis
is performed with \textsc{psrchive} \citep{hsm04} and \textsc{tempo2}
\citep{hem06,ehm06}.

To constrain their radio spectra, newly discovered pulsars are also
observed using the 76-m Lovell telescope at Jodrell Bank at 1532\,MHz
and occasionally at 330\,MHz. For these observations the ROACH backend
\citep{bjk+16} is used, providing 400\,MHz of bandwidth centred at
1532\,MHz and 64\,MHz of bandwidth at 330\,MHz. Pulsars that are
visible at 1532\,MHz are observed as part of the regular pulsar timing
programme of the Lovell telescope. For some pulsars, follow-up
observations at 1.4\,GHz were also obtained with the Nan\c cay Radio
Telescope \citep{gsl+16, cfg+17}.

\begin{figure}
  \includegraphics[width=\columnwidth]{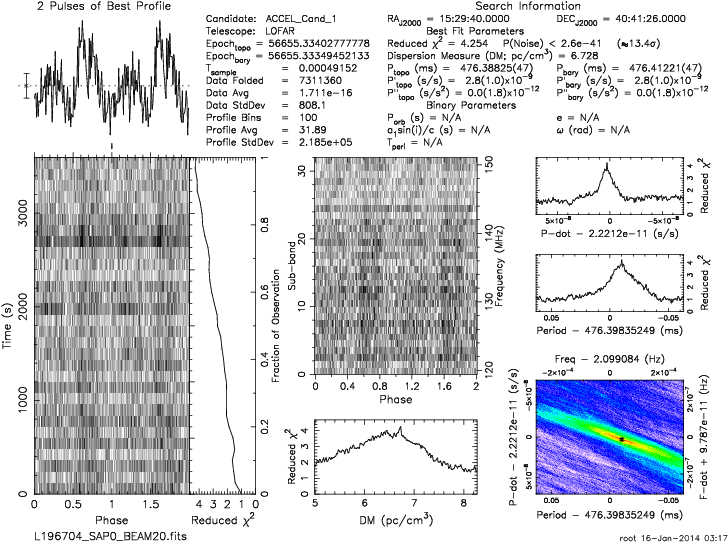}
  \includegraphics[width=\columnwidth]{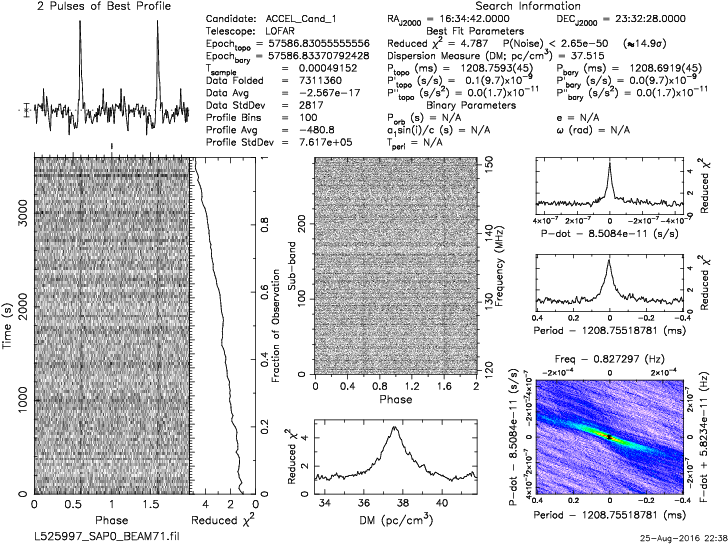}
  \includegraphics[width=\columnwidth]{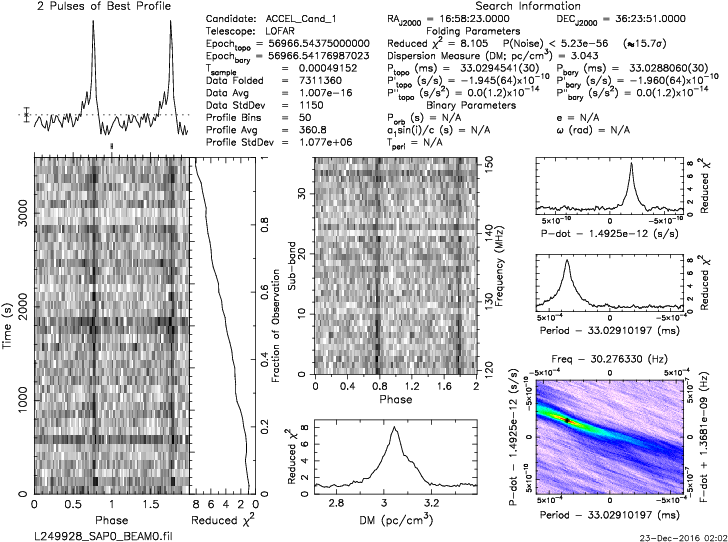}
  \caption{Examples of standard \textsc{presto} diagnostic plots for
    periodic pulsar candidates. These plots are for LOTAAS discovered
    pulsars J1529$+$40, J1635$+$23 and the binary MSP J1658$+$36.}
  \label{fig:discovery_plots}
\end{figure}

\section{Analysis}
\label{sec:analysis}

In terms of data volume, the LOTAAS survey is the largest pulsar
survey performed to date, as the 1953 LOTAAS pointings produce 8\,PB
of raw 8-bit archived data. The management and the processing of such
a huge amount of data is not a trivial task, and can pose a challenge
for even the largest high performance computing (HPC) facilities
available worldwide. For this reason, although at the beginning of the
survey the HPC cluster of the Jodrell Bank pulsar group was used
(\texttt{HYDRUS}; 552 CPU cores, 728\,GB RAM), the main workhorse of
the survey's processing since 2014 has been the Dutch National
Supercomputer
\texttt{Cartesius}\footnote{\url{https://userinfo.surfsara.nl/systems/cartesius}}
(over 44500\,CPU cores, 115\,TB RAM), managed by
\texttt{SURFSara}\footnote{\url{https://www.surf.nl/en/about-surf/subsidiaries/surfsara}}. The
storage of the raw data and the processed results uses the
\texttt{SARA} Long Term Archive (LTA), where a significant fraction of the LOFAR
LTA is also hosted. The co-locality of the \texttt{SARA} LTA and
\texttt{Cartesius} provides increased efficiency, required for various
reprocessing runs of the stored data, with transfer speeds over
1\,GB\,s$^{-1}$.

The processing time required to reduce that amount of raw data is also
immense. For the first processing runs, $10\mathrm{M}$ CPU-hours were
requested in 2013 in order to streamline the LOTAAS pipeline and get
first results. This was followed by two $25\mathrm{M}$ CPU-hours
requests, in 2015 and 2017 respectively.  These $60\mathrm{M}$
CPU-hours have been granted by the Netherlands Organisation for
Scientific Research (NWO) via proposal submission. Out of these
$50\mathrm{M}$ CPU-hours, $30\mathrm{M}$ have been used to process all
acquired data at the time of writing (January 2019), and reprocess the
early LOTAAS data (acquired before May 2015), which had been processed
only with the prototype LOTAAS pipeline.

The LOTAAS processing pipeline is based on the
\textsc{presto}\footnote{\url{https://www.cv.nrao.edu/~sransom/presto}}
\citep{ran01} pulsar search software suite, with additional code
written by members of the LOTAAS group for single-pulse searches
\citep{mhl+18} and candidate classification
\citep{lsc+16,tls+18}. Currently it is a purely CPU-based pipeline;
however, in the near future some parts will be replaced by GPU
implementations. Here we present a detailed breakdown of the
LOTAAS pipeline.

\subsection{Pre-processing}

The 16.9\,TB of raw data for each 1-hr LOTAAS pointing are stored in
\texttt{HDF5}
format\footnote{\url{https://github.com/nextgen-astrodata/DAL}} on the
central processing cluster. Here, the LOFAR Pulsar Pipeline (PuLP;
\citealt{kvh+16}) requantises the raw data from 32-bit floating point
values to scaled and offset 8-bit integers, while also storing the
scales and offsets. This step reduces the data volume by a factor of
4. The requantised output is stored in \texttt{PSRFITS} format
\citep{hsm04}, yielding an 18\,GB file for each of the 222 beams. The
pipeline then identifies and creates radio frequency interference
(RFI) masks, and performs a series of tests concerning the data
quality (i.e.\ RFI statistics, packet loss in the correlator and
beamformer). Finally, the pipeline folds the data for the TABs that
contain known radio pulsars. The requantised \texttt{PSRFITS} files,
RFI masks, and associated metadata are then ingested into the
LTA. Observations for which the data quality tests fail are marked for
re-observation at the end of the survey. Once the data have been
validated to be on the LTA, they become available for downloading to
\texttt{Cartesius}.

On the \texttt{Cartesius} supercomputer, the \texttt{PSRFITS} files
are converted to \texttt{SIGPROC} filterbank format using an adapted
version of \textsc{presto}'s \texttt{psrfits2fil.py}. This version was
tailored to LOTAAS data in order to control the desired block size
(the minimum amount of data that can be processed in \textsc{presto}),
which was set to 512 spectral channels per sub-integration. This
particular block size was selected in order to minimise the effects of
dispersive smearing, which lowers the detection signal-to-noise for
pulsars at higher $\mathrm{DM}$s. Subsequently, new RFI masks are created
using 2.5\,s time integrations and the default settings of
\texttt{rfifind} (for more details, see \citealt{coo17}). The size of
these time segments was selected based on the typical RFI properties,
in order to avoid excessive data masking. Typically, the fraction of
data flagged as RFI is 10\% for the TABs, and 15\% for the
incoherently summed station beams.

\subsection{Dedispersion}
\label{ss:dedispersion}

The next step is the creation of the dedispersed time series for each
beam. The DM range $0-546.5$\,pc\,cm$^{-3}$ is searched; higher DMs
are not searched because of the significant dispersion smearing and
scattering at these low observing frequencies. At a frequency of
135\,MHz, which is the centre frequency of the survey, the
intra-channel dispersion smearing at a $\mathrm{DM}=50$\,pc\,cm$^{-3}$
is $\sim 2$\,ms, whereas for $\mathrm{DM}>100$\,pc\,cm$^{-3}$
scattering likely dominates (see Fig.\,\ref{fig:sensitivity}). The
dedispersion plan was created using the \textsc{presto}
\texttt{DDplan.py} tool and consists of 10120 DM-trials (Table
\ref{tab:ddplan}). This dedispersion plan is aimed at minimizing the
dispersive smearing over the selected DM range, while optimizing the
processing efficiency when using parallelised dedispersion
routines. The \texttt{mpiprepsubband} tool from \textsc{presto} is
used to create incoherently dedispersed time-series from the RFI
masked filterbank files. This tool uses the sub-band dedispersion
technique, which uses a piece-wise linear approximation to the
quadratic dispersion law.

\begin{table}[ht]
  \caption{The dedispersion plan used by the LOTAAS survey, showing
    the ranges in dispersion measure between DM$_\mathrm{min}$ and
    DM$_\mathrm{max}$, with a DM step-size of $\delta$DM, yielding
    $N_\mathrm{DM}$ dispersion measure trials. The downsampling factor
    is denoted by $d$, and indicates by what factor the temporal
    resolution is averaged to yield the sampling time
    $t_\mathrm{samp}$.\label{tab:ddplan}} \centering
  \begin{tabular}{rrrrrr}
    \hline\hline
    \multicolumn{1}{l}{$\mathrm{DM}_\mathrm{min}$} &
    \multicolumn{1}{l}{$\mathrm{DM}_\mathrm{max}$} &
    \multicolumn{1}{l}{$\delta\mathrm{DM}$} &
    \multicolumn{1}{l}{$N_\mathrm{DM}$} &
    \multicolumn{1}{l}{$d$} &
    \multicolumn{1}{l}{$t_\mathrm{samp}$} \\
    \multicolumn{1}{l}{(pc\,cm$^{-3}$)} &
    \multicolumn{1}{l}{(pc\,cm$^{-3}$)} &
    \multicolumn{1}{l}{(pc\,cm$^{-3}$)}
    & & &
    \multicolumn{1}{l}{(ms)} \\
    \hline
      0.00  & 40.48 & 0.01 & 4048 & 1 & 0.49152 \\
     40.48 & 141.68 & 0.05 & 2024 & 2 & 0.98304 \\
    141.68 & 546.48 & 0.10 & 4048 & 4 & 1.96605 \\
    \hline 
  \end{tabular}
\end{table}

\subsection{Periodicity search}

In order to look for periodic signals within the dedispersed time
series, their power spectra are computed by applying a discrete fast
Fourier transform (FFT) using the \texttt{realfft} tool. In order to
improve our sensitivity to long-period pulsars, a red noise removal
procedure is also applied to the power spectra using
\texttt{rednoise}. The whitened power spectra are now searched for
periodicities using \texttt{accelsearch} \citep{rem02}, which detects
the most significant periodic signals and uses harmonic summing to
recover the power at multiples of the spin frequency. No acceleration
searches are currently performed; \texttt{accelsearch} only searches
at zero acceleration, due to the significant processing time that
would otherwise be required.  LOTAAS is thus not sensitive to pulsars
in tight binary systems, unless these are fortuitously observed at a
preferential, low-acceleration orbital phase. If a frequency has
spectral significance in excess of $2\sigma$, it is marked as a
candidate and the corresponding harmonics up to the 16th are summed to
increase the detection significance.

Once the list of candidates for all the DM-trials has been compiled, a
sifting procedure to reduce the number of candidates by selecting only
the most likely is performed. The sifting strategy in place is the one
developed for the LOTAS pilot survey \citep{clh+14} and is described
in detail in \cite{coe13}. Candidates with $P < 2$\,ms or $P > 15$\,s,
along with candidates with $\mathrm{DM} < 0.5$\,pc\,cm$^{-3}$ are
rejected. We note that PSR\,J0250+5854, which has $P=23.5$\,s
\citep{tbc+18}, was discovered through its $P<15$\,s harmonics. Also
rejected are those candidates that are not detected in
neighbouring DM trials.  Candidates with similar DMs
and harmonically related periods are grouped, and only the instance with the
highest signal-to-noise is kept. From the significantly reduced
candidate list, only those with $5\sigma$ detections are folded.

\begin{figure*}
  \includegraphics[width=\textwidth]{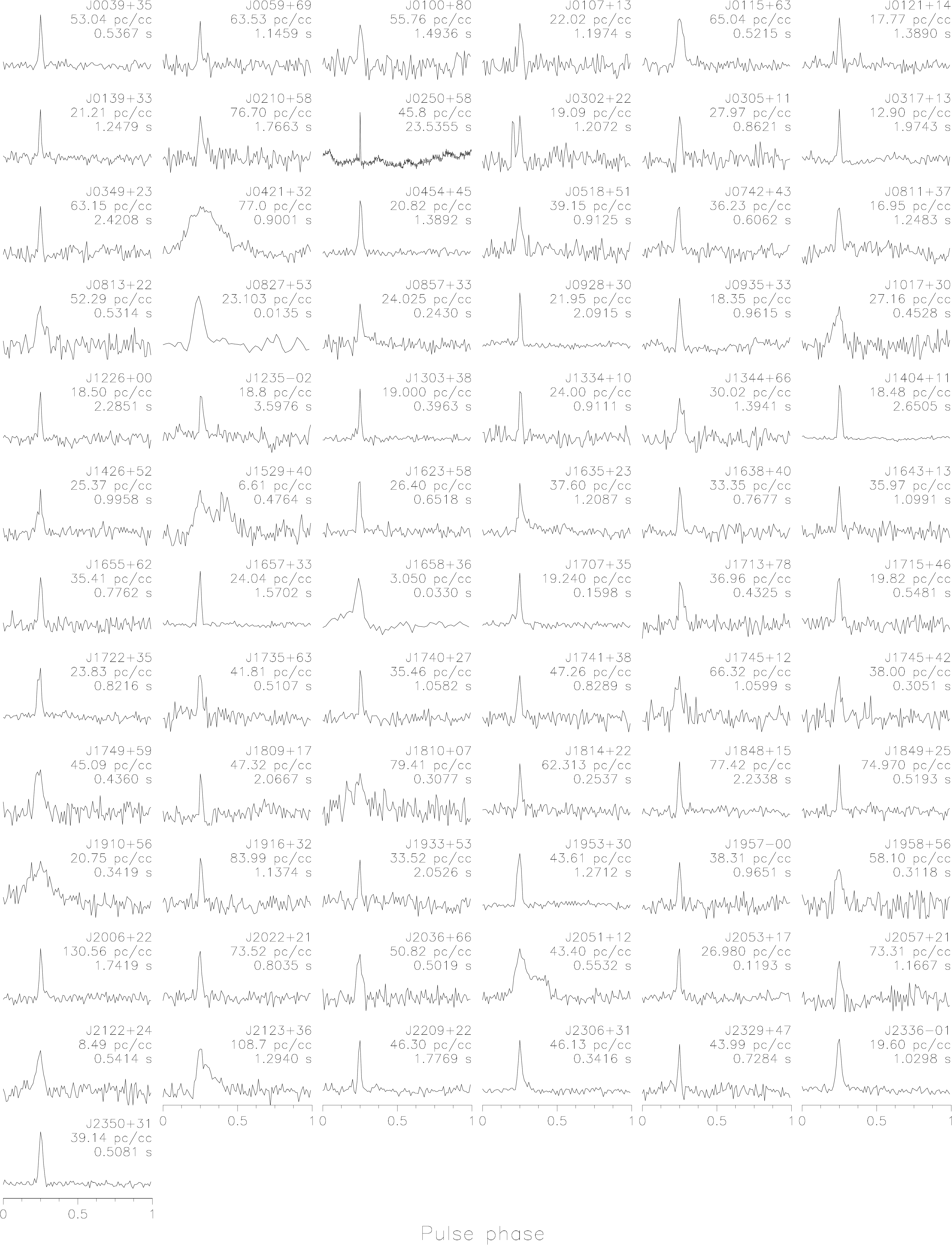}
  \caption{Pulse profiles of the LOTAAS pulsar discoveries. Shown here
    are the discovery profiles using 100 pulse profile bins. The pulse
    profiles are rotated to place the peak at pulse phase
    $\phi=0.25$. The pulsar name, dispersion measure (in
    pc\,cm$^{-3}$) and spin period (in seconds) are quoted for each
    pulsar.}
  \label{fig:profiles}
\end{figure*}

\subsection{Candidate folding}

Folding is performed using the \texttt{prepfold} tool, which creates
candidate files and diagnostic plots such as those shown in
Fig.\,\ref{fig:discovery_plots} for further inspection. In the folding
analysis we use 100 pulse phase bins (50 for $P<50$\,ms), 288
sub-bands, 40 sub-integrations, and a running mean subtraction applied
on the input filterbank file. The latter assisted in the discovery of
PSR\,J0317+13, the observation of which was characterised by strong
broadband and narrowband RFI, but had as a side effect the creation of
artefacts during the folding: in other words, signal `dips' around the
profiles of bright sources with narrow profile widths, which changes
the detection significance (see e.g.\ PSR\,J1635+23 in
Fig.\,\ref{fig:discovery_plots}).

\subsection{Candidate classification}

Each pointing generates approximately 20000 candidates, and on
completion the survey will have produced approximately 40 million
candidates. Since visual inspection of all the candidates would be
impossible, the machine learning classifiers described in
\citet{lsc+16} and \citet{tls+18} are used to reduce the candidate
list to a reasonable number for visual inspection. These classifiers
use the statistics (mean, variance, skewness and kurtosis) of the
pulse profile and the DM curve (see Fig.\,\ref{fig:discovery_plots};
signal-to-noise as a function of DM). The \citet{tls+18} classifier
expands on this approach by calculating the correlation coefficient
between each sub-band and the profile, as well as the correlation
coefficient between each sub-integration and the profile. The
classifier uses the statistics of correlation coefficient
distributions, in addition to the statistics of the profile and DM
curve, to automatically classify periodicity candidates. This approach
reduces the number of candidates per pointing to $\sim220$, i.e.\ one
per beam on average.

\subsection{Single-pulse search}

A search for bright impulsive signals in the dedispersed time series
is performed using \texttt{single\_pulse\_search.py} from
\textsc{presto}. The script convolves box-car functions of variable
widths between $0.5-100$\,ms with the time series at different DMs.  A
value for the signal-to-noise is calculated for each bin of the
convolved time series.  If the signal-to-noise of one bin within a set
of user specified time ranges is above a threshold of $5\sigma$, the
characteristics of the event are stored -- i.e.\ arrival time with
respect to the beginning of the time series, DM, width and
signal-to-noise.  Due to the large parameter space searched and the
geographical area where the Superterp is located, a large number of events
correspond to RFI. During a typical 1-hr LOTAAS observation, $\sim
10^8$ events are detected above a signal-to-noise ratio of 5. In order
to lower the number of events generated by RFI, the \textsc{L-sps}
automated classifier has been developed \citep{mhl+18}. It is a
machine-learning-based classifier that uses the same Very Fast
Decision Tree algorithm as the classifier for the periodicity search
\citep{tls+18}. The classifier uses 5 scores based on the width and
signal-to-noise of the signal as a function of DM.  At the end of the
classification, it produces detailed diagnostic plots to help quickly
identify real signals. Typically, a few tens of diagnostic plots need
to be visually inspected for each pointing.

\section{Results}
\label{sec:results}

\subsection{Discoveries}
\label{ssec:discoveries}

As of January 2019, the LOTAAS survey has discovered and confirmed 73
pulsars; in this overview paper, we present their properties. An up to
date list of LOTAAS discoveries is maintained
online\footnote{\url{http://www.astron.nl/lotaas}}.

The pulse profiles of these 73 pulsars are shown in
Fig.\,\ref{fig:profiles}. The profiles are from the discovery
observations, and hence representative of the data used to search for and
find these pulsars. Throughout the paper, we will refer to the
LOTAAS discoveries by their discovery name, which is based on the
right ascension and declination determined from the follow-up gridding
observations. We report the pulsar name with two digits in
declination. In the future, when timing solutions for these pulsars
are presented, these names will be superseded by names based on their
more precise timing positions.

The properties of the LOTAAS discovered pulsars are listed in
Table\,\ref{tab:discoveries}. The discovery spin period $P$,
$\mathrm{DM}$ and the position ($\alpha_\mathrm{J2000}$,
$\delta_\mathrm{J2000}$) based on the best localization from the
confirmation observations using the full LOFAR core. Here, we have
assumed a positional uncertainty of $3\arcmin$ in both
$\alpha_\mathrm{J2000}$ and $\delta_\mathrm{J2000}$, which should
encompass possible systematic uncertainties of order $1\arcmin$ due to
ionospheric beam jitter, as well as the localization uncertainty due
to gridding. Distance estimates based on the dispersion measure are
computed using the NE2001 model for the Galactic electron distribution
\citep{cl02}.

\begin{table*}
  \centering
  \footnotesize
  \caption{LOTAAS pulsar discoveries and their properties; spin period
    $P$, dispersion measure $\mathrm{DM}$, position in equatorial and
    Galactic coordinates, as well as the distance as predicted by the
    NE2001 model, and the pulse FWHM. Pulsar names are defined based
    on the TAB position of full LOFAR core follow-up
    observations. Here, we use the full core TAB FWHM as the
    positional uncertainty. If the pulsar is mentioned elsewhere with
    a different name, the alternate identification is provided in the
    comments. Uncertainties on measurements are shown in brackets
    corresponding to the least significant digit.}
  \label{tab:discoveries}
  \begin{tabular}{lllllrrrrl}
  \hline\hline
  PSR & \multicolumn{1}{c}{$P$} & \multicolumn{1}{c}{$\mathrm{DM}$} & \multicolumn{1}{c}{$\alpha_\mathrm{J2000}$} & \multicolumn{1}{c}{$\delta_\mathrm{J2000}$} & \multicolumn{1}{c}{$l$} & \multicolumn{1}{c}{$b$} & $d_\mathrm{NE2001}$ & $w_{50}$ & Comments \\
  & \multicolumn{1}{c}{(s)} & \multicolumn{1}{c}{(pc\,cm$^{-3}$)} & & & \multicolumn{1}{c}{(\degr)} & \multicolumn{1}{c}{(\degr)} & (kpc) & (ms) & \\
  \hline
  J0039$+$35   &  $0.5367$ &   $53.04(2)$ &        $00^\mathrm{h}39\fm1(0\fm2)$ &    $+35\degr45\arcmin(3\arcmin)$ &     $120.11$ &     $-27.04$ &   $4.16$ &   $11$ &  \\
J0059$+$69   &  $1.1459$ &   $63.53(5)$ &        $00^\mathrm{h}59\fm5(0\fm6)$ &    $+69\degr55\arcmin(3\arcmin)$ &     $123.63$ &       $7.07$ &   $2.36$ &   $23$ &  \\
J0100$+$80   &  $1.4936$ &  $55.76(13)$ &        $01^\mathrm{h}00\fm3(1\fm2)$ &    $+80\degr22\arcmin(3\arcmin)$ &     $123.32$ &      $17.52$ &   $2.62$ &   $60$ &  \\
J0107$+$13   &  $1.1974$ &   $22.02(8)$ &        $01^\mathrm{h}07\fm6(0\fm2)$ &    $+13\degr25\arcmin(3\arcmin)$ &     $128.95$ &     $-49.25$ &   $1.01$ &   $36$ &  \\
J0115$+$63   &  $0.5215$ &   $65.04(7)$ &        $01^\mathrm{h}15\fm6(0\fm4)$ &    $+63\degr24\arcmin(3\arcmin)$ &     $125.63$ &       $0.67$ &   $2.23$ &   $31$ &  \\
J0121$+$14   &  $1.3890$ &   $17.77(9)$ &        $01^\mathrm{h}22\fm0(0\fm2)$ &    $+14\degr16\arcmin(3\arcmin)$ &     $134.01$ &     $-47.94$ &   $0.79$ &   $42$ & P \\
J0139$+$33   &  $1.2479$ &   $21.21(6)$ &        $01^\mathrm{h}40\fm0(0\fm2)$ &    $+33\degr37\arcmin(3\arcmin)$ &     $134.38$ &     $-28.17$ &   $0.98$ &   $25$ & RRAT, \textsc{L-sps}, P \\
J0210$+$58   &  $1.7663$ &  $76.70(12)$ &        $02^\mathrm{h}11\fm0(0\fm4)$ &    $+58\degr44\arcmin(3\arcmin)$ &     $133.11$ &      $-2.55$ &   $2.51$ &   $53$ &  \\
J0250$+$58   & $23.5355$ &  $45.8(1.1)$ &        $02^\mathrm{h}50\fm3(0\fm4)$ &    $+58\degr54\arcmin(3\arcmin)$ &     $137.77$ &      $-0.50$ &   $1.67$ &  $471$ & J0250$+$5854 \\
J0302$+$22   &  $1.2072$ &  $19.09(11)$ &        $03^\mathrm{h}02\fm5(0\fm2)$ &    $+22\degr50\arcmin(3\arcmin)$ &     $158.44$ &     $-30.84$ &   $0.75$ &   $48$ & J0301$+$20, \textsc{L-sps}, P \\
J0305$+$11   &  $0.8621$ &   $27.97(6)$ &        $03^\mathrm{h}05\fm1(0\fm2)$ &    $+11\degr23\arcmin(3\arcmin)$ &     $167.46$ &     $-39.66$ &   $1.10$ &   $26$ &  \\
J0317$+$13   &  $1.9743$ &   $12.90(4)$ &        $03^\mathrm{h}17\fm9(0\fm2)$ &    $+13\degr29\arcmin(3\arcmin)$ &     $168.75$ &     $-36.03$ &   $0.49$ &   $20$ & \textsc{L-sps}, P \\
J0349$+$23   &  $2.4208$ &  $63.15(11)$ &        $03^\mathrm{h}49\fm9(0\fm2)$ &    $+23\degr41\arcmin(3\arcmin)$ &     $167.42$ &     $-23.38$ &   $3.30$ &   $48$ &  \\
J0421$+$32   &  $0.9001$ &  $77.0(0.5)$ &        $04^\mathrm{h}21\fm4(0\fm2)$ &    $+32\degr54\arcmin(3\arcmin)$ &     $165.83$ &     $-11.96$ &   $2.62$ &  $207$ & P \\
J0454$+$45   &  $1.3892$ &   $20.82(6)$ &        $04^\mathrm{h}54\fm9(0\fm3)$ &    $+45\degr28\arcmin(3\arcmin)$ &     $160.71$ &       $1.22$ &   $0.78$ &   $28$ & \textsc{L-sps} \\
J0518$+$51   &  $0.9125$ &   $39.15(6)$ &        $05^\mathrm{h}18\fm3(0\fm3)$ &    $+51\degr25\arcmin(3\arcmin)$ &     $158.26$ &       $7.88$ &   $1.35$ &   $27$ &  \\
J0742$+$43   &  $0.6062$ &   $36.23(4)$ &        $07^\mathrm{h}42\fm6(0\fm3)$ &    $+43\degr33\arcmin(3\arcmin)$ &     $175.54$ &      $27.19$ &   $1.34$ &   $18$ &  \\
J0811$+$37   &  $1.2483$ &  $16.95(11)$ &        $08^\mathrm{h}11\fm2(0\fm3)$ &    $+37\degr28\arcmin(3\arcmin)$ &     $183.67$ &      $31.22$ &   $0.59$ &   $50$ & P \\
J0813$+$22   &  $0.5314$ &   $52.29(5)$ &        $08^\mathrm{h}13\fm9(0\fm2)$ &    $+22\degr01\arcmin(3\arcmin)$ &     $200.89$ &      $27.48$ &   $2.59$ &   $21$ &  \\
J0827$+$53   &  $0.0135$ &  $23.103(1)$ &        $08^\mathrm{h}27\fm8(0\fm3)$ &    $+53\degr00\arcmin(3\arcmin)$ &     $165.46$ &      $35.64$ &   $0.90$ &    $1$ & Binary MSP \\
J0857$+$33   &  $0.2430$ & $24.025(16)$ &        $08^\mathrm{h}57\fm1(0\fm2)$ &    $+33\degr48\arcmin(3\arcmin)$ &     $190.15$ &      $39.72$ &   $0.89$ &    $7$ &  \\
J0928$+$30   &  $2.0915$ &   $21.95(9)$ &        $09^\mathrm{h}29\fm0(0\fm2)$ &    $+30\degr38\arcmin(3\arcmin)$ &     $195.82$ &      $45.91$ &   $0.84$ &   $42$ & P \\
J0935$+$33   &  $0.9615$ &   $18.35(6)$ &        $09^\mathrm{h}35\fm1(0\fm2)$ &    $+33\degr11\arcmin(3\arcmin)$ &     $192.36$ &      $47.52$ &   $0.68$ &   $29$ & P \\
J1017$+$30   &  $0.4528$ &   $27.16(6)$ &        $10^\mathrm{h}17\fm6(0\fm2)$ &    $+30\degr10\arcmin(3\arcmin)$ &     $198.48$ &      $56.27$ &   $1.23$ &   $27$ &  \\
J1226$+$00   &  $2.2851$ &  $18.50(10)$ &        $12^\mathrm{h}26\fm2(0\fm2)$ &    $+00\degr03\arcmin(3\arcmin)$ &     $289.27$ &      $62.26$ &   $0.87$ &   $46$ &  \\
J1235$-$02   &  $3.5976$ &  $18.8(0.2)$ &        $12^\mathrm{h}35\fm9(0\fm2)$ &    $-02\degr05\arcmin(3\arcmin)$ &     $295.03$ &      $60.54$ &   $0.88$ &  $108$ &  \\
J1303$+$38   &  $0.3963$ &  $19.000(9)$ &        $13^\mathrm{h}03\fm3(0\fm3)$ &    $+38\degr13\arcmin(3\arcmin)$ &     $111.08$ &      $78.62$ &   $1.76$ &    $4$ &  \\
J1334$+$10   &  $0.9111$ &   $24.00(4)$ &        $13^\mathrm{h}34\fm5(0\fm2)$ &    $+10\degr05\arcmin(3\arcmin)$ &     $335.77$ &      $70.16$ &   $2.53$ &   $18$ &  \\
J1344$+$66   &  $1.3941$ &  $30.02(16)$ &        $13^\mathrm{h}43\fm9(0\fm5)$ &    $+66\degr33\arcmin(3\arcmin)$ &     $114.90$ &      $49.75$ &   $1.83$ &   $70$ & J1340$+$65, \textsc{L-sps} \\
J1404$+$11   &  $2.6505$ &  $18.48(12)$ &        $14^\mathrm{h}04\fm6(0\fm2)$ &    $+11\degr57\arcmin(3\arcmin)$ &     $355.02$ &      $67.10$ &   $1.41$ &   $53$ & J1404$+$1159, \textsc{L-sps}, P \\
J1426$+$52   &  $0.9958$ &   $25.37(2)$ &        $14^\mathrm{h}27\fm0(0\fm3)$ &    $+52\degr10\arcmin(3\arcmin)$ &      $93.89$ &      $59.23$ &   $1.41$ &   $10$ &  \\
J1529$+$40   &  $0.4764$ &   $6.61(16)$ &        $15^\mathrm{h}29\fm2(0\fm3)$ &    $+40\degr49\arcmin(3\arcmin)$ &      $66.21$ &      $54.90$ &   $0.68$ &   $71$ & P \\
J1623$+$58   &  $0.6518$ &   $26.40(3)$ &        $16^\mathrm{h}23\fm8(0\fm4)$ &    $+58\degr49\arcmin(3\arcmin)$ &      $89.19$ &      $41.82$ &   $1.48$ &   $13$ &  \\
J1635$+$23   &  $1.2087$ &   $37.60(5)$ &        $16^\mathrm{h}35\fm1(0\fm2)$ &    $+23\degr31\arcmin(3\arcmin)$ &      $41.98$ &      $39.74$ &   $4.84$ &   $24$ & P \\
J1638$+$40   &  $0.7677$ &   $33.35(4)$ &        $16^\mathrm{h}38\fm8(0\fm3)$ &    $+40\degr05\arcmin(3\arcmin)$ &      $63.77$ &      $41.76$ &   $2.41$ &   $15$ &  \\
J1643$+$13   &  $1.0991$ &   $35.97(7)$ &        $16^\mathrm{h}43\fm8(0\fm2)$ &    $+13\degr25\arcmin(3\arcmin)$ &      $31.01$ &      $34.29$ &   $2.06$ &   $33$ & P \\
J1655$+$62   &  $0.7762$ &   $35.41(5)$ &        $16^\mathrm{h}55\fm9(0\fm4)$ &    $+62\degr02\arcmin(3\arcmin)$ &      $91.92$ &      $37.18$ &   $2.35$ &   $23$ &  \\
J1657$+$33   &  $1.5702$ &   $24.04(7)$ &        $16^\mathrm{h}57\fm7(0\fm2)$ &    $+33\degr03\arcmin(3\arcmin)$ &      $55.32$ &      $37.14$ &   $1.40$ &   $31$ & P \\
J1658$+$36   &  $0.0330$ &   $3.050(2)$ &        $16^\mathrm{h}58\fm4(0\fm2)$ &    $+36\degr30\arcmin(3\arcmin)$ &      $59.62$ &      $37.59$ &   $0.49$ &    $1$ & J1658$+$3630, binary MSP \\
J1707$+$35   &  $0.1598$ &  $19.240(7)$ &        $17^\mathrm{h}07\fm0(0\fm2)$ &    $+35\degr56\arcmin(3\arcmin)$ &      $59.26$ &      $35.78$ &   $1.17$ &    $3$ &  \\
J1713$+$78   &  $0.4325$ &   $36.96(3)$ &        $17^\mathrm{h}13\fm5(1\fm0)$ &    $+78\degr09\arcmin(3\arcmin)$ &     $110.29$ &      $31.47$ &   $2.12$ &   $13$ &  \\
J1715$+$46   &  $0.5481$ &   $19.82(5)$ &        $17^\mathrm{h}15\fm8(0\fm3)$ &    $+46\degr03\arcmin(3\arcmin)$ &      $71.86$ &      $35.36$ &   $1.17$ &   $22$ &  \\
J1722$+$35   &  $0.8216$ &   $23.83(6)$ &        $17^\mathrm{h}22\fm1(0\fm2)$ &    $+35\degr18\arcmin(3\arcmin)$ &      $59.24$ &      $32.65$ &   $1.41$ &   $25$ & P \\
J1735$+$63   &  $0.5107$ &   $41.81(3)$ &        $17^\mathrm{h}35\fm1(0\fm4)$ &    $+63\degr19\arcmin(3\arcmin)$ &      $92.71$ &      $32.55$ &   $3.18$ &   $15$ &  \\
J1740$+$27   &  $1.0582$ &   $35.46(5)$ &        $17^\mathrm{h}40\fm5(0\fm2)$ &    $+27\degr13\arcmin(3\arcmin)$ &      $51.46$ &      $26.72$ &   $1.99$ &   $21$ & P \\
J1741$+$38   &  $0.8289$ &   $47.26(6)$ &        $17^\mathrm{h}41\fm2(0\fm3)$ &    $+38\degr54\arcmin(3\arcmin)$ &      $64.28$ &      $29.70$ &   $3.74$ &   $25$ &  \\
J1745$+$12   &  $1.0599$ &  $66.32(17)$ &        $17^\mathrm{h}45\fm7(0\fm2)$ &    $+12\degr51\arcmin(3\arcmin)$ &      $37.36$ &      $20.30$ &   $2.98$ &   $74$ &  \\
J1745$+$42   &  $0.3051$ &   $38.00(3)$ &        $17^\mathrm{h}45\fm8(0\fm3)$ &    $+42\degr53\arcmin(3\arcmin)$ &      $68.97$ &      $29.63$ &   $2.36$ &   $12$ &  \\
J1749$+$59   &  $0.4360$ &   $45.09(6)$ &        $17^\mathrm{h}49\fm6(0\fm4)$ &    $+59\degr51\arcmin(3\arcmin)$ &      $88.59$ &      $30.90$ &   $4.05$ &   $26$ &  \\
J1809$+$17   &  $2.0667$ &   $47.32(9)$ &        $18^\mathrm{h}09\fm1(0\fm2)$ &    $+17\degr04\arcmin(3\arcmin)$ &      $43.84$ &      $16.87$ &   $2.27$ &   $41$ &  \\
J1810$+$07   &  $0.3077$ &   $79.41(9)$ &        $18^\mathrm{h}10\fm7(0\fm2)$ &    $+07\degr03\arcmin(3\arcmin)$ &      $34.65$ &      $12.25$ &   $3.00$ &   $40$ &  \\
J1814$+$22   &  $0.2537$ & $62.313(11)$ &        $18^\mathrm{h}14\fm6(0\fm2)$ &    $+22\degr23\arcmin(3\arcmin)$ &      $49.50$ &      $17.77$ &   $3.27$ &    $5$ &  \\
J1848$+$15   &  $2.2338$ &  $77.42(10)$ &        $18^\mathrm{h}48\fm9(0\fm2)$ &    $+15\degr17\arcmin(3\arcmin)$ &      $46.33$ &       $7.45$ &   $3.29$ &   $45$ & J1849$+$15, \textsc{L-sps}, P \\
J1849$+$25   &  $0.5193$ & $74.970(12)$ &        $18^\mathrm{h}49\fm8(0\fm2)$ &    $+25\degr58\arcmin(3\arcmin)$ &      $56.22$ &      $11.86$ &   $3.88$ &    $5$ &  \\
J1910$+$56   &  $0.3419$ &  $20.75(12)$ &        $19^\mathrm{h}10\fm7(0\fm4)$ &    $+56\degr55\arcmin(3\arcmin)$ &      $87.58$ &      $19.95$ &   $1.52$ &   $51$ &  \\
J1916$+$32   &  $1.1374$ &   $83.99(5)$ &        $19^\mathrm{h}16\fm1(0\fm2)$ &    $+32\degr24\arcmin(3\arcmin)$ &      $64.63$ &       $9.43$ &   $4.46$ &   $23$ &  \\
J1933$+$53   &  $2.0526$ &   $33.52(9)$ &        $19^\mathrm{h}33\fm0(0\fm3)$ &    $+53\degr32\arcmin(3\arcmin)$ &      $85.54$ &      $15.75$ &   $2.18$ &   $41$ &  \\
J1953$+$30   &  $1.2712$ &  $43.61(11)$ &        $19^\mathrm{h}53\fm8(0\fm2)$ &    $+30\degr13\arcmin(3\arcmin)$ &      $66.58$ &       $1.32$ &   $3.12$ &   $51$ &  \\

  \hline
  \end{tabular}
\end{table*}

\setcounter{table}{1}
\begin{table*}
  \centering
  \footnotesize
  \caption{Continued.}
  \begin{tabular}{lllllrrrrl}
  \hline\hline
  PSR & \multicolumn{1}{c}{$P$} & \multicolumn{1}{c}{$\mathrm{DM}$} & \multicolumn{1}{c}{$\alpha_\mathrm{J2000}$} & \multicolumn{1}{c}{$\delta_\mathrm{J2000}$} & \multicolumn{1}{c}{$l$} & \multicolumn{1}{c}{$b$} & $d_\mathrm{NE2001}$ & $w_{50}$ & Comments \\
  & \multicolumn{1}{c}{(s)} & \multicolumn{1}{c}{(pc\,cm$^{-3}$)} & & & \multicolumn{1}{c}{(\degr)} & \multicolumn{1}{c}{(\degr)} & (kpc) & (ms) & \\
  \hline
  J1957$-$00   &  $0.9651$ &   $38.31(6)$ &        $19^\mathrm{h}57\fm6(0\fm2)$ &    $-00\degr01\arcmin(3\arcmin)$ &      $40.63$ &     $-14.69$ &   $1.85$ &   $29$ &  \\
J1958$+$56   &  $0.3118$ &   $58.10(4)$ &        $19^\mathrm{h}58\fm0(0\fm4)$ &    $+56\degr49\arcmin(3\arcmin)$ &      $90.22$ &      $14.01$ &   $3.27$ &   $16$ &  \\
J2006$+$22   &  $1.7419$ &  $130.56(8)$ &        $20^\mathrm{h}06\fm6(0\fm2)$ &    $+22\degr04\arcmin(3\arcmin)$ &      $61.14$ &      $-5.42$ &   $5.73$ &   $35$ &  \\
J2022$+$21   &  $0.8035$ &   $73.52(5)$ &        $20^\mathrm{h}22\fm4(0\fm2)$ &    $+21\degr11\arcmin(3\arcmin)$ &      $62.42$ &      $-8.98$ &   $3.92$ &   $24$ &  \\
J2036$+$66   &  $0.5019$ &   $50.82(4)$ &        $20^\mathrm{h}36\fm8(0\fm5)$ &    $+66\degr44\arcmin(3\arcmin)$ &     $101.41$ &      $15.25$ &   $2.68$ &   $20$ &  \\
J2051$+$12   &  $0.5532$ &  $43.40(11)$ &        $20^\mathrm{h}51\fm4(0\fm2)$ &    $+12\degr48\arcmin(3\arcmin)$ &      $59.35$ &     $-19.44$ &   $2.53$ &   $50$ & P \\
J2053$+$17   &  $0.1193$ &  $26.980(5)$ &        $20^\mathrm{h}53\fm8(0\fm2)$ &    $+17\degr18\arcmin(3\arcmin)$ &      $63.55$ &     $-17.25$ &   $1.91$ &    $2$ & J2053$+$1718 \\
J2057$+$21   &  $1.1667$ &   $73.31(5)$ &        $20^\mathrm{h}57\fm8(0\fm2)$ &    $+21\degr26\arcmin(3\arcmin)$ &      $67.56$ &     $-15.49$ &   $4.73$ &   $23$ &  \\
J2122$+$24   &  $0.5414$ &    $8.49(5)$ &        $21^\mathrm{h}22\fm7(0\fm2)$ &    $+24\degr24\arcmin(3\arcmin)$ &      $73.80$ &     $-17.96$ &   $0.75$ &   $22$ &  \\
J2123$+$36   &  $1.2940$ & $108.7(0.3)$ &        $21^\mathrm{h}23\fm8(0\fm2)$ &    $+36\degr24\arcmin(3\arcmin)$ &      $82.92$ &      $-9.86$ &   $6.18$ &  $116$ &  \\
J2209$+$22   &  $1.7769$ &   $46.30(8)$ &        $22^\mathrm{h}09\fm9(0\fm2)$ &    $+21\degr17\arcmin(3\arcmin)$ &      $79.92$ &     $-27.78$ &   $3.45$ &   $36$ & P \\
J2306$+$31   &  $0.3416$ &   $46.13(2)$ &        $23^\mathrm{h}06\fm2(0\fm2)$ &    $+31\degr23\arcmin(3\arcmin)$ &      $97.96$ &     $-26.33$ &   $3.00$ &   $10$ &  \\
J2329$+$47   &  $0.7284$ &   $43.99(3)$ &        $23^\mathrm{h}29\fm6(0\fm3)$ &    $+47\degr42\arcmin(3\arcmin)$ &     $108.96$ &     $-12.92$ &   $2.17$ &   $15$ &  \\
J2336$-$01   &  $1.0298$ &   $19.60(9)$ &        $23^\mathrm{h}36\fm6(0\fm2)$ &    $-01\degr51\arcmin(3\arcmin)$ &      $84.43$ &     $-59.02$ &   $0.92$ &   $41$ &  \\
J2350$+$31   &  $0.5081$ &   $39.14(3)$ &        $23^\mathrm{h}50\fm7(0\fm2)$ &    $+31\degr39\arcmin(3\arcmin)$ &     $108.11$ &     $-29.45$ &   $2.25$ &   $15$ & P \\

  \hline
  \end{tabular}
  \tablefoot{Notes: \textsc{L-sps} denotes pulsars found by the
    \textsc{L-sps} single pulse algorithm. Pulsars denoted with P have
    been presented as independent discoveries by the Puschino
    telescope.}
\end{table*}

Of these LOTAAS discoveries, 7 have been presented by \citet{mhl+18},
as these pulsars were discovered through their single-pulse emission
using the \textsc{L-sps} single pulse classifier presented in that
paper. We briefly repeat their properties here. PSR\,J0139+33 is a
RRAT that is not detectable in periodicity searches. PSRs\,J0302+22,
J0317+13 and J1848+15 display strong pulse-to-pulse variability, and
the first two were first discovered in the single pulse search, but
are also detectable in periodicity searches. PSR\,J1848+15 and the
remaining pulsars from \citet{mhl+18}, J0454+45, J1344+66 and
J1404+11, were bright enough to be detected in both single pulse and
periodicity searches. We note that \citet{mhl+18} referred to
PSRs\,J0302+20, J1344+66 and J1848+15 as J0301+20, J1340+65 and
J1849+15, respectively. These pulsars and their aliases are listed in
Table\,\ref{tab:discoveries} and identified with the \textsc{L-sps}
comment.

PSR\,J0250+5854, the pulsar with the 23.5\,s spin period, has been
presented previously in \citet{tbc+18}. This spin period is a factor 2 longer
than any previously known radio pulsar. Based on timing measurements
over a 2.2-yr time span, and a position determined from LOFAR
imaging, the spin period derivative was determined to be
$\dot{P}=2.7\times10^{-14}$, placing this pulsar beyond the
conventional pulsar deathline in the $P$-$\dot{P}$ diagram.

One of the LOTAAS discoveries, PSR\,J1404+11 matched in position and
spin period with PSR\,J1404+12 discovered by \citet{cha03} in an
Arecibo drift-scan survey at 430\,MHz. However, \citet{cha03} reported
$\mathrm{DM}=25$\,pc\,cm$^{-3}$, significantly offset from that found
by LOTAAS, $\mathrm{DM}=18.48$\,pc\,cm$^{-3}$. This pulsar has since
been independently confirmed by \citet{ttm18} and \citet{bfrs18}, who
confirm the 18.48\,pc\,cm$^{-3}$ dispersion measure. Another LOTAAS
discovery, PSR\,J2053+17 matches the position, spin period and
$\mathrm{DM}$ of an unconfirmed candidate found by \citet{rtj+96},
also using Arecibo at 430\,MHz. This rediscovery has since been
independently confirmed by \citet{bfrs18}.

During the preparation of this manuscript, some of the LOTAAS pulsars
were presented as independent discoveries by \citet{ttol16,ttk+17} and
\citet{ttm18} with the Puschino Telescope at 111\,MHz. \citet{ttol16}
presents seven new pulsars, of which four match the sky position and
spin period of the LOTAAS discovered pulsars PSRs\,J0302+22,
J0421$+$32, J0928$+$30 and J1722$+$35. Due to the small bandwidth
(2.5\,MHz) used by \citet{ttol16}, their $\mathrm{DM}$s are uncertain
by several DM units, but consistent with the LOTAAS
$\mathrm{DM}$s. The same telescope is used by \citet{ttk+17}, who
report 18 new pulsars, of which ten match the sky position, spin
period, and $\mathrm{DM}$ of LOTAAS pulsars (PSRs\,J0121+14, J0811+37,
J0935+33, J1529+40, J1635+23, J1638+40, J1657+33, J1953+30, J2051+12
and J2350+31). Again using the same telescope and instrumentation, 25
RRAT candidates, discovered through their single pulses, were
published by \citet{ttm18}. Although the Puschino sky positions and
$\mathrm{DM}$s have considerable uncertainties, it appears
\citet{ttm18} detect single pulses from PSRs\,J0139+33, J0317+13,
J1404+11, J1740+27, J1848+15, J2051+12 and J2209+22 (note that
J2051+12 is also presented by
\citealt{ttk+17}). Table\,\ref{tab:discoveries} indicates these
pulsars with P.

Among the pulsars discovered via LOTAAS are two millisecond pulsars,
both in binary systems. PSR\,J1658$+$36 was first discovered through
reclassification of the candidates from early survey data with the new
classifier~\citep{tls+18}. The pulsar has a spin period of 33.0\,ms
and a $\mathrm{DM}$ of 3.05\,pc\,cm$^{-3}$, suggesting that it is
relatively nearby at predicted distances of $225-500$\,pc based on the
Galactic electron density models \citep{cl02,ymw17}. It is found to be
in a binary system with an orbital period of 3.0\,d and a minimum
companion mass of 0.87\,M$_\odot$ (assuming a neutron star mass of
1.4\,M$_\odot$). This, combined with the low eccentricity of the orbit
suggest that the binary companion is most likely a carbon-oxygen white
dwarf. PSR\,J0827+53 is the second binary MSP discovered in the LOTAAS
survey. It has a spin period of 13.52\,ms and a
$\mathrm{DM}=23.103$\,pc\,cm$^{-3}$. Galactic models of the electron
density predict distances of $d=0.90$\,kpc \citep{cl02} and 1.57\,kpc
\citep{ymw17}, respectively. Pulse period measurements show that
PSR\,J0827+53 is in a circular 5.90\,d orbit around a companion with a
minimum mass of 0.69\,M$_\odot$, again assuming a canonical pulsar
mass of 1.4\,M$_\odot$. This suggests that PSR\,J0827+53 is also
likely to have a carbon-oxygen white dwarf, making both PSR\,J0827+53
and J1658+36 intermediate-mass binary pulsars.

Timing observations of all LOTAAS discovered pulsars are
ongoing. These will determine their spin period derivatives and
precise sky positions, and in the case of the two binary systems, the
binary parameters. These results will be presented in future
publications.

\subsection{Redetection of known pulsars}
\label{ssec:redetections}

We have thus far detected 311 individual known pulsars in the LOTAAS
observations. These redetections are listed in
Table~\ref{table:redetection}. Some of the pulsars detected with
LOTAAS were discovered recently by surveys centred at low observing
frequencies. A total of 47 of our redetections were originally
discovered by the Green Bank Northern Celestial Cap Survey (GBNCC;
\citealt{slr+14}). A further 6 were discovered by the AO327 Survey
\citep{dsm+13}. Eleven pulsars that have recently been discovered by
ongoing pulsar surveys and are not yet in the ATNF pulsar catalogue
\citep{mht+05} are listed in the table as well.

To determine the sensitivity of LOTAAS, we have compared the measured
flux density of the detected pulsars with the expected flux density of
the pulsars from literature values. The folded profiles are flux
calibrated using the method detailed by \citet{kvh+16}. The calibrated
flux densities of the redetections are provided in
Table~\ref{table:redetection}.

The measured flux densities ($S_{135}^\mathrm{uncor}$) assume that the
pulsars are located at the centre of the TAB in which they were
detected. However, most of the detections are offset from the centres
of the TABs ($\theta_\mathrm{TAB}$), which results in reduced
sensitivity. The angular distance of a pulsar from the centre of the
SAP ($\theta_\mathrm{SAP}$) in which the pulsar is detected also
reduces the sensitivity. Here, we applied a correction factor
($S_{135}^\times$) to the measured flux densities that takes into
account these offsets in order to obtain more accurate pulsar flux
densities. For this, we modeled the TAB and SAP beams as a
$\sinc^{2}(\theta)$ function with full-width half-maximum at the
central observing frequency of LOTAAS. The maximum correction factors
were set to 3.33 and 5 for the TABs and SAPs, respectively. We did not
consider the evolution of the beam size with observing frequency. The
corrected flux densities $S_{135}^\mathrm{corr}=S_{135}^\times\times
S_{135}^\mathrm{uncor}$ of the detections are listed in
Table~\ref{table:redetection}.

The expected flux densities ($S_{135}^\mathrm{exp}$) of the pulsars
detected at 135\,MHz are calculated by extrapolating values from the
pulsar catalogue \citep{mht+05} using the mean flux densities at
150\,MHz ($S_{150}$) or 400\,MHz ($S_{400}$) if the other is
unavailable and using the spectral indices in the catalogue or $\alpha =
-1.4$ \citep{blv13} if not provided. The majority of the $S_{150}$
flux densities in the catalogue are from
\citet{bkk+16,kvh+16,fjmi16,bmj+16}, while spectral indices are taken
from \citet{lylg95,kl01,jml+09,bkk+16,bmj+16,fjmi16,jsk+18}.

\begin{figure}
  \includegraphics[width=\columnwidth]{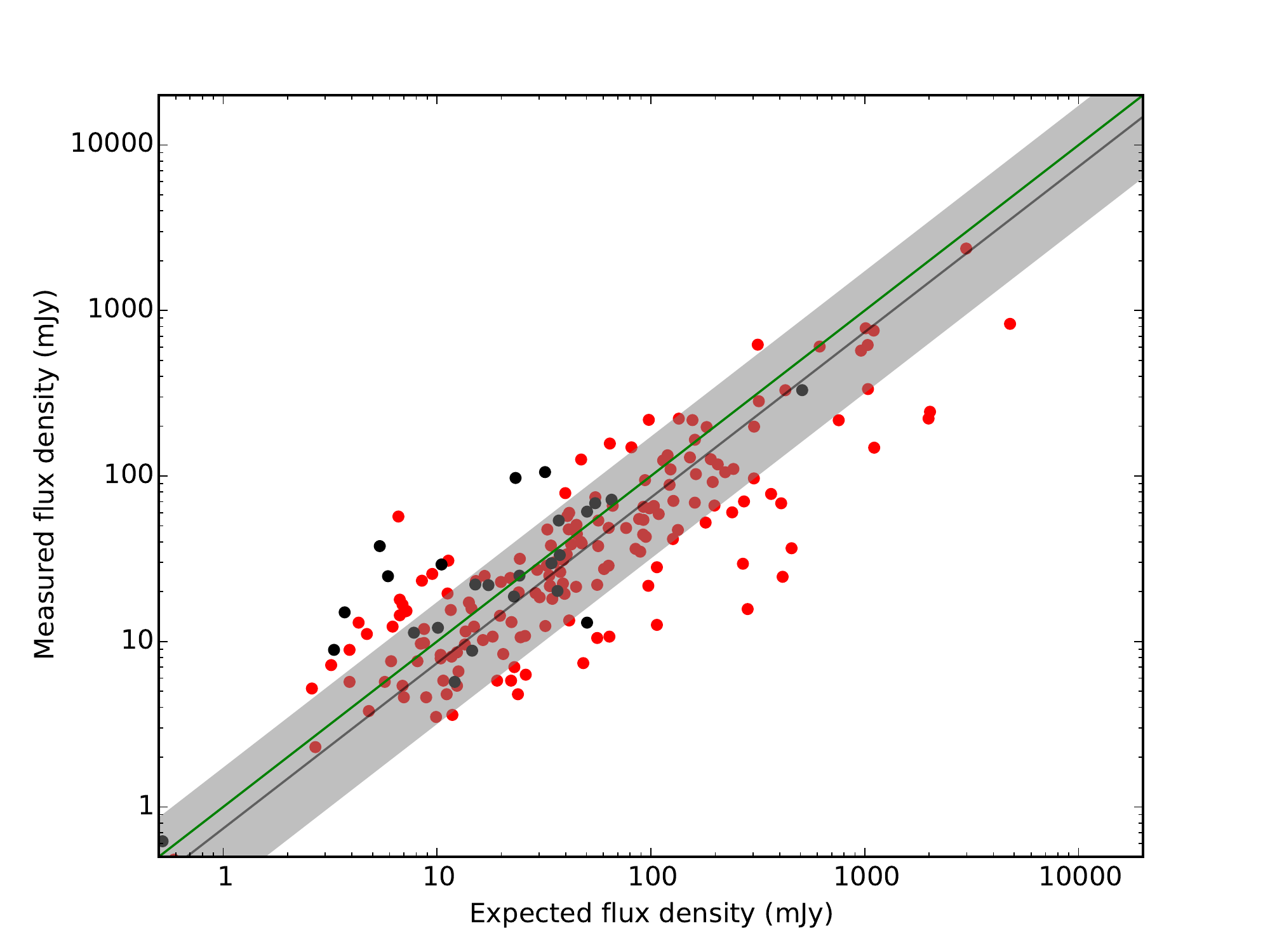}
  \caption{The measured flux densities ($S_{135}^\mathrm{cor}$) of
    LOTAAS redetections against expected flux densities
    ($S_{135}^\mathrm{exp}$) extrapolated from the ATNF pulsar
    catalogue. The red points are pulsars with a measured flux density
    at 150\,MHz ($S_{150}$), while the black points are pulsars with
    flux densities only measured at 400\,MHz ($S_{400}$). The green
    line indicates the point where the measured flux density is equal
    to the expected flux density.  The grey area indicates the
    $1\sigma$ region of calculated ratio of
    $S_{135}^\mathrm{cor}/S_{135}^\mathrm{exp}=0.74^{+0.99}_{-0.42}$
    between measured flux density and the expected flux density.}
  \label{fig:known_pulsar_flux_comparison}
\end{figure}

A comparison of the measured flux densities of the detected pulsars
and the expected values is shown in
Fig.\,\ref{fig:known_pulsar_flux_comparison}. Using all the known pulsars
that we redetected in the LOTAAS survey, and for which we can correct
our flux density measurements as well as predict flux densities based
on literature values, we find that the ratio between the measured and
expected flux density
$S_{135}^\mathrm{cor}/S_{135}^\mathrm{exp}=0.74^{+0.99}_{-0.42}$. We
conclude that, given the assumptions involved, LOTAAS is redetecting
known pulsars at roughly their expected flux densities, on average.

The measured flux densities of redetections of known pulsars also
validate the minimum expected sensitivity limit of 1.2\,mJy, as the
faintest redetections have 135\,MHz flux densities around 2 to 3\,mJy
(see Fig.\,\ref{fig:known_pulsar_flux_comparison}). When accurate
timing positions are available for the LOTAAS discovered pulsars, it
will be possible to perform the beam corrections and determine their
flux densities more accurately. These will be presented in future
publications.

\begin{figure}
  \includegraphics[width=\columnwidth]{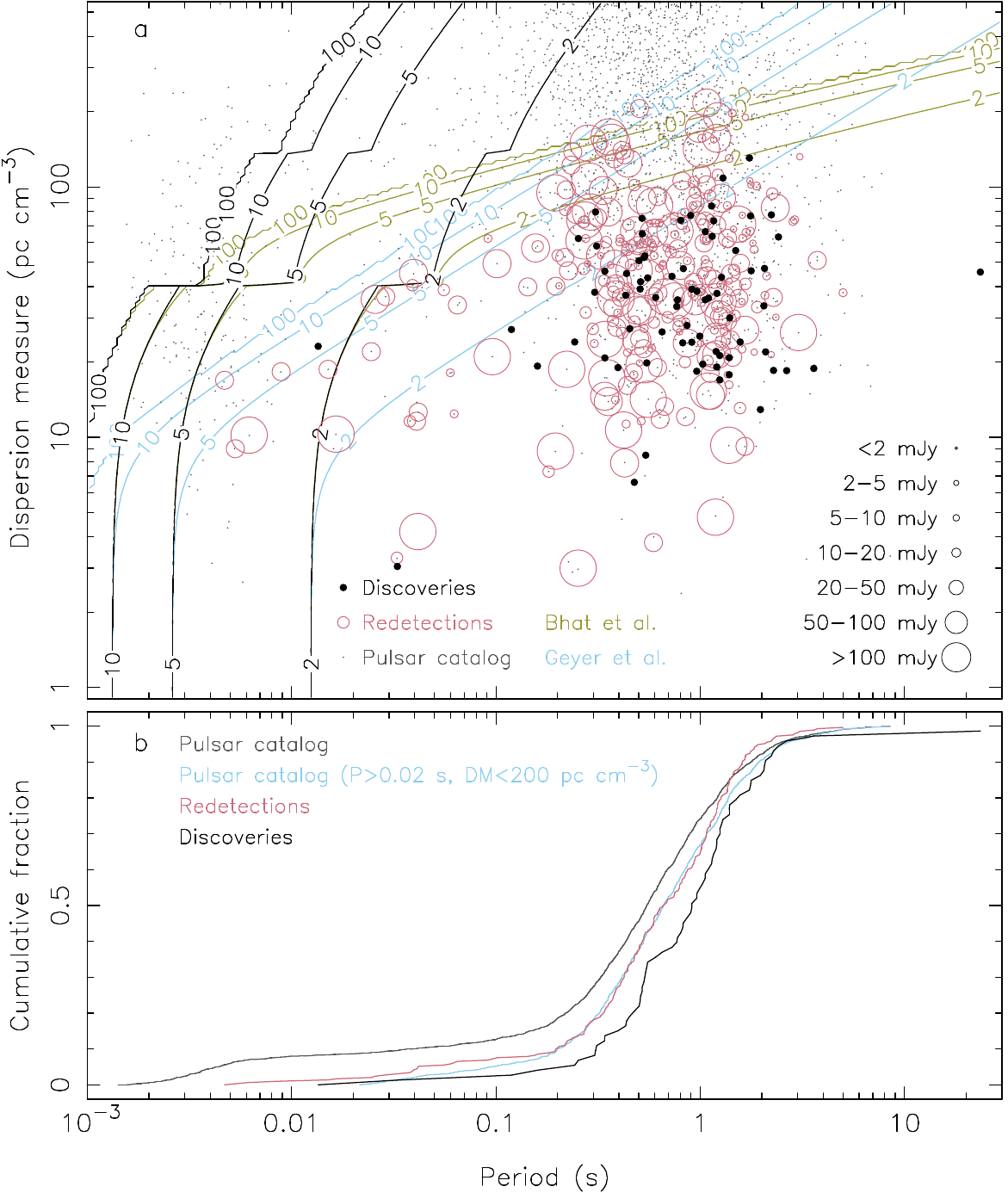}
  \caption{\textit{(top)} The spin period and $\mathrm{DM}$ of
    pulsars that have been discovered (black dots) and redetected (red
    circles) in the LOTAAS survey. The size of the circles for the
    redetected pulsars denotes their flux density. Also plotted as
    thin grey points are known pulsars from the ATNF pulsar catalogue
    (excluding pulsars in globular clusters). The solid lines denote
    the sensitivity limits (in mJy) when taking into account smearing
    due to dispersion (black), as well as scattering, using the
    \citet{bcc+04} (green) and \citet{gkv+17} (blue) scattering
    relations. \textit{(bottom)} Cumulative spin period distributions
    for the pulsars discovered (black) and redetected (red) in the
    LOTAAS survey. These can be compared to the spin period
    distribution of pulsars from the ATNF catalogue (dark grey, again
    excluding pulsars in globular clusters), and when keeping
    pulsars with $P>0.02$\,s and $\mathrm{DM}<200$\,pc\,cm$^{-3}$
    (light grey). The latter distribution can represent the
    distribution of pulsars that have been redetected in the LOTAAS
    survey. However, the LOTAAS discoveries tend to have longer spin
    periods, on average.}
  \label{fig:period_dm}
\end{figure}

\section{Discussion}
\label{sec:discussion}

\subsection{Discovery parameter space}

The LOTAAS survey has, to date, discovered 73 radio pulsars. The spin
periods of these pulsars range from 13.52\,ms to 23.5\,s, and their
$\mathrm{DM}$s from 3.05\,pc\,cm$^{-3}$ to
130.56\,pc\,cm$^{-3}$. Known pulsars are redetected in the LOTAAS
survey with spin periods from 4.2\,ms to 5.0\,s and $\mathrm{DM}$s
from 3.0\,pc\,cm$^{-3}$ to 217.0\,pc\,cm$^{-3}$. In
Fig.\,\ref{fig:period_dm}a we show the spin period and $\mathrm{DM}$
of the LOTAAS discoveries, as well as known pulsars redetected in
LOTAAS observations, in comparison to our predicted sensitivity
limits. These sensitivity limits are for the optimal case of sky
locations with a low sky temperature ($T_\mathrm{sky}=200$\,K) and a
narrow pulse profile (3\% of the spin period); for wider pulse
profiles and higher sky temperatures the sensitivity limits will be
higher. We find that the LOTAAS discoveries and redetections fall
within the area demarcated by the expected sensitivity limits.

Figure\,\ref{fig:period_dm}a also shows that the survey is limited by
scattering; no new or known pulsars are detected above
$\mathrm{DM}\approx220$\,pc\,cm$^{-3}$. This observed limit is close
to what one would expect from pulse profile broadening due to
scattering. This is true when using either the \citet{bcc+04} or
\citet{gkv+17} scattering versus $\mathrm{DM}$ relations.

\subsection{Spin period distributions}

Despite the coarse time resolution of the survey, LOTAAS has proven to
be sensitive to recycled MSPs, as two of the discoveries are mildly
recycled MSPs with periods of 13\,ms and 33\,ms, respectively. Among the
redetections, 6 fully recycled MSPs ($P<10$\,ms) are detected. All of
these are at low $\mathrm{DM}$ ($\mathrm{DM}\la20$\,pc\,cm$^{-3}$),
high Galactic latitude ($|b|\ga20\degr$), and intrinsically bright
($S_{135}^\mathrm{cor}$ or $S_{135}^\mathrm{exp}\ga25$\,mJy).  This allows for their
detection despite temporal smearing from dispersion and scattering.

The cumulative histograms of the spin period distribution of the
pulsars discovered and redetected in the survey
(Fig.\,\ref{fig:period_dm}b) show that they have longer spin periods,
on average, compared with the currently known population. Among the
LOTAAS discoveries is PSR\,J0250+5854, whose 23.5-s spin period is
more than twice as long as the next slowest spinning radio pulsar
known \citep{tbc+18}. To a large extent, the slower spin periods of LOTAAS
discoveries are due to the selection biases of the LOTAAS survey,
i.e.\ reduced sensitivity to short period pulsars and pulsars at high
$\mathrm{DM}$s. Masking these selection biases by placing simple
limits on period and $\mathrm{DM}$, we find that we can reproduce the
spin period distribution of the redetected known pulsars with that of
all known pulsars with spin periods $P>20$\,ms and
$\mathrm{DM}<200$\,pc\,cm$^{-3}$.

Since both the newly discovered pulsars as well as the redetected
known pulsars will have the same selection biases, it is surprising
that the discovered pulsars tend to have longer spin periods. A
two-sided Kolmogorov-Smirnov test shows that there is only a 10\%
probability that both samples are drawn from the same parent
distribution. The origin of this difference is unclear. The 1-hr
integrations improve the sensitivity of LOTAAS to long period pulsars,
compared with surveys using shorter integrations. Foremost, more pulses
will be present in longer observations \citep{tbc+18}, and longer
observations also reduce the impact of pulse-to-pulse variation
\citep{slr+14}. Furthermore, the low radio observing frequency of LOTAAS
provides additional benefits in that low-$\mathrm{DM}$ pulsars can be
more easily identified from zero-$\mathrm{DM}$ RFI due to the larger
$\mathrm{DM}$-induced sweeps. Similarly, at low radio observing frequencies
any unmasked zero-$\mathrm{DM}$ RFI will be spread over more time
samples of the dedispersed time-series, and hence the red noise in the
power spectrum due to these variations will be at lower spin
frequencies, further improving the sensitivity to long period
pulsars. However, the presence of red noise in the power spectrum will
lower the sensitivity to long period pulsars
\citep{hkr17}. Characterizing the impact of these effects as selection
biases on the observed spin period distribution is non-trivial and
will require injecting synthetic pulsar signals into the real data,
e.g.\ as was done for PALFA by \citet{lbh+15}. Measurements of the
spin period derivatives and spectral indices of the LOTAAS discovered
pulsar population may provide additional tests to determine whether
this population has longer spin periods.

\subsection{Maximum dispersion measures}

The measured DMs of the LOTAAS discoveries allowed us to probe the
approximate distribution of these pulsars in the Galaxy via distance
estimates using the NE2001 \citep{cl02} and YMW16 \citep{ymw17} models
for the Galactic electron density distribution. The distance estimates
shown in Table~\ref{tab:discoveries} are from NE2001, and suggest that
all LOTAAS discoveries are relatively nearby, with the furthest being
at a distance of 6.18\,kpc. On the other hand, for the lines-of-sight
of 14 out of the 73 LOTAAS discoveries, the YMW16 model predicts
maximum DM values that are smaller than the observed pulsar DMs,
suggesting that these pulsars are extragalactic. We investigate this
discrepancy in the following paragraphs.

\begin{figure}
  \centering
  \includegraphics[width=\linewidth]{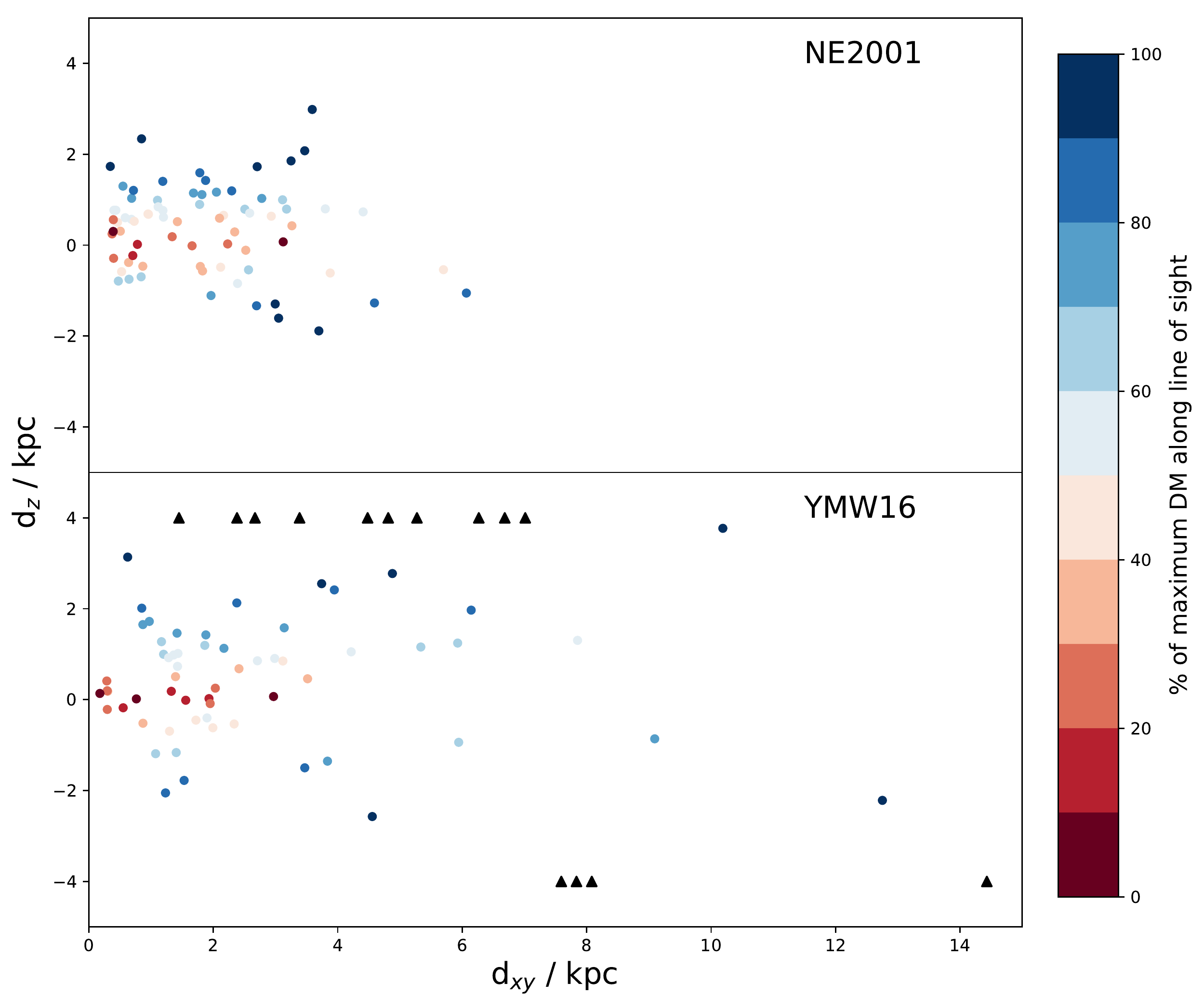} \caption{The
    distances of LOTAAS discoveries above/below the Galactic plane
    versus their distance along the Galactic plane, as predicted by the
    NE2001 \citep{cl02} and YMW16 \citep{ymw17} models for the
    Galactic free electron density distribution. The colour of each point denotes
    the DM of the pulsar compared to the maximum predicted DM along the
    particular line-of-sight. The
    pulsars for which the YMW16 model predicts maximum DMs less than
    the observed DMs are plotted with black triangles, where the
    distances to the pulsars are fixed at a limit of 4\,kpc above or
    below the Galactic plane.}  \label{fig:LOTAASheight}
\end{figure}

As most of the LOTAAS discoveries are found at high Galactic
latitudes, we first looked at the distribution of these pulsars in
terms of their distances from the Galactic
plane. Figure~\ref{fig:LOTAASheight} shows the distance of each pulsar
above or below the Galactic plane versus the distances to the pulsars
from Earth along the plane based on the predictions made by the two
electron density models. The colours of the circles denote the
percentage of maximum DM value expected from free electrons in the
Galaxy along the line-of-sight to the pulsar. The black triangles are
pulsars with measured DMs that are greater than the maximum-expected
DM from the Galaxy. We note that for the NE2001 model, most of the
pulsars are predicted to be within 2\,kpc of the Galactic plane, while
the distribution of the distances of the pulsars away from the
Galactic plane is more spread out according to the YMW16 model. While
the distances predicted by the NE2001 model are more reasonable in our
sample, the model is known to underestimate distances to pulsars at
high Galactic latitudes~\citep[e.g.][]{lfl+06,cbv+09}. Attempts have
been made to revise the electron density models away from the Galactic
plane~\citep[e.g.][]{gmcm08,sw09,sch12}.

The YMW16 model, however, seems to underestimate the electron density
away from the Galactic plane. It is unclear if this effect is
systematic or random. As the electron density models are fits to the
mean electron density distribution, fluctuations in the real electron
density distribution will lead to both over as well as under-estimated
distances. We found that there is no specific direction where the
pulsars are found to have distances overestimated by the YMW16
model. The discrepancy compared to the YMW16 model most likely
originates from the thin and thick disc components used to model the
electron density away from the Galactic plane, where the actual
electron density is likely to be higher than the modelled value. For
the 14 pulsars with DMs in excess of the maximum DM predicted by the
YWM16 model, the mean excess is about 10\%, suggesting that the
electron density is at least 10\% higher than predicted by the YMW16
model. The discrepancy impacts 14 out of 55 pulsars with
$|b|>15\degr$.

\subsection{Candidate matching}

In an all-sky survey such as LOTAAS, many candidates correspond to the
redetection of known pulsars. These redetections are useful for
testing the search pipeline, veryfing that the sensitivity of the
survey is as expected, and are used in training the candidate
classifiers. However, they can also be problematic. In the LOTAAS
survey, PSR\,B0329+54, the brightest pulsar in the Northern sky, was
detected at least 1400 times in 141 beams out of a possible 222 beams
of a single pointing. These detections were of both its fundamental
spin period, but also many harmonics of the spin period. As a result,
bright known pulsars can make up a significant fraction of candidates
generated in the survey, potentially limiting the ability to recognise
new pulsars \citep{lsl+18}.

Identifying that a candidate corresponds to a known pulsar can also be
problematic. When a candidate pulsar is detected, the position, period
and $\mathrm{DM}$ are used as key identifiers to associate it with a
known pulsar in the pulsar catalogue and pulsar discovery lists (such as
web pages related to ongoing surveys). Although the combination of
these identifiers is normally enough to determine a match/non-match,
they can sometimes be incorrect or misleading. For example, the
optimised $\mathrm{DM}$ of a candidate may not match the $\mathrm{DM}$
in the pulsar catalogue for an associated source. This is particularly
true for $\mathrm{DM}$ values that were obtained at high frequencies.

The discovery of PSR\,J2053+17 in the LOTAAS survey demonstrates the
importance of accurate $\mathrm{DM}$ measurements of known
pulsars. The candidate period and $\mathrm{DM}$ of PSR\,J2053+17
showed remarkable resemblance to that of the 1/6th harmonic of
PSR\,B0329+54 with a period ratio of 5.99. PSR\,B0329+54 is located
81\fdg5 away from the beam in which PSR\,J2053+17 was detected but, as
discussed above, PSR\,B0329+54 is easily detected by LOFAR and
sidelobe detections are common. Using the catalogue parameters alone, it
was difficult to determine the difference between these
sources. However, as PSR\,B0329+54 had been previously detected by
LOTAAS it was possible to compare the LOFAR measurements of
$\mathrm{DM}$ and determine that they were definitely different
sources. The precise and accurate measurement of the $\mathrm{DM}$ of
both pulsars was made possible within the LOTAAS survey because of its
large bandwidth and low observing frequency. We compared the detected
$\mathrm{DM}$ values of the known pulsars with the values from the
ATNF pulsar catalogue \citep{mht+05} before updates that used LOFAR
derived $\mathrm{DM}$s from \citet{bkk+16}, and found that 12 of the
pulsars have a difference in $\mathrm{DM}$ value of more than
2\,pc\,cm$^{-3}$. Since then, five of the pulsars have been updated
with the improved $\mathrm{DM}$ values.

Another example of the issues associated with candidate matching
uncertainty is highlighted by the discovery of PSR\,J1740+27
($P=1.0582$\,s, $\mathrm{DM}=35.46\pm0.04$\,pc\,cm$^{-3}$). The spin
period of this pulsar matched that of known pulsar PSR\,J1746+2540 to
within a period ratio of 1.00002, and with an angular separation of
just $2\degr$ between the discovery beam and position of the
previously known pulsar. When considering the sidelobes of LOFAR, it
was possible that these two sources were the same. The original
discovery of PSR\,J1746+2540 by \citet{fcwa95} using Arecibo at
430\,MHz reported $\mathrm{DM}=50\pm8$\,pc\,cm$^{-3}$, which was
improved to $\mathrm{DM}=51.5\pm0.2$\,pc\,cm$^{-3}$ by \citet{lwf+04}
from follow-up timing observations. These relatively large
$\mathrm{DM}$ uncertainties complicate the unambiguous identification
of a candidate as a new pulsar. For comparison, the LOFAR census by
\citet{bkk+16} constrained the dispersion measure to
$\mathrm{DM}=51.2044\pm0.0033$\,pc\,cm$^{-3}$. A full sky campaign to
determine accurate $\mathrm{DM}$ values of all currently known pulsars
is required before surveys are conducted with telescopes like the
Square Kilometre Array (SKA; \citealt{bbg+15, kbk+15}) in order to
enable the necessary automatic filtering of known pulsars and
facilitate the unambiguous identification of new pulsar candidates.

This process led us to question how likely it was for two pulsars to
have the same period to such precision. This was answered by
considering the well-known `birthday problem' using Stirling’s
approximation for large factorials. We determined that only 380
pulsars are required for a 50\% probability of two pulsars sharing the
same period to 4 decimal places and 1000 pulsars for 99.9\%
probability. The number of known pulsars now exceeds 2500 so it is
statistically certain that at least two pulsars will share the same
period to this precision. As we move forward to the SKA this problem
may become ever more acute.

\section{Conclusions}
\label{sec:conclusions}

The LOFAR Tied-Array All-Sky Survey (LOTAAS) is a 135-MHz pulsar and
fast transient survey covering the entire Northern hemisphere. By
using the LOFAR Superterp, the dense central part of the LOFAR core,
in combination with the tied-array beamforming capabilities of LOFAR,
LOTAAS is able to reach sensitivities of 1 to 5\,mJy for a large
instantaneous field of view, allowing for long integration times. This
is the deepest survey to date for radio pulsars and fast transients at
such low observing frequencies ($<200$\,MHz).

As of January 2019, over 90\% of the survey observations have been
obtained, and first-pass processing using the pulsar search pipeline
described in \S\,\ref{sec:analysis} is progressing. LOTAAS has so far
discovered 73 radio pulsars, whose basic parameters are presented
here. Furthermore, a total of 311 previously known pulsars have been
redetected in LOTAAS survey pointings. Despite the relatively coarse
time resolution of $492$\,$\upmu$s, LOTAAS is sensitive to mildly
recycled MSPs, and has discovered two binary MSPs. Due to the long
integration times of 1\,hr, LOTAAS is particularly sensitive to
long-spin period pulsars, which is demonstrated by the discovery of
the 23.5-s spin period pulsar J0250+5854 \citep{tbc+18}. The observed
spin period distribution of LOTAAS pulsar discoveries is skewed
towards longer spin periods in comparison to the known pulsar
population. Selection biases may explain this difference, though we
can not rule out that the LOTAAS-discovered pulsars have, on average,
longer spin periods.

The observations of known pulsars redetected in the LOTAAS survey are
used to assess the survey sensitivity and flux density scale. We find
that, on average, LOTAAS is recovering known pulsars at the expected
flux densities, and is reaching the expected sensitivity limit.

While the current periodicity pipeline has proven successful in
discovering pulsars, the computational requirements to process the
large data volume produced by LOTAAS has meant that currently only an
un-accelerated FFT-based periodicity search has been performed. With
computational capabilities ever increasing, and new algorithms being
developed, we plan to reprocess the LOTAAS survey in several
ways. First, we plan to use recent implementations of the fast folding
algorithm (FFA; \citealt{stae69,kml+09,cbc+17,pkr+18}) to further
improve the sensitivity to long period pulsars. Secondly,
GPU-accelerated algorithms to search for pulsars in binary systems,
taking into account acceleration (e.g.\ \citealt{dat+18}), have become
available, which may enable searching the LOTAAS survey for
accelerated pulsars. Reprocessing the Parkes Multi-beam Survey
\citep{mlc+01} has led to many new pulsar discoveries, long after the
data was taken (e.g.\ \citealt{kek+13,eklk13}). Given that all
LOTAAS data, after a 1\,yr proprietary period, is publicly available
in the LOFAR long-term archive, it may reveal many more pulsars in the
future.

\begin{acknowledgements}
  This paper is based (in part) on data obtained with the
  International LOFAR Telescope (ILT) under project codes LC0\_034,
  LC1\_052, LT2\_003, LC3\_014, LC4\_030, LT5\_004, LC9\_023 and
  LT10\_005. LOFAR \citep{hwg+13} is the Low Frequency Array designed
  and constructed by ASTRON. It has observing, data processing, and
  data storage facilities in several countries, that are owned by
  various parties (each with their own funding sources), and that are
  collectively operated by the ILT foundation under a joint scientific
  policy. The ILT resources have benefitted from the following recent
  major funding sources: CNRS-INSU, Observatoire de Paris and
  Universit\'{e} d'Orl\'{e}ans, France; BMBF, MIWF-NRW, MPG, Germany; Science
  Foundation Ireland (SFI), Department of Business, Enterprise and
  Innovation (DBEI), Ireland; NWO, The Netherlands; The Science and
  Technology Facilities Council, UK. We thank the operators of LOFAR
  for performing the LOTAAS observations. This work was carried out on
  the Dutch national e-infrastructure with the support of the SURF
  Cooperative. We thank the SURFsara staff for their patience and
  expertise. The research was supported by a NWO Science subsidy for
  the use of the National Computer Facilities (project 16676). JWTH,
  CB, DM, SS and VIK acknowledge funding from an NWO Vidi fellowship
  and from the European Research Council (ERC) under the European Union's
  Seventh Framework Programme (FP/2007-2013) / ERC Starting Grant
  agreement nr. 337062 (`DRAGNET'; PI: Hessels). BWS and SS
  acknowledge funding from the ERC under
  the European Union’s Horizon 2020 research and innovation programme
  (grant agreement nr. 694745). JvL acknowledges funding from the
  ERC under the European Union’s Seventh
  Framework Programme (FP/2007-2013) / ERC Grant Agreement nr. 617199
  (`ALERT'), and from Vici research programme `ARGO' with project
  nr. 639.043.815, financed by NWO.
\end{acknowledgements}

\bibliographystyle{aa}

\begin{thebibliography}{115}
\expandafter\ifx\csname natexlab\endcsname\relax\def\natexlab#1{#1}\fi

\bibitem[{{Antoniadis} {et~al.}(2013){Antoniadis}, {Freire}, {Wex}, {Tauris},
  {Lynch}, {van Kerkwijk}, {Kramer}, {Bassa}, {Dhillon}, {Driebe}, {Hessels},
  {Kaspi}, {Kondratiev}, {Langer}, {Marsh}, {McLaughlin}, {Pennucci}, {Ransom},
  {Stairs}, {van Leeuwen}, {Verbiest}, \& {Whelan}}]{afw+13}
{Antoniadis}, J., {Freire}, P.~C.~C., {Wex}, N., {et~al.} 2013, Science, 340,
  448

\bibitem[{{Archibald} {et~al.}(2018){Archibald}, {Gusinskaia}, {Hessels},
  {Deller}, {Kaplan}, {Lorimer}, {Lynch}, {Ransom}, \& {Stairs}}]{agh+18}
{Archibald}, A.~M., {Gusinskaia}, N.~V., {Hessels}, J.~W.~T., {et~al.} 2018,
  \nat, 559, 73

\bibitem[{{Archibald} {et~al.}(2009){Archibald}, {Stairs}, {Ransom}, {Kaspi},
  {Kondratiev}, {Lorimer}, {McLaughlin}, {Boyles}, {Hessels}, {Lynch}, {van
  Leeuwen}, {Roberts}, {Jenet}, {Champion}, {Rosen}, {Barlow}, {Dunlap}, \&
  {Remillard}}]{asr+09}
{Archibald}, A.~M., {Stairs}, I.~H., {Ransom}, S.~M., {et~al.} 2009, Science,
  324, 1411

\bibitem[{{Bassa} {et~al.}(2016){Bassa}, {Janssen}, {Karuppusamy}, {Kramer},
  {Lee}, {Liu}, {McKee}, {Perrodin}, {Purver}, {Sanidas}, {Smits}, \&
  {Stappers}}]{bjk+16}
{Bassa}, C.~G., {Janssen}, G.~H., {Karuppusamy}, R., {et~al.} 2016, \mnras,
  456, 2196

\bibitem[{{Bassa} {et~al.}(2017{\natexlab{a}}){Bassa}, {Pleunis}, \&
  {Hessels}}]{bph17a}
{Bassa}, C.~G., {Pleunis}, Z., \& {Hessels}, J.~W.~T. 2017{\natexlab{a}},
  Astronomy and Computing, 18, 40

\bibitem[{{Bassa} {et~al.}(2017{\natexlab{b}}){Bassa}, {Pleunis}, {Hessels},
  {Ferrara}, {Breton}, {Gusinskaia}, {Kondratiev}, {Sanidas}, {Nieder},
  {Clark}, {Li}, {van Amesfoort}, {Burnett}, {Camilo}, {Michelson}, {Ransom},
  {Ray}, \& {Wood}}]{bph+17b}
{Bassa}, C.~G., {Pleunis}, Z., {Hessels}, J.~W.~T., {et~al.}
  2017{\natexlab{b}}, \apjl, 846, L20

\bibitem[{{Bassa} {et~al.}(2018){Bassa}, {Pleunis}, {Hessels}, {Ferrara},
  {Kondratiev}, {Sanidas}, {Lyne}, {Stappers}, {Ransom}, \& {Fermi Pulsar
  Search Consortium}}]{bph+18}
{Bassa}, C.~G., {Pleunis}, Z., {Hessels}, J.~W.~T., {et~al.} 2018, in IAU
  Symposium, Vol. 337, Pulsar Astrophysics the Next Fifty Years, ed.
  P.~{Weltevrede}, B.~B.~P. {Perera}, L.~L. {Preston}, \& S.~{Sanidas}, 33--36

\bibitem[{{Bates} {et~al.}(2013){Bates}, {Lorimer}, \& {Verbiest}}]{blv13}
{Bates}, S.~D., {Lorimer}, D.~R., \& {Verbiest}, J.~P.~W. 2013, \mnras, 431,
  1352

\bibitem[{{Bell} {et~al.}(2016){Bell}, {Murphy}, {Johnston}, {Kaplan}, {Croft},
  {Hancock}, {Callingham}, {Zic}, {Dobie}, {Swiggum}, {Rowlinson},
  {Hurley-Walker}, {Offringa}, {Bernardi}, {Bowman}, {Briggs}, {Cappallo},
  {Deshpande}, {Gaensler}, {Greenhill}, {Hazelton}, {Johnston-Hollitt},
  {Lonsdale}, {McWhirter}, {Mitchell}, {Morales}, {Morgan}, {Oberoi}, {Ord},
  {Prabu}, {Shankar}, {Srivani}, {Subrahmanyan}, {Tingay}, {Wayth}, {Webster},
  {Williams}, \& {Williams}}]{bmj+16}
{Bell}, M.~E., {Murphy}, T., {Johnston}, S., {et~al.} 2016, \mnras, 461, 908

\bibitem[{{Bhat} {et~al.}(2004){Bhat}, {Cordes}, {Camilo}, {Nice}, \&
  {Lorimer}}]{bcc+04}
{Bhat}, N.~D.~R., {Cordes}, J.~M., {Camilo}, F., {Nice}, D.~J., \& {Lorimer},
  D.~R. 2004, \apj, 605, 759

\bibitem[{{Bhattacharyya} {et~al.}(2016){Bhattacharyya}, {Cooper}, {Malenta},
  {Roy}, {Chengalur}, {Keith}, {Kudale}, {McLaughlin}, {Ransom}, {Ray}, \&
  {Stappers}}]{bcm+16}
{Bhattacharyya}, B., {Cooper}, S., {Malenta}, M., {et~al.} 2016, \apj, 817, 130

\bibitem[{{Bilous} {et~al.}(2016){Bilous}, {Kondratiev}, {Kramer}, {Keane},
  {Hessels}, {Stappers}, {Malofeev}, {Sobey}, {Breton}, {Cooper}, {Falcke},
  {Karastergiou}, {Michilli}, {Os{\l}owski}, {Sanidas}, {ter Veen}, {van
  Leeuwen}, {Verbiest}, {Weltevrede}, {Zarka}, {Grie{\ss}meier}, {Serylak},
  {Bell}, {Broderick}, {Eisl{\"o}ffel}, {Markoff}, \& {Rowlinson}}]{bkk+16}
{Bilous}, A.~V., {Kondratiev}, V.~I., {Kramer}, M., {et~al.} 2016, \aap, 591,
  A134

\bibitem[{{Boyles} {et~al.}(2013){Boyles}, {Lynch}, {Ransom}, {Stairs},
  {Lorimer}, {McLaughlin}, {Hessels}, {Kaspi}, {Kondratiev}, {Archibald},
  {Berndsen}, {Cardoso}, {Cherry}, {Epstein}, {Karako-Argaman}, {McPhee},
  {Pennucci}, {Roberts}, {Stovall}, \& {van Leeuwen}}]{blr+13}
{Boyles}, J., {Lynch}, R.~S., {Ransom}, S.~M., {et~al.} 2013, \apj, 763, 80

\bibitem[{{Braun} {et~al.}(2015){Braun}, {Bourke}, {Green}, {Keane}, \&
  {Wagg}}]{bbg+15}
{Braun}, R., {Bourke}, T., {Green}, J.~A., {Keane}, E., \& {Wagg}, J. 2015,
  Advancing Astrophysics with the Square Kilometre Array (AASKA14), 174

\bibitem[{{Brentjens} \& {de Bruyn}(2005)}]{bb05}
{Brentjens}, M.~A. \& {de Bruyn}, A.~G. 2005, \aap, 441, 1217

\bibitem[{{Brinkman} {et~al.}(2018){Brinkman}, {Freire}, {Rankin}, \&
  {Stovall}}]{bfrs18}
{Brinkman}, C., {Freire}, P.~C.~C., {Rankin}, J., \& {Stovall}, K. 2018,
  \mnras, 474, 2012

\bibitem[{{Broekema} {et~al.}(2018){Broekema}, {Mol}, {Nijboer}, {van
  Amesfoort}, {Brentjens}, {Loose}, {Klijn}, \& {Romein}}]{bmn+18}
{Broekema}, P.~C., {Mol}, J.~J.~D., {Nijboer}, R., {et~al.} 2018, Astronomy and
  Computing, 23, 180

\bibitem[{{Cameron} {et~al.}(2017){Cameron}, {Barr}, {Champion}, {Kramer}, \&
  {Zhu}}]{cbc+17}
{Cameron}, A.~D., {Barr}, E.~D., {Champion}, D.~J., {Kramer}, M., \& {Zhu},
  W.~W. 2017, \mnras, 468, 1994

\bibitem[{{Camilo} {et~al.}(2006){Camilo}, {Ransom}, {Halpern}, {Reynolds},
  {Helfand}, {Zimmerman}, \& {Sarkissian}}]{crh+06}
{Camilo}, F., {Ransom}, S.~M., {Halpern}, J.~P., {et~al.} 2006, \nat, 442, 892

\bibitem[{{Chandler}(2003)}]{cha03}
{Chandler}, A.~M. 2003, PhD thesis, CALIFORNIA INSTITUTE OF TECHNOLOGY

\bibitem[{{Chatterjee} {et~al.}(2009){Chatterjee}, {Brisken}, {Vlemmings},
  {Goss}, {Lazio}, {Cordes}, {Thorsett}, {Fomalont}, {Lyne}, \&
  {Kramer}}]{cbv+09}
{Chatterjee}, S., {Brisken}, W.~F., {Vlemmings}, W.~H.~T., {et~al.} 2009, \apj,
  698, 250

\bibitem[{{Coenen}(2013)}]{coe13}
{Coenen}, T. 2013, PhD thesis, The University of Amsterdam

\bibitem[{{Coenen} {et~al.}(2014){Coenen}, {van Leeuwen}, {Hessels},
  {Stappers}, {Kondratiev}, {Alexov}, {Breton}, {Bilous}, {Cooper}, {Falcke},
  {Fallows}, {Gajjar}, {Grie{\ss}meier}, {Hassall}, {Karastergiou}, {Keane},
  {Kramer}, {Kuniyoshi}, {Noutsos}, {Os{\l}owski}, {Pilia}, {Serylak},
  {Schrijvers}, {Sobey}, {ter Veen}, {Verbiest}, {Weltevrede}, {Wijnholds},
  {Zagkouris}, {van Amesfoort}, {Anderson}, {Asgekar}, {Avruch}, {Bell},
  {Bentum}, {Bernardi}, {Best}, {Bonafede}, {Breitling}, {Broderick},
  {Br{\"u}ggen}, {Butcher}, {Ciardi}, {Corstanje}, {Deller}, {Duscha},
  {Eisl{\"o}ffel}, {Fender}, {Ferrari}, {Frieswijk}, {Garrett}, {de Gasperin},
  {de Geus}, {Gunst}, {Hamaker}, {Heald}, {Hoeft}, {van der Horst}, {Juette},
  {Kuper}, {Law}, {Mann}, {McFadden}, {McKay-Bukowski}, {McKean}, {Munk},
  {Orru}, {Paas}, {Pandey-Pommier}, {Polatidis}, {Reich}, {Renting},
  {R{\"o}ttgering}, {Rowlinson}, {Scaife}, {Schwarz}, {Sluman}, {Smirnov},
  {Swinbank}, {Tagger}, {Tang}, {Tasse}, {Thoudam}, {Toribio}, {Vermeulen},
  {Vocks}, {van Weeren}, {Wucknitz}, {Zarka}, \& {Zensus}}]{clh+14}
{Coenen}, T., {van Leeuwen}, J., {Hessels}, J.~W.~T., {et~al.} 2014, \aap, 570,
  A60

\bibitem[{{Cognard} {et~al.}(2017){Cognard}, {Freire}, {Guillemot}, {Theureau},
  {Tauris}, {Wex}, {Graikou}, {Kramer}, {Stappers}, {Lyne}, {Bassa},
  {Desvignes}, \& {Lazarus}}]{cfg+17}
{Cognard}, I., {Freire}, P.~C.~C., {Guillemot}, L., {et~al.} 2017, \apj, 844,
  128

\bibitem[{{Cooper}(2017)}]{coo17}
{Cooper}, S. 2017, PhD thesis, The University of Manchester

\bibitem[{{Cordes} \& {Lazio}(2002)}]{cl02}
{Cordes}, J.~M. \& {Lazio}, T.~J.~W. 2002, ArXiv Astrophysics e-prints
  [\eprint{astro-ph/0207156}]

\bibitem[{{Cromartie} {et~al.}(2016){Cromartie}, {Camilo}, {Kerr}, {Deneva},
  {Ransom}, {Ray}, {Ferrara}, {Michelson}, \& {Wood}}]{cck+16}
{Cromartie}, H.~T., {Camilo}, F., {Kerr}, M., {et~al.} 2016, \apj, 819, 34

\bibitem[{{Demorest} {et~al.}(2010){Demorest}, {Pennucci}, {Ransom}, {Roberts},
  \& {Hessels}}]{dpr+10}
{Demorest}, P.~B., {Pennucci}, T., {Ransom}, S.~M., {Roberts}, M.~S.~E., \&
  {Hessels}, J.~W.~T. 2010, \nat, 467, 1081

\bibitem[{{Deneva} {et~al.}(2013){Deneva}, {Stovall}, {McLaughlin}, {Bates},
  {Freire}, {Martinez}, {Jenet}, \& {Bagchi}}]{dsm+13}
{Deneva}, J.~S., {Stovall}, K., {McLaughlin}, M.~A., {et~al.} 2013, \apj, 775,
  51

\bibitem[{{Detweiler}(1979)}]{det79}
{Detweiler}, S. 1979, \apj, 234, 1100

\bibitem[{{Dewey} {et~al.}(1985){Dewey}, {Taylor}, {Weisberg}, \&
  {Stokes}}]{dtws85}
{Dewey}, R.~J., {Taylor}, J.~H., {Weisberg}, J.~M., \& {Stokes}, G.~H. 1985,
  \apjl, 294, L25

\bibitem[{{Dimoudi} {et~al.}(2018){Dimoudi}, {Adamek}, {Thiagaraj}, {Ransom},
  {Karastergiou}, \& {Armour}}]{dat+18}
{Dimoudi}, S., {Adamek}, K., {Thiagaraj}, P., {et~al.} 2018, \apjs, 239, 28

\bibitem[{{Eatough} {et~al.}(2013){Eatough}, {Kramer}, {Lyne}, \&
  {Keith}}]{eklk13}
{Eatough}, R.~P., {Kramer}, M., {Lyne}, A.~G., \& {Keith}, M.~J. 2013, \mnras,
  431, 292

\bibitem[{{Edwards} {et~al.}(2006){Edwards}, {Hobbs}, \& {Manchester}}]{ehm06}
{Edwards}, R.~T., {Hobbs}, G.~B., \& {Manchester}, R.~N. 2006, \mnras, 372,
  1549

\bibitem[{{Faucher-Gigu{\`e}re} \& {Kaspi}(2006)}]{fk06}
{Faucher-Gigu{\`e}re}, C.-A. \& {Kaspi}, V.~M. 2006, \apj, 643, 332

\bibitem[{{Foster} {et~al.}(1995){Foster}, {Cadwell}, {Wolszczan}, \&
  {Anderson}}]{fcwa95}
{Foster}, R.~S., {Cadwell}, B.~J., {Wolszczan}, A., \& {Anderson}, S.~B. 1995,
  \apj, 454, 826

\bibitem[{{Frail} {et~al.}(2016){Frail}, {Jagannathan}, {Mooley}, \&
  {Intema}}]{fjmi16}
{Frail}, D.~A., {Jagannathan}, P., {Mooley}, K.~P., \& {Intema}, H.~T. 2016,
  \apj, 829, 119

\bibitem[{{Gaensler} {et~al.}(2008){Gaensler}, {Madsen}, {Chatterjee}, \&
  {Mao}}]{gmcm08}
{Gaensler}, B.~M., {Madsen}, G.~J., {Chatterjee}, S., \& {Mao}, S.~A. 2008,
  \pasa, 25, 184

\bibitem[{{Geyer} {et~al.}(2017){Geyer}, {Karastergiou}, {Kondratiev},
  {Zagkouris}, {Kramer}, {Stappers}, {Grie{\ss}meier}, {Hessels}, {Michilli},
  {Pilia}, \& {Sobey}}]{gkv+17}
{Geyer}, M., {Karastergiou}, A., {Kondratiev}, V.~I., {et~al.} 2017, \mnras,
  470, 2659

\bibitem[{{Guillemot} {et~al.}(2016){Guillemot}, {Smith}, {Laffon}, {Janssen},
  {Cognard}, {Theureau}, {Desvignes}, {Ferrara}, \& {Ray}}]{gsl+16}
{Guillemot}, L., {Smith}, D.~A., {Laffon}, H., {et~al.} 2016, \aap, 587, A109

\bibitem[{{Haslam} {et~al.}(1981){Haslam}, {Klein}, {Salter}, {Stoffel},
  {Wilson}, {Cleary}, {Cooke}, \& {Thomasson}}]{hks+81}
{Haslam}, C.~G.~T., {Klein}, U., {Salter}, C.~J., {et~al.} 1981, \aap, 100, 209

\bibitem[{{Haslam} {et~al.}(1982){Haslam}, {Salter}, {Stoffel}, \&
  {Wilson}}]{hssw82}
{Haslam}, C.~G.~T., {Salter}, C.~J., {Stoffel}, H., \& {Wilson}, W.~E. 1982,
  \aaps, 47, 1

\bibitem[{{Hassall} {et~al.}(2012){Hassall}, {Stappers}, {Hessels}, {Kramer},
  {Alexov}, {Anderson}, {Coenen}, {Karastergiou}, {Keane}, {Kondratiev},
  {Lazaridis}, {van Leeuwen}, {Noutsos}, {Serylak}, {Sobey}, {Verbiest},
  {Weltevrede}, {Zagkouris}, {Fender}, {Wijers}, {B{\"a}hren}, {Bell},
  {Broderick}, {Corbel}, {Daw}, {Dhillon}, {Eisl{\"o}ffel}, {Falcke},
  {Grie{\ss}meier}, {Jonker}, {Law}, {Markoff}, {Miller-Jones}, {Osten}, {Rol},
  {Scaife}, {Scheers}, {Schellart}, {Spreeuw}, {Swinbank}, {ter Veen}, {Wise},
  {Wijnands}, {Wucknitz}, {Zarka}, {Asgekar}, {Bell}, {Bentum}, {Bernardi},
  {Best}, {Bonafede}, {Boonstra}, {Brentjens}, {Brouw}, {Br{\"u}ggen},
  {Butcher}, {Ciardi}, {Garrett}, {Gerbers}, {Gunst}, {van Haarlem}, {Heald},
  {Hoeft}, {Holties}, {de Jong}, {Koopmans}, {Kuniyoshi}, {Kuper}, {Loose},
  {Maat}, {Masters}, {McKean}, {Meulman}, {Mevius}, {Munk}, {Noordam},
  {Orr{\'u}}, {Paas}, {Pandey-Pommier}, {Pandey}, {Pizzo}, {Polatidis},
  {Reich}, {R{\"o}ttgering}, {Sluman}, {Steinmetz}, {Sterks}, {Tagger}, {Tang},
  {Tasse}, {Vermeulen}, {van Weeren}, {Wijnholds}, \& {Yatawatta}}]{hsh+12}
{Hassall}, T.~E., {Stappers}, B.~W., {Hessels}, J.~W.~T., {et~al.} 2012, \aap,
  543, A66

\bibitem[{{Hessels} {et~al.}(2008){Hessels}, {Ransom}, {Kaspi}, {Roberts},
  {Champion}, \& {Stappers}}]{hrk+08}
{Hessels}, J.~W.~T., {Ransom}, S.~M., {Kaspi}, V.~M., {et~al.} 2008, in
  American Institute of Physics Conference Series, Vol. 983, 40 Years of
  Pulsars: Millisecond Pulsars, Magnetars and More, ed. C.~{Bassa}, Z.~{Wang},
  A.~{Cumming}, \& V.~M. {Kaspi}, 613--615

\bibitem[{{Hessels} {et~al.}(2011){Hessels}, {Roberts}, {McLaughlin}, {Ray},
  {Bangale}, {Ransom}, {Kerr}, {Camilo}, \& {Decesar}}]{hrm+11}
{Hessels}, J.~W.~T., {Roberts}, M.~S.~E., {McLaughlin}, M.~A., {et~al.} 2011,
  in American Institute of Physics Conference Series, Vol. 1357, American
  Institute of Physics Conference Series, ed. M.~{Burgay}, N.~{D'Amico},
  P.~{Esposito}, A.~{Pellizzoni}, \& A.~{Possenti}, 40--43

\bibitem[{{Hewish} {et~al.}(1968){Hewish}, {Bell}, {Pilkington}, {Scott}, \&
  {Collins}}]{hbp+68}
{Hewish}, A., {Bell}, S.~J., {Pilkington}, J.~D.~H., {Scott}, P.~F., \&
  {Collins}, R.~A. 1968, \nat, 217, 709

\bibitem[{{Hobbs} {et~al.}(2010){Hobbs}, {Archibald}, {Arzoumanian}, {Backer},
  {Bailes}, {Bhat}, {Burgay}, {Burke-Spolaor}, {Champion}, {Cognard}, {Coles},
  {Cordes}, {Demorest}, {Desvignes}, {Ferdman}, {Finn}, {Freire}, {Gonzalez},
  {Hessels}, {Hotan}, {Janssen}, {Jenet}, {Jessner}, {Jordan}, {Kaspi},
  {Kramer}, {Kondratiev}, {Lazio}, {Lazaridis}, {Lee}, {Levin}, {Lommen},
  {Lorimer}, {Lynch}, {Lyne}, {Manchester}, {McLaughlin}, {Nice}, {Oslowski},
  {Pilia}, {Possenti}, {Purver}, {Ransom}, {Reynolds}, {Sanidas}, {Sarkissian},
  {Sesana}, {Shannon}, {Siemens}, {Stairs}, {Stappers}, {Stinebring},
  {Theureau}, {van Haasteren}, {van Straten}, {Verbiest}, {Yardley}, \&
  {You}}]{haa+10}
{Hobbs}, G., {Archibald}, A., {Arzoumanian}, Z., {et~al.} 2010, Classical and
  Quantum Gravity, 27, 084013

\bibitem[{{Hobbs} {et~al.}(2012){Hobbs}, {Coles}, {Manchester}, {Keith},
  {Shannon}, {Chen}, {Bailes}, {Bhat}, {Burke-Spolaor}, {Champion},
  {Chaudhary}, {Hotan}, {Khoo}, {Kocz}, {Levin}, {Oslowski}, {Preisig}, {Ravi},
  {Reynolds}, {Sarkissian}, {van Straten}, {Verbiest}, {Yardley}, \&
  {You}}]{hcm+12}
{Hobbs}, G., {Coles}, W., {Manchester}, R.~N., {et~al.} 2012, \mnras, 427, 2780

\bibitem[{{Hobbs} {et~al.}(2006){Hobbs}, {Edwards}, \& {Manchester}}]{hem06}
{Hobbs}, G.~B., {Edwards}, R.~T., \& {Manchester}, R.~N. 2006, \mnras, 369, 655

\bibitem[{{Hotan} {et~al.}(2004){Hotan}, {van Straten}, \&
  {Manchester}}]{hsm04}
{Hotan}, A.~W., {van Straten}, W., \& {Manchester}, R.~N. 2004, \pasa, 21, 302

\bibitem[{{Jankowski} {et~al.}(2018){Jankowski}, {van Straten}, {Keane},
  {Bailes}, {Barr}, {Johnston}, \& {Kerr}}]{jsk+18}
{Jankowski}, F., {van Straten}, W., {Keane}, E.~F., {et~al.} 2018, \mnras, 473,
  4436

\bibitem[{{Johnston} \& {Karastergiou}(2017)}]{jk17}
{Johnston}, S. \& {Karastergiou}, A. 2017, \mnras, 467, 3493

\bibitem[{{Joshi} {et~al.}(2009){Joshi}, {McLaughlin}, {Lyne}, {Ludovici},
  {Pawar}, {Faulkner}, {Lorimer}, {Kramer}, \& {Davies}}]{jml+09}
{Joshi}, B.~C., {McLaughlin}, M.~A., {Lyne}, A.~G., {et~al.} 2009, \mnras, 398,
  943

\bibitem[{{Kawash} {et~al.}(2018){Kawash}, {McLaughlin}, {Kaplan}, {DeCesar},
  {Levin}, {Lorimer}, {Lynch}, {Stovall}, {Swiggum}, {Fonseca}, {Archibald},
  {Banaszak}, {Biwer}, {Boyles}, {Cui}, {Dartez}, {Day}, {Ernst}, {Ford},
  {Flanigan}, {Heatherly}, {Hessels}, {Hinojosa}, {Jenet}, {Karako-Argaman},
  {Kaspi}, {Kondratiev}, {Leake}, {Lunsford}, {Martinez}, {Mata}, {Matheny},
  {Mcewen}, {Mingyar}, {Orsini}, {Ransom}, {Roberts}, {Rohr}, {Siemens},
  {Spiewak}, {Stairs}, {van Leeuwen}, {Walker}, \& {Wells}}]{klk+18}
{Kawash}, A.~M., {McLaughlin}, M.~A., {Kaplan}, D.~L., {et~al.} 2018, \apj,
  857, 131

\bibitem[{{Keane} {et~al.}(2015){Keane}, {Bhattacharyya}, {Kramer}, {Stappers},
  {Keane}, {Bhattacharyya}, {Kramer}, {Stappers}, {Bates}, {Burgay},
  {Chatterjee}, {Champion}, {Eatough}, {Hessels}, {Janssen}, {Lee}, {van
  Leeuwen}, {Margueron}, {Oertel}, {Possenti}, {Ransom}, {Theureau}, \&
  {Torne}}]{kbk+15}
{Keane}, E., {Bhattacharyya}, B., {Kramer}, M., {et~al.} 2015, Advancing
  Astrophysics with the Square Kilometre Array (AASKA14), 40

\bibitem[{{Knispel} {et~al.}(2013){Knispel}, {Eatough}, {Kim}, {Keane},
  {Allen}, {Anderson}, {Aulbert}, {Bock}, {Crawford}, {Eggenstein}, {Fehrmann},
  {Hammer}, {Kramer}, {Lyne}, {Machenschalk}, {Miller}, {Papa}, {Rastawicki},
  {Sarkissian}, {Siemens}, \& {Stappers}}]{kek+13}
{Knispel}, B., {Eatough}, R.~P., {Kim}, H., {et~al.} 2013, \apj, 774, 93

\bibitem[{{Kondratiev} {et~al.}(2009){Kondratiev}, {McLaughlin}, {Lorimer},
  {Burgay}, {Possenti}, {Turolla}, {Popov}, \& {Zane}}]{kml+09}
{Kondratiev}, V.~I., {McLaughlin}, M.~A., {Lorimer}, D.~R., {et~al.} 2009,
  \apj, 702, 692

\bibitem[{{Kondratiev} {et~al.}(2016){Kondratiev}, {Verbiest}, {Hessels},
  {Bilous}, {Stappers}, {Kramer}, {Keane}, {Noutsos}, {Os{\l}owski}, {Breton},
  {Hassall}, {Alexov}, {Cooper}, {Falcke}, {Grie{\ss}meier}, {Karastergiou},
  {Kuniyoshi}, {Pilia}, {Sobey}, {ter Veen}, {van Leeuwen}, {Weltevrede},
  {Bell}, {Broderick}, {Corbel}, {Eisl{\"o}ffel}, {Markoff}, {Rowlinson},
  {Swinbank}, {Wijers}, {Wijnands}, \& {Zarka}}]{kvh+16}
{Kondratiev}, V.~I., {Verbiest}, J.~P.~W., {Hessels}, J.~W.~T., {et~al.} 2016,
  \aap, 585, A128

\bibitem[{{Kramer} {et~al.}(2006){Kramer}, {Lyne}, {O'Brien}, {Jordan}, \&
  {Lorimer}}]{klo+06}
{Kramer}, M., {Lyne}, A.~G., {O'Brien}, J.~T., {Jordan}, C.~A., \& {Lorimer},
  D.~R. 2006, Science, 312, 549

\bibitem[{{Kuzmin} \& {Losovsky}(2001)}]{kl01}
{Kuzmin}, A.~D. \& {Losovsky}, B.~Y. 2001, \aap, 368, 230

\bibitem[{{Lawson} {et~al.}(1987){Lawson}, {Mayer}, {Osborne}, \&
  {Parkinson}}]{lmop87}
{Lawson}, K.~D., {Mayer}, C.~J., {Osborne}, J.~L., \& {Parkinson}, M.~L. 1987,
  \mnras, 225, 307

\bibitem[{{Lazarus} {et~al.}(2015){Lazarus}, {Brazier}, {Hessels},
  {Karako-Argaman}, {Kaspi}, {Lynch}, {Madsen}, {Patel}, {Ransom}, {Scholz},
  {Swiggum}, {Zhu}, {Allen}, {Bogdanov}, {Camilo}, {Cardoso}, {Chatterjee},
  {Cordes}, {Crawford}, {Deneva}, {Ferdman}, {Freire}, {Jenet}, {Knispel},
  {Lee}, {van Leeuwen}, {Lorimer}, {Lyne}, {McLaughlin}, {Siemens}, {Spitler},
  {Stairs}, {Stovall}, \& {Venkataraman}}]{lbh+15}
{Lazarus}, P., {Brazier}, A., {Hessels}, J.~W.~T., {et~al.} 2015, \apj, 812, 81

\bibitem[{{Lewandowski} {et~al.}(2015){Lewandowski}, {Ro{\.z}ko}, {Kijak}, \&
  {Melikidze}}]{lrkm15}
{Lewandowski}, W., {Ro{\.z}ko}, K., {Kijak}, J., \& {Melikidze}, G.~I. 2015,
  \apj, 808, 18

\bibitem[{{Lewandowski} {et~al.}(2004){Lewandowski}, {Wolszczan}, {Feiler},
  {Konacki}, \& {So{\l}tysi{\'n}ski}}]{lwf+04}
{Lewandowski}, W., {Wolszczan}, A., {Feiler}, G., {Konacki}, M., \&
  {So{\l}tysi{\'n}ski}, T. 2004, \apj, 600, 905

\bibitem[{{Lorimer}(2011)}]{lor11}
{Lorimer}, D.~R. 2011, in High-Energy Emission from Pulsars and their Systems,
  ed. D.~F. {Torres} \& N.~{Rea}, 21

\bibitem[{{Lorimer} {et~al.}(2007){Lorimer}, {Bailes}, {McLaughlin},
  {Narkevic}, \& {Crawford}}]{lbm+07}
{Lorimer}, D.~R., {Bailes}, M., {McLaughlin}, M.~A., {Narkevic}, D.~J., \&
  {Crawford}, F. 2007, Science, 318, 777

\bibitem[{{Lorimer} {et~al.}(2006){Lorimer}, {Faulkner}, {Lyne}, {Manchester},
  {Kramer}, {McLaughlin}, {Hobbs}, {Possenti}, {Stairs}, {Camilo}, {Burgay},
  {D'Amico}, {Corongiu}, \& {Crawford}}]{lfl+06}
{Lorimer}, D.~R., {Faulkner}, A.~J., {Lyne}, A.~G., {et~al.} 2006, \mnras, 372,
  777

\bibitem[{{Lorimer} \& {Kramer}(2012)}]{lk12}
{Lorimer}, D.~R. \& {Kramer}, M. 2012, {Handbook of Pulsar Astronomy}

\bibitem[{{Lorimer} {et~al.}(1995){Lorimer}, {Yates}, {Lyne}, \&
  {Gould}}]{lylg95}
{Lorimer}, D.~R., {Yates}, J.~A., {Lyne}, A.~G., \& {Gould}, D.~M. 1995,
  \mnras, 273, 411

\bibitem[{{Lynch} {et~al.}(2013){Lynch}, {Boyles}, {Ransom}, {Stairs},
  {Lorimer}, {McLaughlin}, {Hessels}, {Kaspi}, {Kondratiev}, {Archibald},
  {Berndsen}, {Cardoso}, {Cherry}, {Epstein}, {Karako-Argaman}, {McPhee},
  {Pennucci}, {Roberts}, {Stovall}, \& {van Leeuwen}}]{lbr+13}
{Lynch}, R.~S., {Boyles}, J., {Ransom}, S.~M., {et~al.} 2013, \apj, 763, 81

\bibitem[{{Lynch} {et~al.}(2018){Lynch}, {Swiggum}, {Kondratiev}, {Kaplan},
  {Stovall}, {Fonseca}, {Roberts}, {Levin}, {DeCesar}, {Cui}, {Cenko},
  {Gatkine}, {Archibald}, {Banaszak}, {Biwer}, {Boyles}, {Chawla}, {Dartez},
  {Day}, {Ford}, {Flanigan}, {Hessels}, {Hinojosa}, {Jenet}, {Karako-Argaman},
  {Kaspi}, {Leake}, {Lunsford}, {Martinez}, {Mata}, {McLaughlin}, {Noori},
  {Ransom}, {Rohr}, {Siemens}, {Spiewak}, {Stairs}, {van Leeuwen}, {Walker}, \&
  {Wells}}]{lsk+18}
{Lynch}, R.~S., {Swiggum}, J.~K., {Kondratiev}, V.~I., {et~al.} 2018, \apj,
  859, 93

\bibitem[{{Lyon} {et~al.}(2016){Lyon}, {Stappers}, {Cooper}, {Brooke}, \&
  {Knowles}}]{lsc+16}
{Lyon}, R.~J., {Stappers}, B.~W., {Cooper}, S., {Brooke}, J.~M., \& {Knowles},
  J.~D. 2016, \mnras, 459 [\eprint[arXiv]{1603.05166}]

\bibitem[{{Lyon} {et~al.}(2018){Lyon}, {Stappers}, {Levin}, {Mickaliger}, \&
  {Scaife}}]{lsl+18}
{Lyon}, R.~J., {Stappers}, B.~W., {Levin}, L., {Mickaliger}, M.~B., \&
  {Scaife}, A. 2018, arXiv e-prints [\eprint[arXiv]{1810.06012}]

\bibitem[{{Manchester}(1972)}]{man72}
{Manchester}, R.~N. 1972, \apj, 172, 43

\bibitem[{{Manchester} {et~al.}(2005){Manchester}, {Hobbs}, {Teoh}, \&
  {Hobbs}}]{mht+05}
{Manchester}, R.~N., {Hobbs}, G.~B., {Teoh}, A., \& {Hobbs}, M. 2005, \aj, 129,
  1993

\bibitem[{{Manchester} {et~al.}(2001){Manchester}, {Lyne}, {Camilo}, {Bell},
  {Kaspi}, {D'Amico}, {McKay}, {Crawford}, {Stairs}, {Possenti}, {Kramer}, \&
  {Sheppard}}]{mlc+01}
{Manchester}, R.~N., {Lyne}, A.~G., {Camilo}, F., {et~al.} 2001, \mnras, 328,
  17

\bibitem[{{Manchester} {et~al.}(1996){Manchester}, {Lyne}, {D'Amico}, {Bailes},
  {Johnston}, {Lorimer}, {Harrison}, {Nicastro}, \& {Bell}}]{mla+96}
{Manchester}, R.~N., {Lyne}, A.~G., {D'Amico}, N., {et~al.} 1996, \mnras, 279,
  1235

\bibitem[{{Maron} {et~al.}(2000){Maron}, {Kijak}, {Kramer}, \&
  {Wielebinski}}]{mkkw00}
{Maron}, O., {Kijak}, J., {Kramer}, M., \& {Wielebinski}, R. 2000, \aaps, 147,
  195

\bibitem[{{McLaughlin} {et~al.}(2006){McLaughlin}, {Lyne}, {Lorimer}, {Kramer},
  {Faulkner}, {Manchester}, {Cordes}, {Camilo}, {Possenti}, {Stairs}, {Hobbs},
  {D'Amico}, {Burgay}, \& {O'Brien}}]{mll+06}
{McLaughlin}, M.~A., {Lyne}, A.~G., {Lorimer}, D.~R., {et~al.} 2006, \nat, 439,
  817

\bibitem[{{Michilli} {et~al.}(2018){Michilli}, {Hessels}, {Lyon}, {Tan},
  {Bassa}, {Cooper}, {Kondratiev}, {Sanidas}, {Stappers}, \& {van
  Leeuwen}}]{mhl+18}
{Michilli}, D., {Hessels}, J.~W.~T., {Lyon}, R.~J., {et~al.} 2018, \mnras, 480,
  3457

\bibitem[{{Mol} \& {Romein}(2011)}]{mr11}
{Mol}, J.~D. \& {Romein}, J.~W. 2011, ArXiv e-prints
  [\eprint[arXiv]{1105.0661}]

\bibitem[{{Parent} {et~al.}(2018){Parent}, {Kaspi}, {Ransom}, {Krasteva},
  {Patel}, {Scholz}, {Brazier}, {McLaughlin}, {Boyce}, {Zhu}, {Pleunis},
  {Allen}, {Bogdanov}, {Caballero}, {Camilo}, {Camuccio}, {Chatterjee},
  {Cordes}, {Crawford}, {Deneva}, {Ferdman}, {Freire}, {Hessels}, {Jenet},
  {Knispel}, {Lazarus}, {van Leeuwen}, {Lyne}, {Lynch}, {Seymour}, {Siemens},
  {Stairs}, {Stovall}, \& {Swiggum}}]{pkr+18}
{Parent}, E., {Kaspi}, V.~M., {Ransom}, S.~M., {et~al.} 2018, \apj, 861, 44

\bibitem[{{Pilia} {et~al.}(2016){Pilia}, {Hessels}, {Stappers}, {Kondratiev},
  {Kramer}, {van Leeuwen}, {Weltevrede}, {Lyne}, {Zagkouris}, {Hassall},
  {Bilous}, {Breton}, {Falcke}, {Grie{\ss}meier}, {Keane}, {Karastergiou},
  {Kuniyoshi}, {Noutsos}, {Os{\l}owski}, {Serylak}, {Sobey}, {ter Veen},
  {Alexov}, {Anderson}, {Asgekar}, {Avruch}, {Bell}, {Bentum}, {Bernardi},
  {B{\^i}rzan}, {Bonafede}, {Breitling}, {Broderick}, {Br{\"u}ggen}, {Ciardi},
  {Corbel}, {de Geus}, {de Jong}, {Deller}, {Duscha}, {Eisl{\"o}ffel},
  {Fallows}, {Fender}, {Ferrari}, {Frieswijk}, {Garrett}, {Gunst}, {Hamaker},
  {Heald}, {Horneffer}, {Jonker}, {Juette}, {Kuper}, {Maat}, {Mann}, {Markoff},
  {McFadden}, {McKay-Bukowski}, {Miller-Jones}, {Nelles}, {Paas},
  {Pandey-Pommier}, {Pietka}, {Pizzo}, {Polatidis}, {Reich}, {R{\"o}ttgering},
  {Rowlinson}, {Schwarz}, {Smirnov}, {Steinmetz}, {Stewart}, {Swinbank},
  {Tagger}, {Tang}, {Tasse}, {Thoudam}, {Toribio}, {van der Horst},
  {Vermeulen}, {Vocks}, {van Weeren}, {Wijers}, {Wijnands}, {Wijnholds},
  {Wucknitz}, \& {Zarka}}]{phs+16}
{Pilia}, M., {Hessels}, J.~W.~T., {Stappers}, B.~W., {et~al.} 2016, \aap, 586,
  A92

\bibitem[{{Pleunis} {et~al.}(2017){Pleunis}, {Bassa}, {Hessels}, {Kondratiev},
  {Camilo}, {Cognard}, {Grie{\ss}meier}, {Stappers}, {van Amesfoort}, \&
  {Sanidas}}]{pbh+17}
{Pleunis}, Z., {Bassa}, C.~G., {Hessels}, J.~W.~T., {et~al.} 2017, \apjl, 846,
  L19

\bibitem[{{Rajwade} {et~al.}(2016){Rajwade}, {Lorimer}, \& {Anderson}}]{rla16}
{Rajwade}, K., {Lorimer}, D.~R., \& {Anderson}, L.~D. 2016, \mnras, 455, 493

\bibitem[{{Ransom}(2001)}]{ran01}
{Ransom}, S.~M. 2001, PhD thesis, Harvard University

\bibitem[{{Ransom} {et~al.}(2002){Ransom}, {Eikenberry}, \&
  {Middleditch}}]{rem02}
{Ransom}, S.~M., {Eikenberry}, S.~S., \& {Middleditch}, J. 2002, \aj, 124, 1788

\bibitem[{{Ransom} {et~al.}(2014){Ransom}, {Stairs}, {Archibald}, {Hessels},
  {Kaplan}, {van Kerkwijk}, {Boyles}, {Deller}, {Chatterjee},
  {Schechtman-Rook}, {Berndsen}, {Lynch}, {Lorimer}, {Karako-Argaman}, {Kaspi},
  {Kondratiev}, {McLaughlin}, {van Leeuwen}, {Rosen}, {Roberts}, \&
  {Stovall}}]{rsa+14}
{Ransom}, S.~M., {Stairs}, I.~H., {Archibald}, A.~M., {et~al.} 2014, \nat, 505,
  520

\bibitem[{{Ray} {et~al.}(2012){Ray}, {Abdo}, {Parent}, {Bhattacharya},
  {Bhattacharyya}, {Camilo}, {Cognard}, {Theureau}, {Ferrara}, {Harding},
  {Thompson}, {Freire}, {Guillemot}, {Gupta}, {Roy}, {Hessels}, {Johnston},
  {Keith}, {Shannon}, {Kerr}, {Michelson}, {Romani}, {Kramer}, {McLaughlin},
  {Ransom}, {Roberts}, {Saz Parkinson}, {Ziegler}, {Smith}, {Stappers},
  {Weltevrede}, \& {Wood}}]{rap+12}
{Ray}, P.~S., {Abdo}, A.~A., {Parent}, D., {et~al.} 2012, ArXiv e-prints
  [\eprint[arXiv]{1205.3089}]

\bibitem[{{Ray} {et~al.}(1996){Ray}, {Thorsett}, {Jenet}, {van Kerkwijk},
  {Kulkarni}, {Prince}, {Sandhu}, \& {Nice}}]{rtj+96}
{Ray}, P.~S., {Thorsett}, S.~E., {Jenet}, F.~A., {et~al.} 1996, \apj, 470, 1103

\bibitem[{{Ray} {et~al.}(2017){Ray}, {Wood}, \& {Wolff}}]{rww17}
{Ray}, P.~S., {Wood}, K.~S., \& {Wolff}, M.~T. 2017, arXiv e-prints
  [\eprint[arXiv]{1711.08507}]

\bibitem[{{Rickett}(1970)}]{ric70}
{Rickett}, B.~J. 1970, \mnras, 150, 67

\bibitem[{{Rickett}(1977)}]{ric77}
{Rickett}, B.~J. 1977, \araa, 15, 479

\bibitem[{{Rickett}(1990)}]{ric90}
{Rickett}, B.~J. 1990, \araa, 28, 561

\bibitem[{{Savage} \& {Wakker}(2009)}]{sw09}
{Savage}, B.~D. \& {Wakker}, B.~P. 2009, in American Institute of Physics
  Conference Series, Vol. 1135, American Institute of Physics Conference
  Series, ed. M.~E. {van Steenberg}, G.~{Sonneborn}, H.~W. {Moos}, \& W.~P.
  {Blair}, 46--48

\bibitem[{{Schnitzeler}(2012)}]{sch12}
{Schnitzeler}, D.~H.~F.~M. 2012, \mnras, 427, 664

\bibitem[{{Sobey} {et~al.}(2019){Sobey}, {Bilous}, {Grie{\ss}meier}, {Hessels},
  {Karastergiou}, {Keane}, {Kondratiev}, {Kramer}, {Michilli}, {Noutsos},
  {Pilia}, {Polzin}, {Stappers}, {Tan}, {van Leeuwen}, {Verbiest},
  {Weltevrede}, {Heald}, {Alves}, {Carretti}, {En{\ss}lin}, {Haverkorn},
  {Iacobelli}, {Reich}, \& {Van Eck}}]{sbg+19}
{Sobey}, C., {Bilous}, A.~V., {Grie{\ss}meier}, J.-M., {et~al.} 2019, \mnras,
  484, 3646

\bibitem[{{Spitler} {et~al.}(2014){Spitler}, {Cordes}, {Hessels}, {Lorimer},
  {McLaughlin}, {Chatterjee}, {Crawford}, {Deneva}, {Kaspi}, {Wharton},
  {Allen}, {Bogdanov}, {Brazier}, {Camilo}, {Freire}, {Jenet},
  {Karako-Argaman}, {Knispel}, {Lazarus}, {Lee}, {van Leeuwen}, {Lynch},
  {Ransom}, {Scholz}, {Siemens}, {Stairs}, {Stovall}, {Swiggum},
  {Venkataraman}, {Zhu}, {Aulbert}, \& {Fehrmann}}]{sch+14}
{Spitler}, L.~G., {Cordes}, J.~M., {Hessels}, J.~W.~T., {et~al.} 2014, \apj,
  790, 101

\bibitem[{{Staelin}(1969)}]{stae69}
{Staelin}, D.~H. 1969, IEEE Proceedings, 57, 724

\bibitem[{{Stappers} {et~al.}(2011){Stappers}, {Hessels}, {Alexov}, {Anderson},
  {Coenen}, {Hassall}, {Karastergiou}, {Kondratiev}, {Kramer}, {van Leeuwen},
  {Mol}, {Noutsos}, {Romein}, {Weltevrede}, {Fender}, {Wijers}, {B{\"a}hren},
  {Bell}, {Broderick}, {Daw}, {Dhillon}, {Eisl{\"o}ffel}, {Falcke},
  {Griessmeier}, {Law}, {Markoff}, {Miller-Jones}, {Scheers}, {Spreeuw},
  {Swinbank}, {Ter Veen}, {Wise}, {Wucknitz}, {Zarka}, {Anderson}, {Asgekar},
  {Avruch}, {Beck}, {Bennema}, {Bentum}, {Best}, {Bregman}, {Brentjens}, {van
  de Brink}, {Broekema}, {Brouw}, {Br{\"u}ggen}, {de Bruyn}, {Butcher},
  {Ciardi}, {Conway}, {Dettmar}, {van Duin}, {van Enst}, {Garrett}, {Gerbers},
  {Grit}, {Gunst}, {van Haarlem}, {Hamaker}, {Heald}, {Hoeft}, {Holties},
  {Horneffer}, {Koopmans}, {Kuper}, {Loose}, {Maat}, {McKay-Bukowski},
  {McKean}, {Miley}, {Morganti}, {Nijboer}, {Noordam}, {Norden}, {Olofsson},
  {Pandey-Pommier}, {Polatidis}, {Reich}, {R{\"o}ttgering}, {Schoenmakers},
  {Sluman}, {Smirnov}, {Steinmetz}, {Sterks}, {Tagger}, {Tang}, {Vermeulen},
  {Vermaas}, {Vogt}, {de Vos}, {Wijnholds}, {Yatawatta}, \& {Zensus}}]{sha+11}
{Stappers}, B.~W., {Hessels}, J.~W.~T., {Alexov}, A., {et~al.} 2011, \aap, 530,
  A80

\bibitem[{{Stinebring} {et~al.}(2001){Stinebring}, {McLaughlin}, {Cordes},
  {Becker}, {Goodman}, {Kramer}, {Sheckard}, \& {Smith}}]{smc+01}
{Stinebring}, D.~R., {McLaughlin}, M.~A., {Cordes}, J.~M., {et~al.} 2001,
  \apjl, 549, L97

\bibitem[{{Stovall} {et~al.}(2014){Stovall}, {Lynch}, {Ransom}, {Archibald},
  {Banaszak}, {Biwer}, {Boyles}, {Dartez}, {Day}, {Ford}, {Flanigan}, {Garcia},
  {Hessels}, {Hinojosa}, {Jenet}, {Kaplan}, {Karako-Argaman}, {Kaspi},
  {Kondratiev}, {Leake}, {Lorimer}, {Lunsford}, {Martinez}, {Mata},
  {McLaughlin}, {Roberts}, {Rohr}, {Siemens}, {Stairs}, {van Leeuwen},
  {Walker}, \& {Wells}}]{slr+14}
{Stovall}, K., {Lynch}, R.~S., {Ransom}, S.~M., {et~al.} 2014, \apj, 791, 67

\bibitem[{{Swiggum} {et~al.}(2014){Swiggum}, {Lorimer}, {McLaughlin}, {Bates},
  {Champion}, {Ransom}, {Lazarus}, {Brazier}, {Hessels}, {Nice}, {Ellis},
  {Senty}, {Allen}, {Bhat}, {Bogdanov}, {Camilo}, {Chatterjee}, {Cordes},
  {Crawford}, {Deneva}, {Freire}, {Jenet}, {Karako-Argaman}, {Kaspi},
  {Knispel}, {Lee}, {van Leeuwen}, {Lynch}, {Lyne}, {Scholz}, {Siemens},
  {Stairs}, {Stappers}, {Stovall}, {Venkataraman}, \& {Zhu}}]{slm+14}
{Swiggum}, J.~K., {Lorimer}, D.~R., {McLaughlin}, M.~A., {et~al.} 2014, \apj,
  787, 137

\bibitem[{{Tan} {et~al.}(2018{\natexlab{a}}){Tan}, {Bassa}, {Cooper},
  {Dijkema}, {Esposito}, {Hessels}, {Kondratiev}, {Kramer}, {Michilli},
  {Sanidas}, {Shimwell}, {Stappers}, {van Leeuwen}, {Cognard},
  {Grie{\ss}meier}, {Karastergiou}, {Keane}, {Sobey}, \& {Weltevrede}}]{tbc+18}
{Tan}, C.~M., {Bassa}, C.~G., {Cooper}, S., {et~al.} 2018{\natexlab{a}}, \apj,
  866, 54

\bibitem[{{Tan} {et~al.}(2018{\natexlab{b}}){Tan}, {Lyon}, {Stappers},
  {Cooper}, {Hessels}, {Kondratiev}, {Michilli}, \& {Sanidas}}]{tls+18}
{Tan}, C.~M., {Lyon}, R.~J., {Stappers}, B.~W., {et~al.} 2018{\natexlab{b}},
  \mnras, 474, 4571

\bibitem[{{Tauris} {et~al.}(2015){Tauris}, {Kaspi}, {Breton}, {Deller},
  {Keane}, {Kramer}, {Lorimer}, {McLaughlin}, {Possenti}, {Ray}, {Stappers}, \&
  {Weltevrede}}]{tkb+15}
{Tauris}, T.~M., {Kaspi}, V.~M., {Breton}, R.~P., {et~al.} 2015, Advancing
  Astrophysics with the Square Kilometre Array (AASKA14), 39

\bibitem[{{Thornton} {et~al.}(2013){Thornton}, {Stappers}, {Bailes},
  {Barsdell}, {Bates}, {Bhat}, {Burgay}, {Burke-Spolaor}, {Champion}, {Coster},
  {D'Amico}, {Jameson}, {Johnston}, {Keith}, {Kramer}, {Levin}, {Milia}, {Ng},
  {Possenti}, \& {van Straten}}]{tsb+13}
{Thornton}, D., {Stappers}, B., {Bailes}, M., {et~al.} 2013, Science, 341, 53

\bibitem[{{Tyul'bashev} {et~al.}(2017){Tyul'bashev}, {Tyul'bashev}, {Kitaeva},
  {Chernyshova}, {Malofeev}, {Chashei}, {Shishov}, {Dagkesamanskii},
  {Klimenko}, {Nikitin}, \& {Nikitina}}]{ttk+17}
{Tyul'bashev}, S.~A., {Tyul'bashev}, V.~S., {Kitaeva}, M.~A., {et~al.} 2017,
  Astronomy Reports, 61, 848

\bibitem[{{Tyul'bashev} {et~al.}(2018){Tyul'bashev}, {Tyul'bashev}, \&
  {Malofeev}}]{ttm18}
{Tyul'bashev}, S.~A., {Tyul'bashev}, V.~S., \& {Malofeev}, V.~M. 2018, \aap,
  618, A70

\bibitem[{{Tyul'bashev} {et~al.}(2016){Tyul'bashev}, {Tyul'bashev}, {Oreshko},
  \& {Logvinenko}}]{ttol16}
{Tyul'bashev}, S.~A., {Tyul'bashev}, V.~S., {Oreshko}, V.~V., \& {Logvinenko},
  S.~V. 2016, Astronomy Reports, 60, 220

\bibitem[{{van Haarlem} {et~al.}(2013){van Haarlem}, {Wise}, {Gunst}, {Heald},
  {McKean}, {Hessels}, {de Bruyn}, {Nijboer}, {Swinbank}, {Fallows},
  {Brentjens}, {Nelles}, {Beck}, {Falcke}, {Fender}, {H{\"o}randel},
  {Koopmans}, {Mann}, {Miley}, {R{\"o}ttgering}, {Stappers}, {Wijers},
  {Zaroubi}, {van den Akker}, {Alexov}, {Anderson}, {Anderson}, {van Ardenne},
  {Arts}, {Asgekar}, {Avruch}, {Batejat}, {B{\"a}hren}, {Bell}, {Bell}, {van
  Bemmel}, {Bennema}, {Bentum}, {Bernardi}, {Best}, {B{\^i}rzan}, {Bonafede},
  {Boonstra}, {Braun}, {Bregman}, {Breitling}, {van de Brink}, {Broderick},
  {Broekema}, {Brouw}, {Br{\"u}ggen}, {Butcher}, {van Cappellen}, {Ciardi},
  {Coenen}, {Conway}, {Coolen}, {Corstanje}, {Damstra}, {Davies}, {Deller},
  {Dettmar}, {van Diepen}, {Dijkstra}, {Donker}, {Doorduin}, {Dromer}, {Drost},
  {van Duin}, {Eisl{\"o}ffel}, {van Enst}, {Ferrari}, {Frieswijk}, {Gankema},
  {Garrett}, {de Gasperin}, {Gerbers}, {de Geus}, {Grie{\ss}meier}, {Grit},
  {Gruppen}, {Hamaker}, {Hassall}, {Hoeft}, {Holties}, {Horneffer}, {van der
  Horst}, {van Houwelingen}, {Huijgen}, {Iacobelli}, {Intema}, {Jackson},
  {Jelic}, {de Jong}, {Juette}, {Kant}, {Karastergiou}, {Koers}, {Kollen},
  {Kondratiev}, {Kooistra}, {Koopman}, {Koster}, {Kuniyoshi}, {Kramer},
  {Kuper}, {Lambropoulos}, {Law}, {van Leeuwen}, {Lemaitre}, {Loose}, {Maat},
  {Macario}, {Markoff}, {Masters}, {McFadden}, {McKay-Bukowski}, {Meijering},
  {Meulman}, {Mevius}, {Middelberg}, {Millenaar}, {Miller-Jones}, {Mohan},
  {Mol}, {Morawietz}, {Morganti}, {Mulcahy}, {Mulder}, {Munk}, {Nieuwenhuis},
  {van Nieuwpoort}, {Noordam}, {Norden}, {Noutsos}, {Offringa}, {Olofsson},
  {Omar}, {Orr{\'u}}, {Overeem}, {Paas}, {Pandey-Pommier}, {Pandey}, {Pizzo},
  {Polatidis}, {Rafferty}, {Rawlings}, {Reich}, {de Reijer}, {Reitsma},
  {Renting}, {Riemers}, {Rol}, {Romein}, {Roosjen}, {Ruiter}, {Scaife}, {van
  der Schaaf}, {Scheers}, {Schellart}, {Schoenmakers}, {Schoonderbeek},
  {Serylak}, {Shulevski}, {Sluman}, {Smirnov}, {Sobey}, {Spreeuw}, {Steinmetz},
  {Sterks}, {Stiepel}, {Stuurwold}, {Tagger}, {Tang}, {Tasse}, {Thomas},
  {Thoudam}, {Toribio}, {van der Tol}, {Usov}, {van Veelen}, {van der Veen},
  {ter Veen}, {Verbiest}, {Vermeulen}, {Vermaas}, {Vocks}, {Vogt}, {de Vos},
  {van der Wal}, {van Weeren}, {Weggemans}, {Weltevrede}, {White}, {Wijnholds},
  {Wilhelmsson}, {Wucknitz}, {Yatawatta}, {Zarka}, {Zensus}, \& {van
  Zwieten}}]{hwg+13}
{van Haarlem}, M.~P., {Wise}, M.~W., {Gunst}, A.~W., {et~al.} 2013, \aap, 556,
  A2

\bibitem[{{van Heerden} {et~al.}(2017){van Heerden}, {Karastergiou}, \&
  {Roberts}}]{hkr17}
{van Heerden}, E., {Karastergiou}, A., \& {Roberts}, S.~J. 2017, \mnras, 467,
  1661

\bibitem[{{van Straten} \& {Bailes}(2011)}]{sb10}
{van Straten}, W. \& {Bailes}, M. 2011, \pasa, 28, 1

\bibitem[{{Verbiest} {et~al.}(2016){Verbiest}, {Lentati}, {Hobbs}, {van
  Haasteren}, {Demorest}, {Janssen}, {Wang}, {Desvignes}, {Caballero}, {Keith},
  {Champion}, {Arzoumanian}, {Babak}, {Bassa}, {Bhat}, {Brazier}, {Brem},
  {Burgay}, {Burke-Spolaor}, {Chamberlin}, {Chatterjee}, {Christy}, {Cognard},
  {Cordes}, {Dai}, {Dolch}, {Ellis}, {Ferdman}, {Fonseca}, {Gair},
  {Garver-Daniels}, {Gentile}, {Gonzalez}, {Graikou}, {Guillemot}, {Hessels},
  {Jones}, {Karuppusamy}, {Kerr}, {Kramer}, {Lam}, {Lasky}, {Lassus},
  {Lazarus}, {Lazio}, {Lee}, {Levin}, {Liu}, {Lynch}, {Lyne}, {Mckee},
  {McLaughlin}, {McWilliams}, {Madison}, {Manchester}, {Mingarelli}, {Nice},
  {Os{\l}owski}, {Palliyaguru}, {Pennucci}, {Perera}, {Perrodin}, {Possenti},
  {Petiteau}, {Ransom}, {Reardon}, {Rosado}, {Sanidas}, {Sesana}, {Shaifullah},
  {Shannon}, {Siemens}, {Simon}, {Smits}, {Spiewak}, {Stairs}, {Stappers},
  {Stinebring}, {Stovall}, {Swiggum}, {Taylor}, {Theureau}, {Tiburzi},
  {Toomey}, {Vallisneri}, {van Straten}, {Vecchio}, {Wang}, {Wen}, {You},
  {Zhu}, \& {Zhu}}]{vlh+16}
{Verbiest}, J.~P.~W., {Lentati}, L., {Hobbs}, G., {et~al.} 2016, \mnras, 458,
  1267

\bibitem[{{Yao} {et~al.}(2017){Yao}, {Manchester}, \& {Wang}}]{ymw17}
{Yao}, J.~M., {Manchester}, R.~N., \& {Wang}, N. 2017, \apj, 835, 29

\end{thebibliography}

\begin{appendix}
  \section{Additional tables}

\setcounter{table}{0}
\renewcommand{\thetable}{A.\arabic{table}}

\begin{table*}
  \centering
  \footnotesize
  \caption{\label{table:redetection} List of known pulsars detected by
    LOTAAS. The table shows the detected spin period ($P$) and
    $\mathrm{DM}$ of the pulsars, the offset between the known
    position of the pulsars and the centre position of the TAB
    ($\theta_\mathrm{TAB}$) and SAP ($\theta_\mathrm{SAP}$) where the
    pulsar is detected, the signal-to-noise $\sigma$ and estimated
    flux density of the pulsars at 135\,MHz, before
    ($S_{135}^\mathrm{uncor}$) and after ($_{135}^\mathrm{cor}$)
    applying the correction factor ($S_{135}^{\times}$) due to the
    offset in position. Flux densities at 400\,MHz ($S_{400}$) and
    150\,MHz ($S_{150}$), and spectral indices ($\alpha$) as listed in
    PSRCAT v1.59 \citep{mht+05} are also provided, and used to
    estimate the expected flux density at 135\,MHz
    ($S_{135}^\mathrm{exp}$).}
  \begin{tabular}{lrrrrrrrrrrrrl}
    \hline
    PSR & \multicolumn{1}{c}{$P$} & \multicolumn{1}{c}{$\mathrm{DM}$} & \multicolumn{1}{c}{$\theta_\mathrm{TAB}$} & \multicolumn{1}{c}{$\theta_\mathrm{SAP}$} & \multicolumn{1}{c}{$\sigma$} & \multicolumn{1}{c}{$S_{135}^\mathrm{uncor}$} & \multicolumn{1}{c}{$S_{135}^{\times}$} & \multicolumn{1}{c}{$S_{135}^\mathrm{cor}$} & \multicolumn{1}{c}{$S_{400}$} & \multicolumn{1}{c}{$S_{150}$} & \multicolumn{1}{c}{$\alpha$} & \multicolumn{1}{c}{$S_{135}^\mathrm{exp}$} & \multicolumn{1}{c}{Notes} \\
    & \multicolumn{1}{c}{(s)} & \multicolumn{1}{c}{(pc\,cm$^{-3}$)} & \multicolumn{1}{c}{(\degr)} & \multicolumn{1}{c}{(\degr)} & & \multicolumn{1}{c}{(mJy)} & & \multicolumn{1}{c}{(mJy)} & \multicolumn{1}{c}{(mJy)} & \multicolumn{1}{c}{(mJy)} & & \multicolumn{1}{c}{(mJy)} & \\
    \hline
    B0011$+$47 & 1.2407 & 30.3 & 0.1 & 0.7 & 63.6 & 22.2 & 1.4 & 31.1 & 14.0 & 35.0 & $-$1.00 & 38.9 & \\
    B0031$-$07 & 0.9430 & 10.8 & 6.7 & 5.0 & 30.3 & 54.2 & $\ldots$ & $\ldots$ & 52.0 & 640.0 & $-$1.60 & 757.5 & \\
    B0037$+$56 & 1.1182 & 92.6 & 0.0 & 1.2 & 33.6 & 15.3 & 1.3 & 19.4 & 7.5 & 33.0 & $-$1.70 & 39.5 & \\
    B0045$+$33 & 1.2171 & 39.9 & 0.1 & 0.2 & 107.6 & 20.4 & 1.2 & 24.9 & 2.3 & 13.0 & $-$2.40 & 16.7 & \\
    B0052$+$51 & 2.1152 & 44.1 & 0.0 & 2.1 & 16.3 & 3.7 & 2.1 & 7.9 & 3.4 & 9.4 & $-$1.00 & 10.4 & \\
    B0053$+$47 & 0.4720 & 18.1 & 0.1 & 0.7 & 59.9 & 20.5 & 1.1 & 23.3 & 3.0 & 7.5 & $-$1.20 & 8.5 & \\
    B0059$+$65 & 1.6792 & 65.8 & 0.1 & 1.0 & 11.9 & 5.4 & 1.4 & 7.4 & 8.5 & $\ldots$ & $-$1.60 & 48.3 & \\
    B0105$+$65 & 1.2837 & 30.5 & 0.0 & 1.4 & 89.4 & 28.9 & 1.4 & 39.2 & 10.0 & 41.0 & $-$1.40 & 47.5 & \\
    B0105$+$68 & 1.0711 & 61.1 & 0.0 & 2.5 & 7.0 & 2.7 & 3.0 & 8.3 & 3.7 & 8.9 & $-$1.50 & 10.4 & \\
    B0114$+$58 & 0.1014 & 49.4 & 0.0 & 1.2 & 46.8 & 43.9 & 1.2 & 54.3 & 7.6 & 71.3 & $-$2.45 & 92.3 & \\
    B0136$+$57 & 0.2725 & 73.8 & 0.1 & 1.0 & 130.9 & 121.2 & 1.6 & 197.6 & 28.0 & 154.5 & $-$1.57 & 182.3 & \\
    B0138$+$59 & 1.2230 & 34.9 & 0.0 & 1.1 & 289.5 & 84.6 & 1.2 & 102.6 & 49.0 & 138.2 & $-$1.53 & 162.4 & \\
    B0144$+$59 & 0.1963 & 40.1 & 0.0 & 1.4 & 43.3 & 19.3 & 1.4 & 26.3 & 6.6 & 33.1 & $-$1.23 & 37.7 & 3 \\
    B0153$+$39 & 1.8116 & 59.6 & 0.0 & 1.7 & 10.6 & 2.2 & 1.6 & 3.5 & 4.0 & 8.5 & $-$1.40 & 9.9 & \\
    B0154$+$61 & 2.3519 & 30.2 & 0.0 & 1.2 & 10.5 & 3.9 & 1.2 & 4.8 & 6.5 & $\ldots$ & $-$1.20 & 23.9 & 3 \\
    B0226$+$70 & 1.4668 & 46.7 & 0.1 & 0.6 & 45.1 & 10.5 & 1.2 & 12.3 & 2.4 & 5.3 & $-$1.50 & 6.2 & \\
    B0301$+$19 & 1.3876 & 15.6 & 0.9 & 1.5 & 35.1 & 12.1 & $\ldots$ & $\ldots$ & 27.0 & 42.0 & $-$1.40 & 48.7 & \\
    B0320$+$39 & 3.0321 & 26.2 & 0.0 & 2.3 & 269.1 & 43.8 & 2.5 & 109.6 & 34.0 & 92.0 & $-$2.80 & 123.6 & \\
    B0329$+$54 & 0.7145 & 26.8 & 2.2 & 1.2 & 434.2 & 168.8 & $\ldots$ & $\ldots$ & 1500.0 & $\ldots$ & $-$1.60 & 8528.1 & \\
    B0331$+$45 & 0.2692 & 47.1 & 0.1 & 0.1 & 50.6 & 11.7 & 1.1 & 13.4 & 6.0 & 34.0 & $-$1.90 & 41.5 & \\
    B0339$+$53 & 1.9345 & 67.5 & 2.1 & 1.1 & 15.2 & 4.3 & $\ldots$ & $\ldots$ & 3.7 & $\ldots$ & $-$1.90 & 29.1 & \\
    B0355$+$54 & 0.1564 & 57.1 & 0.0 & 2.3 & 68.0 & 35.3 & 2.5 & 88.4 & 46.0 & $\ldots$ & $-$0.90 & 122.3 & \\
    B0402$+$61 & 0.5946 & 65.3 & 0.0 & 1.6 & 69.6 & 36.3 & 1.5 & 55.2 & $\ldots$ & $\ldots$ & $-$1.20 & $\ldots$ & 6 \\
    B0410$+$69 & 0.3907 & 27.5 & 0.0 & 1.4 & 55.4 & 16.3 & 1.3 & 21.7 & 6.4 & 28.0 & $-$1.80 & 33.8 & \\
    B0450$+$55 & 0.3407 & 14.6 & 0.0 & 2.3 & 73.4 & 28.5 & 2.3 & 65.9 & 59.0 & 91.0 & $-$1.20 & 103.3 & \\
    B0523$+$11 & 0.3544 & 79.5 & 0.0 & 1.7 & 43.5 & 37.2 & 1.6 & 59.9 & 19.5 & 34.0 & $-$1.90 & 41.5 & \\
    B0525$+$21 & 3.7456 & 50.9 & 0.0 & 1.4 & 234.7 & 21.8 & 1.4 & 29.5 & 57.0 & 230.0 & $-$1.50 & 269.4 & \\
    B0531$+$21 & 0.0337 & 56.8 & 3.0 & 2.5 & 9.6 & 432.5 & $\ldots$ & $\ldots$ & 550.0 & 7500.0 & $-$3.10 & 10397.0 & \\
    B0540$+$23 & 0.2460 & 77.7 & 0.0 & 1.5 & 25.1 & 15.8 & 1.4 & 22.4 & 29.0 & 36.0 & $-$0.70 & 38.8 & \\
    B0609$+$37 & 0.2980 & 27.1 & 0.0 & 1.0 & 30.8 & 9.1 & 1.2 & 10.6 & 16.0 & 21.0 & $-$1.50 & 24.6 & \\
    B0611$+$22 & 0.3350 & 96.9 & 0.1 & 0.5 & 67.8 & 55.1 & 1.2 & 66.3 & 29.0 & $\ldots$ & $-$1.77 & 198.3 & \\
    B0621$-$04 & 1.0391 & 70.8 & 0.0 & 2.0 & 17.1 & 8.1 & 1.9 & 15.8 & 4.9 & $\ldots$ & $-$1.00 & 14.5 & 3 \\
    B0626$+$24 & 0.4766 & 84.2 & 0.2 & 0.6 & 97.3 & 32.8 & $\ldots$ & $\ldots$ & 31.0 & 73.0 & $-$1.50 & 85.5 & \\
    B0643$+$80 & 1.2144 & 33.3 & 0.0 & 0.9 & 65.6 & 11.6 & 1.1 & 13.1 & 6.5 & 19.0 & $-$1.50 & 22.3 & \\
    B0655$+$64 & 0.1957 & 8.8 & 0.2 & 0.7 & 376.1 & 80.6 & 1.9 & 157.1 & 5.0 & 51.0 & $-$2.20 & 64.3 & \\
    B0656$+$14 & 0.3849 & 14.1 & 0.0 & 1.1 & 16.0 & 7.2 & 1.2 & 8.6 & 6.5 & 11.0 & $-$1.10 & 12.4 & \\
    B0751$+$32 & 1.4423 & 40.0 & 0.2 & 0.7 & 44.8 & 10.5 & 2.2 & 22.9 & 8.0 & 17.0 & $-$1.50 & 19.9 & \\
    B0809$+$74 & 1.2922 & 5.8 & 4.1 & 2.0 & 390.5 & 101.8 & $\ldots$ & $\ldots$ & 79.0 & 300.0 & $-$1.70 & 358.8 & \\
    B0820$+$02 & 0.8649 & 23.8 & 0.0 & 2.0 & 110.6 & 41.0 & 1.9 & 77.9 & 30.0 & $\ldots$ & $-$2.30 & 364.8 & \\
    B0823$+$26 & 0.5307 & 19.5 & 0.0 & 2.0 & 126.3 & 306.1 & 2.0 & 605.6 & 73.0 & 520.0 & $-$1.60 & 615.5 & \\
    B0834$+$06 & 1.2738 & 12.9 & 2.0 & 2.0 & 1266.0 & 286.2 & $\ldots$ & $\ldots$ & 89.0 & $\ldots$ & $-$2.70 & 1671.3 & \\
    B0841$+$80 & 1.6022 & 34.9 & 0.1 & 0.4 & 10.8 & 3.0 & 1.5 & 4.6 & 1.5 & 5.9 & $-$1.60 & 7.0 & \\
    B0917$+$63 & 1.5680 & 13.1 & 0.0 & 1.8 & 29.2 & 6.2 & 1.7 & 10.8 & 5.0 & 22.0 & $-$1.50 & 25.8 & \\
    B0919$+$06 & 0.4306 & 27.3 & 2.6 & 1.1 & 271.6 & 254.2 & $\ldots$ & $\ldots$ & 52.0 & $\ldots$ & $-$1.80 & 367.4 & \\
    B0940$+$16 & 1.0874 & 20.2 & 0.0 & 1.2 & 27.7 & 15.5 & 1.3 & 19.6 & 7.0 & 26.0 & $-$1.00 & 28.9 & \\
    B0943$+$10 & 1.0977 & 15.3 & 0.2 & 0.8 & 597.8 & 156.8 & 2.1 & 330.2 & 4.0 & 440.0 & $\ldots$ & 509.9 & \\
    B0950$+$08 & 0.2531 & 3.0 & 0.2 & 0.4 & 278.5 & 1098.6 & 2.2 & 2368.0 & 400.0 & 2600.0 & $-$1.30 & 2981.7 & \\
    B1112$+$50 & 1.6564 & 9.2 & 0.0 & 1.1 & 286.6 & 22.5 & 1.2 & 27.4 & 12.0 & 50.0 & $-$1.80 & 60.4 & \\
    B1133$+$16 & 1.1879 & 4.8 & 0.0 & 2.7 & 879.9 & 217.8 & 3.5 & 756.3 & 257.0 & 940.0 & $-$1.50 & 1100.9 & \\
    B1237$+$25 & 1.3824 & 9.3 & 0.0 & 1.9 & 301.9 & 67.4 & 1.7 & 117.5 & 110.0 & 170.0 & $-$1.80 & 205.5 & \\
    B1257$+$12 & 0.0062 & 10.2 & 0.2 & 0.7 & 161.6 & 121.6 & 1.8 & 218.4 & 20.0 & 80.9 & $-$1.80 & 97.8 & \\
    B1322$+$83 & 0.6700 & 13.3 & 0.0 & 1.3 & 73.7 & 19.2 & 1.3 & 25.0 & 11.0 & 28.0 & $-$1.70 & 33.5 & \\
    B1508$+$55 & 0.7397 & 19.6 & 3.3 & 4.0 & 95.9 & 13.8 & $\ldots$ & $\ldots$ & 114.0 & 770.0 & $-$2.30 & 981.1 & \\
    B1530$+$27 & 1.1248 & 14.7 & 0.0 & 1.6 & 126.8 & 38.4 & 1.5 & 57.3 & 13.0 & 33.0 & $-$2.00 & 40.7 & \\
    B1534$+$12 & 0.0379 & 11.6 & 0.0 & 0.5 & 24.1 & 14.9 & 1.1 & 15.7 & 36.0 & $\ldots$ & $-$1.90 & 283.5 & 3 \\
    B1540$-$06 & 0.7091 & 18.4 & 6.1 & 3.8 & 35.7 & 14.6 & $\ldots$ & $\ldots$ & 40.0 & $\ldots$ & $-$2.50 & 604.5 & \\
    B1541$+$09 & 0.7484 & 35.0 & 0.0 & 2.2 & 286.6 & 260.7 & 2.2 & 572.2 & 78.0 & 770.0 & $-$2.10 & 960.7 & \\
    B1604$-$00 & 0.4218 & 10.7 & 0.2 & 0.7 & 278.8 & 138.0 & 2.4 & 329.9 & 54.0 & $\ldots$ & $-$1.90 & 425.3 & \\
    B1612$+$07 & 1.2068 & 21.4 & 0.0 & 1.9 & 83.3 & 29.3 & 1.8 & 52.3 & 9.6 & $\ldots$ & $-$2.70 & 180.3 & \\
    B1633$+$24 & 0.4905 & 24.3 & 0.0 & 1.5 & 104.7 & 45.6 & 1.4 & 63.8 & 9.1 & 79.0 & $-$2.10 & 98.6 & \\
    B1642$-$03 & 0.3877 & 35.7 & 0.0 & 1.2 & 933.0 & 658.8 & 1.3 & 830.5 & 393.0 & $\ldots$ & $-$2.30 & 4779.3 & \\
    \hline
  \end{tabular}
\end{table*}

\setcounter{table}{0}
\begin{table*}
  \footnotesize
  \caption{Continued.}
  \begin{tabular}{lrrrrrrrrrrrrl}
    \hline
    PSR & \multicolumn{1}{c}{$P$} & \multicolumn{1}{c}{$\mathrm{DM}$} & \multicolumn{1}{c}{$\theta_\mathrm{TAB}$} & \multicolumn{1}{c}{$\theta_\mathrm{SAP}$} & \multicolumn{1}{c}{$\sigma$} & \multicolumn{1}{c}{$S_{135}^\mathrm{uncor}$} & \multicolumn{1}{c}{$S_{135}^{\times}$} & \multicolumn{1}{c}{$S_{135}^\mathrm{cor}$} & \multicolumn{1}{c}{$S_{400}$} & \multicolumn{1}{c}{$S_{150}$} & \multicolumn{1}{c}{$\alpha$} & \multicolumn{1}{c}{$S_{135}^\mathrm{exp}$} & \multicolumn{1}{c}{Notes} \\
    & \multicolumn{1}{c}{(s)} & \multicolumn{1}{c}{(pc\,cm$^{-3}$)} & \multicolumn{1}{c}{(\degr)} & \multicolumn{1}{c}{(\degr)} & & \multicolumn{1}{c}{(mJy)} & & \multicolumn{1}{c}{(mJy)} & \multicolumn{1}{c}{(mJy)} & \multicolumn{1}{c}{(mJy)} & & \multicolumn{1}{c}{(mJy)} & \\
    \hline
    B1718$-$02 & 0.4777 & 66.9 & 0.2 & 0.2 & 22.9 & 23.7 & 2.5 & 60.3 & 22.0 & $\ldots$ & $-$2.20 & 240.0 & 3 \\
    B1726$-$00 & 0.3860 & 41.0 & 0.0 & 2.0 & 41.7 & 23.9 & 2.0 & 47.2 & 11.0 & $\ldots$ & $-$2.30 & 133.8 & 3 \\
    B1737$+$13 & 0.8031 & 48.6 & 0.0 & 2.9 & 58.6 & 21.6 & 4.4 & 94.4 & 24.0 & 81.0 & $-$1.40 & 93.9 & \\
    B1753$+$52 & 0.7971 & 35.1 & 0.0 & 1.0 & 8.1 & 1.9 & 1.2 & 2.3 & 4.9 & 2.4 & $-$1.10 & 2.7 & \\
    B1758$-$03 & 0.9215 & 120.5 & 2.5 & 1.5 & 10.8 & 17.9 & $\ldots$ & $\ldots$ & 17.0 & $\ldots$ & $-$2.60 & 286.4 & 3 \\
    B1802$+$03 & 0.2187 & 81.0 & 0.0 & 1.1 & 7.9 & 15.3 & 1.2 & 18.7 & 5.0 & $\ldots$ & $\ldots$ & 22.9 & 3 \\
    B1811$+$40 & 0.9311 & 41.6 & 0.1 & 0.2 & 162.3 & 31.7 & 1.2 & 37.7 & 8.0 & 50.0 & $-$1.20 & 56.7 & \\
    B1818$-$04 & 0.5981 & 84.4 & 0.0 & 2.2 & 43.2 & 153.5 & 2.2 & 335.2 & 157.0 & 830.0 & $-$2.10 & 1035.5 & \\
    B1821$+$05 & 0.7529 & 66.8 & 0.0 & 2.7 & 59.4 & 35.3 & 3.5 & 124.1 & 18.0 & $\ldots$ & $-$1.70 & 114.1 & \\
    B1822$+$00 & 0.7790 & 56.6 & 0.1 & 0.4 & 16.7 & 18.8 & 1.2 & 21.7 & 8.0 & $\ldots$ & $-$2.30 & 97.3 & 3 \\
    B1831$-$00 & 0.5210 & 89.0 & 0.2 & 0.2 & 17.6 & 33.6 & 2.9 & 97.3 & 5.1 & $\ldots$ & $\ldots$ & 23.3 & 3 \\
    B1831$-$04 & 0.2901 & 79.4 & 0.0 & 1.9 & 55.6 & 350.4 & 1.8 & 621.6 & 77.0 & $\ldots$ & $-$1.30 & 316.0 & \\
    B1839$+$09 & 0.3813 & 49.2 & 0.0 & 2.2 & 37.2 & 31.2 & 2.2 & 69.0 & 20.0 & 130.0 & $-$2.00 & 160.5 & \\
    B1839$+$56 & 1.6529 & 26.8 & 0.0 & 1.6 & 203.3 & 37.5 & 1.5 & 55.0 & 21.0 & 76.0 & $-$1.40 & 88.1 & \\
    B1842$+$14 & 0.3755 & 41.5 & 0.0 & 0.4 & 345.9 & 207.4 & 1.1 & 222.1 & 20.0 & 110.0 & $-$1.95 & 135.1 & \\
    B1848$+$12 & 1.2053 & 70.6 & 0.0 & 1.6 & 20.6 & 14.4 & 1.5 & 21.4 & 8.0 & 37.0 & $-$1.80 & 44.7 & \\
    B1848$+$13 & 0.3456 & 60.2 & 0.1 & 0.7 & 19.3 & 13.0 & 1.4 & 17.9 & 6.0 & 5.7 & $-$1.50 & 6.7 & \\
    B1859$+$01 & 0.2882 & 105.4 & 0.2 & 0.2 & 46.2 & 152.9 & 1.9 & 283.0 & 13.7 & $\ldots$ & $-$2.90 & 319.7 & 3 \\
    B1905$+$39 & 1.2358 & 30.9 & 0.1 & 0.5 & 119.0 & 43.1 & 1.2 & 53.8 & 23.0 & 46.0 & $-$2.00 & 56.8 & \\
    B1907$+$00 & 1.0170 & 112.8 & 0.0 & 1.1 & 26.4 & 30.2 & 1.2 & 36.3 & 12.0 & $\ldots$ & $-$1.80 & 84.8 & \\
    B1907$+$02 & 0.9898 & 171.7 & 0.0 & 1.7 & 19.0 & 44.5 & 1.6 & 70.2 & 21.0 & $\ldots$ & $-$2.36 & 272.6 & \\
    B1907$+$03 & 2.3303 & 82.7 & 0.1 & 0.4 & 18.5 & 34.7 & 1.5 & 50.7 & 21.0 & $\ldots$ & $-$0.70 & 44.9 & 3 \\
    B1907$+$10 & 0.2836 & 149.9 & 0.0 & 1.0 & 40.8 & 186.9 & 1.2 & 217.2 & 50.0 & $\ldots$ & $-$2.50 & 755.6 & \\
    B1907$-$03 & 0.5046 & 206.0 & 0.0 & 2.3 & 4.6 & 15.5 & 2.4 & 36.6 & 27.0 & $\ldots$ & $-$2.60 & 454.8 & 3 \\
    B1911$-$04 & 0.8259 & 89.4 & 0.0 & 1.7 & 100.0 & 139.0 & 1.6 & 222.5 & 118.0 & $\ldots$ & $-$2.60 & 1987.8 & \\
    B1913$+$167 & 1.6162 & 62.6 & 0.1 & 0.4 & 13.4 & 6.4 & 1.1 & 7.0 & 4.5 & $\ldots$ & $-$1.50 & 23.0 & 3 \\
    B1914$+$09 & 0.2703 & 61.0 & 0.1 & 0.4 & 36.5 & 79.6 & 1.4 & 110.3 & 20.0 & $\ldots$ & $-$2.30 & 243.2 & \\
    B1915$+$13 & 0.1946 & 94.7 & 0.0 & 1.4 & 71.1 & 145.9 & 1.4 & 198.7 & 43.0 & $\ldots$ & $-$1.80 & 303.8 & \\
    B1915$+$22 & 0.4259 & 134.7 & 0.2 & 0.7 & 13.0 & 9.6 & 2.4 & 23.2 & 3.0 & 12.0 & $-$2.22 & 15.2 & \\
    B1917$+$00 & 1.2723 & 90.2 & 0.0 & 2.5 & 31.0 & 32.8 & 2.8 & 92.0 & 16.0 & $\ldots$ & $-$2.30 & 194.6 & \\
    B1918$+$19 & 0.8210 & 153.6 & 0.2 & 0.9 & 28.7 & 29.2 & $\ldots$ & $\ldots$ & 34.0 & $\ldots$ & $-$2.40 & 460.9 & 3 \\
    B1918$+$26 & 0.7855 & 27.7 & 0.0 & 1.9 & 36.9 & 11.3 & 1.7 & 19.8 & 6.0 & 21.0 & $-$1.30 & 24.1 & \\
    B1919$+$14 & 0.6182 & 91.7 & 0.0 & 1.3 & 17.5 & 19.3 & 1.3 & 25.6 & 3.2 & $\ldots$ & $-$1.00 & 9.5 & 3 \\
    B1919$+$21 & 1.3373 & 12.4 & 1.6 & 1.8 & 283.4 & 82.2 & $\ldots$ & $\ldots$ & 57.0 & 1300.0 & $-$1.90 & 1588.1 & \\
    B1920$+$21 & 1.0779 & 217.0 & 0.1 & 0.2 & 64.8 & 60.1 & 1.1 & 68.4 & 30.0 & $\ldots$ & $-$2.40 & 406.7 & \\
    B1923$+$04 & 1.0741 & 102.1 & 0.0 & 1.4 & 20.3 & 18.2 & 1.4 & 24.6 & 22.0 & $\ldots$ & $-$2.70 & 413.1 & \\
    B1926$+$18 & 1.2205 & 111.9 & 0.0 & 1.6 & 11.0 & 7.5 & 1.5 & 11.3 & 1.7 & $\ldots$ & $\ldots$ & 7.8 & 3 \\
    B1929$+$10 & 0.2265 & 3.2 & 2.8 & 3.4 & 60.2 & 65.3 & $\ldots$ & $\ldots$ & 303.0 & 540.0 & $-$1.70 & 645.9 & \\
    B1931$+$24 & 0.8137 & 105.9 & 0.0 & 2.7 & 10.4 & 8.5 & 3.5 & 29.7 & 7.5 & $\ldots$ & $\ldots$ & 34.3 & 3 \\
    B1933$+$16 & 0.3587 & 158.6 & 0.0 & 1.0 & 118.8 & 126.9 & 1.2 & 148.3 & 242.0 & $\ldots$ & $-$1.40 & 1107.2 & 3 \\
    B1935$+$25 & 0.2010 & 53.3 & 0.0 & 1.4 & 11.9 & 12.8 & 1.3 & 17.2 & 6.6 & $\ldots$ & $-$0.70 & 14.1 & 3 \\
    B1942$-$00 & 1.0456 & 59.7 & 0.0 & 2.8 & 18.6 & 10.2 & 3.8 & 38.6 & 6.0 & $\ldots$ & $-$1.80 & 42.4 & 3 \\
    B1944$+$17 & 0.4406 & 16.1 & 0.0 & 1.5 & 35.7 & 30.9 & 1.4 & 44.6 & 40.0 & 41.0 & $-$0.90 & 45.1 & \\
    B1946$+$35 & 0.7173 & 128.9 & 0.3 & 0.9 & 7.0 & 10.7 & $\ldots$ & $\ldots$ & 145.0 & 170.0 & $-$2.20 & 214.3 & \\
    B1949$+$14 & 0.2750 & 31.5 & 0.0 & 1.2 & 31.9 & 12.3 & 1.2 & 15.2 & 6.0 & 6.4 & $-$1.00 & 7.1 & \\
    B1951$+$32 & 0.0395 & 45.2 & 0.0 & 1.7 & 40.2 & 50.6 & 1.6 & 78.8 & 7.0 & $\ldots$ & $-$1.60 & 39.8 & 3 \\
    B1952$+$29 & 0.4267 & 7.9 & 0.0 & 2.1 & 43.5 & 27.5 & 2.1 & 56.9 & 6.6 & $\ldots$ & 0.00 & 6.6 & 3 \\
    B1953$+$50 & 0.5189 & 32.0 & 0.0 & 1.6 & 129.1 & 48.9 & 1.5 & 74.4 & 26.0 & 48.0 & $-$1.30 & 55.0 & \\
    B2016$+$28 & 0.5580 & 14.2 & 0.0 & 2.7 & 153.9 & 180.0 & 3.4 & 618.8 & 314.0 & 810.0 & $-$2.30 & 1032.1 & \\
    B2020$+$28 & 0.3434 & 24.6 & 0.0 & 1.7 & 283.3 & 138.2 & 1.6 & 217.7 & 71.0 & 150.0 & $-$0.40 & 156.5 & \\
    B2021$+$51 & 0.5292 & 22.6 & 0.0 & 0.5 & 75.1 & 44.5 & 1.1 & 48.6 & 77.0 & 59.0 & $-$0.70 & 63.5 & \\
    B2022$+$50 & 0.3726 & 33.0 & 0.2 & 1.1 & 17.5 & 15.4 & 2.2 & 33.6 & 6.5 & 36.0 & $-$1.10 & 40.4 & \\
    B2027$+$37 & 1.2168 & 191.0 & 0.0 & 1.5 & 4.6 & 66.3 & 1.5 & 96.7 & 18.0 & $\ldots$ & $-$2.60 & 303.2 & 3 \\
    B2028$+$22 & 0.6305 & 71.8 & 0.2 & 0.9 & 8.4 & 4.2 & 2.3 & 9.7 & 5.0 & 7.4 & $-$1.20 & 8.4 & \\
    B2034$+$19 & 2.0744 & 36.9 & 0.0 & 1.2 & 22.6 & 4.9 & 1.3 & 6.3 & 2.0 & 20.0 & $-$2.50 & 26.0 & \\
    B2043$-$04 & 1.5469 & 35.9 & 0.0 & 1.8 & 66.5 & 24.9 & 1.7 & 41.6 & 20.0 & $\ldots$ & $-$1.70 & 126.8 & \\
    B2044$+$15 & 1.1383 & 39.8 & 0.3 & 0.7 & 52.0 & 20.3 & $\ldots$ & $\ldots$ & 11.5 & 33.0 & $-$1.60 & 39.1 & \\
    B2045$+$56 & 0.4767 & 101.8 & 0.0 & 1.3 & 39.4 & 14.5 & 1.3 & 18.5 & 4.6 & 25.0 & $-$1.80 & 30.2 & \\
    B2053$+$21 & 0.8152 & 36.4 & 0.0 & 1.2 & 96.5 & 22.9 & 1.3 & 28.7 & 9.0 & 29.0 & $-$1.10 & 32.6 & \\
    B2053$+$36 & 0.2215 & 97.5 & 0.1 & 0.8 & 10.4 & 18.9 & 1.6 & 30.1 & 28.0 & 29.0 & $-$1.90 & 35.4 & \\
    B2106$+$44 & 0.4149 & 140.1 & 0.0 & 0.8 & 9.0 & 24.9 & 1.1 & 28.1 & 26.0 & $\ldots$ & $-$1.30 & 106.7 & \\
    B2110$+$27 & 1.2029 & 25.2 & 0.0 & 1.8 & 217.9 & 41.5 & 1.7 & 70.7 & 18.0 & 100.0 & $-$2.30 & 127.4 & \\
    B2111$+$46 & 1.0147 & 141.6 & 0.2 & 0.9 & 68.6 & 91.7 & 2.7 & 244.5 & 230.0 & $\ldots$ & $-$2.00 & 2019.2 & 3 \\
    B2113$+$14 & 0.4402 & 56.2 & 0.1 & 0.9 & 31.7 & 14.4 & 1.4 & 19.5 & 9.0 & 9.4 & $-$1.70 & 11.2 & \\
    B2122$+$13 & 0.6941 & 30.2 & 0.1 & 0.8 & 19.1 & 7.0 & 1.6 & 11.1 & 4.0 & 4.0 & $-$1.50 & 4.7 & \\
    \hline
  \end{tabular}
\end{table*}

\setcounter{table}{0}
\begin{table*}
  \footnotesize
  \caption{Continued.}
  \begin{tabular}{lrrrrrrrrrrrrl}
    \hline
    PSR & \multicolumn{1}{c}{$P$} & \multicolumn{1}{c}{$\mathrm{DM}$} & \multicolumn{1}{c}{$\theta_\mathrm{TAB}$} & \multicolumn{1}{c}{$\theta_\mathrm{SAP}$} & \multicolumn{1}{c}{$\sigma$} & \multicolumn{1}{c}{$S_{135}^\mathrm{uncor}$} & \multicolumn{1}{c}{$S_{135}^{\times}$} & \multicolumn{1}{c}{$S_{135}^\mathrm{cor}$} & \multicolumn{1}{c}{$S_{400}$} & \multicolumn{1}{c}{$S_{150}$} & \multicolumn{1}{c}{$\alpha$} & \multicolumn{1}{c}{$S_{135}^\mathrm{exp}$} & \multicolumn{1}{c}{Notes} \\
    & \multicolumn{1}{c}{(s)} & \multicolumn{1}{c}{(pc\,cm$^{-3}$)} & \multicolumn{1}{c}{(\degr)} & \multicolumn{1}{c}{(\degr)} & & \multicolumn{1}{c}{(mJy)} & & \multicolumn{1}{c}{(mJy)} & \multicolumn{1}{c}{(mJy)} & \multicolumn{1}{c}{(mJy)} & & \multicolumn{1}{c}{(mJy)} & \\
    \hline
    B2148$+$52 & 0.3322 & 149.1 & 0.0 & 1.3 & 12.4 & 8.3 & 1.3 & 10.7 & 15.6 & $\ldots$ & $-$1.30 & 64.0 & 3 \\
    B2148$+$63 & 0.3801 & 129.7 & 0.0 & 1.7 & 33.9 & 30.9 & 1.6 & 48.5 & 32.0 & 64.0 & $-$1.70 & 76.6 & \\
    B2154$+$40 & 1.5253 & 71.1 & 0.2 & 0.5 & 117.6 & 48.1 & 2.2 & 105.4 & 105.0 & 190.0 & $-$1.50 & 222.5 & \\
    B2210$+$29 & 1.0046 & 74.5 & 0.0 & 1.4 & 37.4 & 10.3 & 1.4 & 14.3 & 6.3 & 17.0 & $-$1.40 & 19.7 & \\
    B2217$+$47 & 0.5385 & 43.5 & 0.0 & 2.4 & 198.7 & 285.2 & 2.7 & 781.2 & 111.0 & 820.0 & $-$1.98 & 1010.2 & \\
    B2224$+$65 & 0.6825 & 36.4 & 0.0 & 1.7 & 155.6 & 81.9 & 1.5 & 126.3 & 22.0 & 160.0 & $-$1.66 & 190.6 & \\
    B2227$+$61 & 0.4431 & 124.6 & 0.0 & 1.1 & 44.8 & 48.2 & 1.2 & 58.9 & 17.0 & 90.0 & $-$1.80 & 108.8 & \\
    B2241$+$69 & 1.6645 & 41.0 & 0.0 & 1.3 & 39.4 & 9.7 & 1.3 & 12.4 & 2.4 & 26.0 & $-$2.00 & 32.1 & \\
    B2255$+$58 & 0.3683 & 151.1 & 0.1 & 0.2 & 30.6 & 140.6 & 1.1 & 149.0 & 34.0 & $\ldots$ & $-$0.80 & 81.1 & \\
    B2303$+$30 & 1.5759 & 49.6 & 0.0 & 0.5 & 162.3 & 32.5 & 1.1 & 34.9 & 24.0 & 70.0 & $-$2.30 & 89.2 & \\
    B2303$+$46 & 1.0665 & 62.0 & 0.0 & 1.3 & 23.5 & 8.3 & 1.3 & 10.7 & 1.9 & 16.0 & $-$1.20 & 18.2 & \\
    B2306$+$55 & 0.4751 & 46.6 & 0.1 & 0.4 & 78.5 & 89.9 & 1.5 & 133.4 & 19.0 & 99.0 & $-$1.80 & 119.7 & \\
    B2310$+$42 & 0.3494 & 17.3 & 0.0 & 2.8 & 126.3 & 33.5 & 3.9 & 129.5 & 89.0 & 130.0 & $-$1.50 & 152.3 & \\
    B2315$+$21 & 1.4447 & 20.9 & 0.0 & 2.3 & 71.3 & 19.0 & 2.3 & 44.3 & 15.0 & 73.0 & $-$2.20 & 92.0 & \\
    B2319$+$60 & 2.2565 & 94.3 & 0.1 & 0.8 & 21.1 & 8.0 & 1.6 & 12.6 & 36.0 & $\ldots$ & $-$1.00 & 106.7 & 3 \\
    B2323$+$63 & 1.4363 & 197.3 & 0.0 & 1.0 & 5.6 & 4.9 & 1.2 & 5.8 & 8.0 & $\ldots$ & $-$0.80 & 19.1 & 3 \\
    B2324$+$60 & 0.2337 & 122.3 & 0.0 & 1.6 & 17.7 & 14.7 & 1.5 & 22.0 & 17.0 & $\ldots$ & $-$1.10 & 56.1 & 3 \\
    B2334$+$61 & 0.4954 & 58.4 & 0.0 & 1.2 & 45.3 & 23.0 & 1.2 & 28.7 & 10.0 & $\ldots$ & $-$1.70 & 63.4 & \\
    B2351$+$61 & 0.9448 & 94.9 & 0.0 & 1.3 & 18.6 & 7.9 & 1.3 & 10.5 & 17.0 & $\ldots$ & $-$1.10 & 56.1 & 3 \\
    J0030$+$0451 & 0.0049 & 4.3 & 0.3 & 1.1 & 60.2 & 253.8 & $\ldots$ & $\ldots$ & 7.9 & 44.9 & $-$1.93 & 55.0 & \\
    J0033$+$57 & 0.3145 & 75.6 & 0.0 & 0.5 & 51.6 & 28.6 & 1.0 & 29.7 & $\ldots$ & $\ldots$ & $\ldots$ & $\ldots$ & 3, 6 \\
    J0033$+$61 & 0.9120 & 37.5 & 0.0 & 0.4 & 21.0 & 6.7 & 1.0 & 6.9 & $\ldots$ & $\ldots$ & $\ldots$ & $\ldots$ & 3, 6 \\
    J0051$+$0423 & 0.3547 & 13.9 & 0.0 & 1.4 & 115.8 & 55.6 & 1.3 & 73.9 & $\ldots$ & $\ldots$ & $\ldots$ & $\ldots$ & 6 \\
    J0104$+$64 & 1.3862 & 42.1 & 0.0 & 1.0 & 12.4 & 4.2 & 1.2 & 4.9 & $\ldots$ & $\ldots$ & $\ldots$ & $\ldots$ & 1, 3, 4, 6 \\
    J0137$+$1654 & 0.4148 & 26.1 & 0.1 & 0.7 & 33.4 & 11.8 & 1.3 & 15.3 & 1.4 & 6.2 & $-$1.40 & 7.2 & \\
    J0139$+$5621 & 1.7753 & 101.9 & 0.1 & 1.0 & 7.7 & 2.4 & 1.5 & 3.6 & 0.7 & $\ldots$ & $-$2.60 & 11.8 & 3 \\
    J0158$+$21 & 0.5053 & 19.8 & 0.0 & 0.6 & 8.0 & 3.1 & 1.1 & 3.4 & $\ldots$ & $\ldots$ & $\ldots$ & $\ldots$ & 2, 3, 6 \\
    J0201$+$7002 & 1.3492 & 21.1 & 0.0 & 0.9 & 10.8 & 3.7 & 1.1 & 4.2 & $\ldots$ & $\ldots$ & $\ldots$ & $\ldots$ & 3, 4, 6 \\
    J0212$+$5222 & 0.3764 & 38.2 & 0.2 & 0.8 & 14.6 & 5.2 & 2.2 & 11.5 & 4.1 & 12.0 & $-$1.20 & 13.6 & \\
    J0214$+$5222 & 0.0246 & 22.0 & 0.1 & 0.8 & 39.4 & 15.0 & 1.7 & 25.0 & 0.9 & 21.0 & $\ldots$ & 24.3 & 1 \\
    J0220$+$36 & 1.0298 & 45.3 & 0.0 & 0.2 & 22.5 & 14.5 & 1.0 & 14.6 & $\ldots$ & $\ldots$ & $\ldots$ & $\ldots$ & 3, 6 \\
    J0242$+$62 & 0.5917 & 3.8 & 0.1 & 0.5 & 60.2 & 30.9 & 1.6 & 48.7 & $\ldots$ & $\ldots$ & $\ldots$ & $\ldots$ & 3 \\
    J0329$+$1654 & 0.8933 & 40.9 & 0.0 & 1.8 & 9.1 & 3.0 & 1.7 & 5.2 & 0.6 & 2.3 & $-$1.30 & 2.6 & \\
    J0338$+$66 & 1.7619 & 66.6 & 0.0 & 0.6 & 35.2 & 8.8 & 1.1 & 9.4 & $\ldots$ & $\ldots$ & $\ldots$ & $\ldots$ & 1, 3, 6 \\
    J0341$+$5711 & 1.8875 & 101.0 & 0.0 & 0.2 & 31.6 & 6.6 & 1.0 & 6.7 & $\ldots$ & $\ldots$ & $\ldots$ & $\ldots$ & 3, 6 \\
    J0348$+$0432 & 0.0391 & 40.5 & 0.1 & 0.7 & 10.3 & 7.6 & 1.4 & 10.8 & $\ldots$ & $\ldots$ & $\ldots$ & $\ldots$ & 3 \\
    J0355$+$28 & 0.3649 & 48.7 & 0.0 & 0.6 & 10.4 & 4.2 & 1.1 & 4.5 & $\ldots$ & $\ldots$ & $\ldots$ & $\ldots$ & 1, 3, 4, 6 \\
    J0358$+$42 & 0.2265 & 46.3 & 0.0 & 0.3 & 61.6 & 21.6 & 1.0 & 21.9 & $\ldots$ & $\ldots$ & $\ldots$ & $\ldots$ & 1, 3, 6 \\
    J0358$+$66 & 0.0915 & 62.3 & 0.0 & 1.0 & 10.8 & 5.1 & 1.2 & 5.9 & $\ldots$ & $\ldots$ & $\ldots$ & $\ldots$ & 1, 3, 6 \\
    J0407$+$1607 & 0.0257 & 35.6 & 0.0 & 1.6 & 63.5 & 47.2 & 1.5 & 71.9 & 10.2 & 56.5 & $\ldots$ & 65.5 & \\
    J0408$+$551 & 1.8376 & 56.3 & 0.0 & 0.9 & 22.6 & 5.8 & 1.1 & 6.6 & $\ldots$ & $\ldots$ & $\ldots$ & $\ldots$ & 3, 6 \\
    J0413$+$58 & 0.6865 & 57.1 & 0.0 & 0.8 & 18.1 & 7.1 & 1.1 & 7.9 & $\ldots$ & $\ldots$ & $\ldots$ & $\ldots$ & 3, 6 \\
    J0417$+$35 & 0.6544 & 48.5 & 0.0 & 1.4 & 11.4 & 3.7 & 1.3 & 5.0 & $\ldots$ & $\ldots$ & $-$2.40 & $\ldots$ & 5, 6 \\
    J0426$+$4933 & 0.9225 & 84.4 & 0.1 & 0.1 & 10.6 & 3.6 & 1.4 & 5.1 & $\ldots$ & $\ldots$ & $\ldots$ & $\ldots$ & 3 \\
    J0435$+$27 & 0.3263 & 53.2 & 0.1 & 1.0 & 13.3 & 3.1 & 1.5 & 4.6 & 2.0 & 7.0 & $-$2.30 & 8.9 & \\
    J0457$+$23 & 0.5049 & 58.8 & 0.1 & 0.8 & 14.4 & 7.5 & 1.2 & 9.2 & $\ldots$ & $\ldots$ & $\ldots$ & $\ldots$ & 2, 3 \\
    J0459$-$0210 & 1.1331 & 21.0 & 0.0 & 1.9 & 88.4 & 35.0 & 1.7 & 60.9 & 11.0 & $\ldots$ & $\ldots$ & 50.3 & 3 \\
    J0517$+$22 & 0.2224 & 18.7 & 0.2 & 0.8 & 58.2 & 45.8 & 2.3 & 105.6 & 7.0 & $\ldots$ & $\ldots$ & 32.0 & 3 \\
    J0519$+$54 & 0.3402 & 42.3 & 0.0 & 0.7 & 27.3 & 10.3 & 1.1 & 11.2 & $\ldots$ & $\ldots$ & $\ldots$ & $\ldots$ & 1, 3, 6 \\
    J0533$+$0402 & 0.9630 & 83.3 & 0.2 & 0.5 & 3.2 & 4.0 & 2.9 & 11.5 & $\ldots$ & $\ldots$ & $\ldots$ & $\ldots$ & 3 \\
    J0538$+$2817 & 0.1432 & 39.9 & 0.2 & 0.7 & 20.5 & 18.2 & 1.8 & 33.3 & 8.2 & $\ldots$ & $\ldots$ & 37.5 & 3 \\
    J0540$+$3207 & 0.5243 & 62.1 & 0.1 & 1.6 & 45.5 & 16.4 & 1.7 & 28.1 & $\ldots$ & $\ldots$ & $\ldots$ & $\ldots$ & 3 \\
    J0546$+$2441 & 2.8439 & 72.9 & 0.2 & 0.2 & 24.0 & 2.8 & 2.0 & 5.7 & 2.6 & $\ldots$ & $\ldots$ & 12.1 & 3 \\
    J0555$+$3948 & 1.1469 & 36.4 & 0.0 & 0.2 & 19.7 & 4.0 & 1.0 & 4.0 & $\ldots$ & $\ldots$ & $\ldots$ & $\ldots$ & 3, 6 \\
    J0608$+$00 & 1.0762 & 48.5 & 0.0 & 0.7 & 17.3 & 10.0 & 1.1 & 10.8 & $\ldots$ & $\ldots$ & $\ldots$ & $\ldots$ & 2, 3, 6 \\
    J0608$+$16 & 0.9458 & 86.1 & 0.0 & 0.9 & 77.1 & 27.8 & 1.1 & 31.4 & $\ldots$ & $\ldots$ & $\ldots$ & $\ldots$ & 3, 4, 6 \\
    J0609$+$2130 & 0.0557 & 38.8 & 0.1 & 0.6 & 20.4 & 10.2 & 1.5 & 15.0 & 0.8 & $\ldots$ & $\ldots$ & 3.7 & 3 \\
    J0610$+$37 & 0.4439 & 39.3 & 0.0 & 1.0 & 50.8 & 14.8 & 1.2 & 17.4 & $\ldots$ & $\ldots$ & $\ldots$ & $\ldots$ & 1, 3, 6 \\
    J0611$+$30 & 1.4121 & 45.2 & 0.0 & 0.7 & 81.2 & 25.1 & 1.1 & 27.2 & $\ldots$ & $\ldots$ & $-$3.20 & $\ldots$ & 6 \\
    J0613$+$3731 & 0.6192 & 19.0 & 0.2 & 0.9 & 46.5 & 11.5 & 2.7 & 30.8 & 1.6 & $\ldots$ & $-$1.80 & 11.3 & 3 \\
    J0621$+$0336 & 0.2700 & 72.5 & 0.1 & 0.6 & 24.4 & 12.6 & 1.6 & 19.8 & $\ldots$ & $\ldots$ & $\ldots$ & $\ldots$ & 3 \\
    J0621$+$1002 & 0.0289 & 36.5 & 0.0 & 1.8 & 6.5 & 18.7 & 1.7 & 31.6 & $\ldots$ & 20.0 & $-$1.90 & 24.4 & \\
    J0627$+$0706 & 0.2379 & 138.2 & 0.0 & 1.6 & 15.8 & 25.8 & 1.5 & 38.0 & 6.0 & $\ldots$ & $-$1.60 & 34.1 & 3 \\
    J0645$+$5158 & 0.0089 & 18.3 & 0.2 & 1.1 & 9.5 & 9.4 & 4.0 & 37.7 & 2.4 & 4.7 & $\ldots$ & 5.4 & 1 \\
    J0645$+$80 & 0.6579 & 49.5 & 0.0 & 0.7 & 12.6 & 4.2 & 1.1 & 4.5 & $\ldots$ & $\ldots$ & $\ldots$ & $\ldots$ & 1, 3, 6 \\
    \hline
  \end{tabular}
\end{table*}

\setcounter{table}{0}
\begin{table*}
  \footnotesize
  \caption{Continued.}
  \begin{tabular}{lrrrrrrrrrrrrl}
    \hline
    PSR & \multicolumn{1}{c}{$P$} & \multicolumn{1}{c}{$\mathrm{DM}$} & \multicolumn{1}{c}{$\theta_\mathrm{TAB}$} & \multicolumn{1}{c}{$\theta_\mathrm{SAP}$} & \multicolumn{1}{c}{$\sigma$} & \multicolumn{1}{c}{$S_{135}^\mathrm{uncor}$} & \multicolumn{1}{c}{$S_{135}^{\times}$} & \multicolumn{1}{c}{$S_{135}^\mathrm{cor}$} & \multicolumn{1}{c}{$S_{400}$} & \multicolumn{1}{c}{$S_{150}$} & \multicolumn{1}{c}{$\alpha$} & \multicolumn{1}{c}{$S_{135}^\mathrm{exp}$} & \multicolumn{1}{c}{Notes} \\
    & \multicolumn{1}{c}{(s)} & \multicolumn{1}{c}{(pc\,cm$^{-3}$)} & \multicolumn{1}{c}{(\degr)} & \multicolumn{1}{c}{(\degr)} & & \multicolumn{1}{c}{(mJy)} & & \multicolumn{1}{c}{(mJy)} & \multicolumn{1}{c}{(mJy)} & \multicolumn{1}{c}{(mJy)} & & \multicolumn{1}{c}{(mJy)} & \\
    \hline
    J0750$+$57 & 1.1749 & 26.8 & 0.0 & 0.2 & 18.2 & 5.3 & 1.0 & 5.4 & $\ldots$ & $\ldots$ & $\ldots$ & $\ldots$ & 1, 3, 6 \\
    J0815$+$4611 & 0.4342 & 11.3 & 0.0 & 0.2 & 35.5 & 9.6 & 1.0 & 9.7 & $\ldots$ & $\ldots$ & $\ldots$ & $\ldots$ & 3, 4, 6 \\
    J0943$+$22 & 0.5330 & 27.2 & 0.1 & 1.1 & 31.6 & 4.2 & 1.3 & 5.4 & 5.5 & 6.3 & $-$0.80 & 6.9 & 5 \\
    J0943$+$41 & 2.2294 & 21.4 & 0.0 & 2.0 & 10.1 & 1.1 & 1.9 & 2.1 & $\ldots$ & $\ldots$ & $\ldots$ & $\ldots$ & 1, 3, 6 \\
    J0947$+$27 & 0.8510 & 28.9 & 0.0 & 1.2 & 29.2 & 5.3 & 1.3 & 6.6 & 1.0 & 10.0 & $-$2.20 & 12.6 & \\
    J1000$+$08 & 0.4404 & 21.8 & 0.0 & 0.6 & 14.6 & 5.1 & 1.1 & 5.4 & $\ldots$ & $\ldots$ & $\ldots$ & $\ldots$ & 3, 4, 6 \\
    J1012$+$5307 & 0.0053 & 9.0 & 0.0 & 2.3 & 50.7 & 19.9 & 2.4 & 47.6 & 30.0 & 34.5 & $-$1.70 & 41.3 & \\
    J1022$+$1001 & 0.0165 & 10.3 & 0.0 & 2.4 & 137.4 & 48.2 & 2.6 & 125.8 & 20.0 & 40.7 & $-$1.40 & 47.2 & \\
    J1101$+$65 & 3.6312 & 19.2 & 0.0 & 0.2 & 15.1 & 1.3 & 1.0 & 1.3 & $\ldots$ & $\ldots$ & $\ldots$ & $\ldots$ & 1, 3, 6 \\
    J1110$+$58 & 0.7933 & 26.4 & 0.0 & 0.7 & 13.4 & 2.5 & 1.1 & 2.7 & $\ldots$ & $\ldots$ & $\ldots$ & $\ldots$ & 1, 3, 6 \\
    J1125$+$7819 & 0.0042 & 11.2 & 0.3 & 1.9 & 117.0 & 75.0 & $\ldots$ & $\ldots$ & 17.1 & $\ldots$ & $\ldots$ & 78.2 & 3 \\
    J1134$+$24 & 1.0021 & 23.7 & 0.0 & 1.0 & 21.1 & 3.7 & 1.2 & 4.3 & $\ldots$ & $\ldots$ & $\ldots$ & $\ldots$ & 3, 4, 6 \\
    J1238$+$2152 & 1.1186 & 18.0 & 0.0 & 1.2 & 131.2 & 21.3 & 1.3 & 27.1 & 2.0 & 27.0 & $-$0.80 & 29.4 & 3 \\
    J1239$+$32 & 0.0047 & 16.9 & 0.0 & 0.6 & 63.7 & 23.2 & 1.1 & 24.7 & $\ldots$ & $\ldots$ & $\ldots$ & $\ldots$ & 1, 3, 4, 6 \\
    J1242$+$39 & 1.3103 & 26.5 & 0.0 & 0.7 & 32.5 & 5.4 & 1.1 & 5.9 & $\ldots$ & $\ldots$ & $\ldots$ & $\ldots$ & 3, 6 \\
    J1246$+$22 & 0.4739 & 17.8 & 0.1 & 0.1 & 18.8 & 4.3 & 1.3 & 5.7 & 29.0 & 4.2 & 0.80 & 3.9 & \\
    J1313$+$0931 & 0.8489 & 12.1 & 0.0 & 2.2 & 71.1 & 19.5 & 2.2 & 42.0 & $\ldots$ & $\ldots$ & $-$2.60 & $\ldots$ & 6 \\
    J1327$+$34 & 0.0415 & 4.2 & 0.0 & 0.0 & 121.3 & 108.4 & 1.0 & 108.4 & $\ldots$ & $\ldots$ & $\ldots$ & $\ldots$ & 1, 3, 4, 6 \\
    J1411$+$25 & 0.0625 & 12.4 & 0.1 & 0.9 & 10.3 & 3.8 & 1.4 & 5.3 & $\ldots$ & $\ldots$ & $\ldots$ & $\ldots$ & 2, 3 \\
    J1434$+$7257 & 0.0417 & 12.6 & 0.2 & 0.8 & 43.8 & 10.8 & 2.3 & 24.8 & 1.3 & $\ldots$ & $\ldots$ & 5.9 & 1, 3 \\
    J1518$+$4904 & 0.0409 & 11.6 & 0.0 & 2.5 & 50.4 & 16.1 & 3.0 & 47.5 & 8.0 & $\ldots$ & $-$1.30 & 32.8 & 3 \\
    J1549$+$2113 & 1.2625 & 24.1 & 0.1 & 0.3 & 36.9 & 10.8 & 1.1 & 11.9 & 0.9 & 7.2 & $-$1.80 & 8.7 & \\
    J1627$+$1419 & 0.4909 & 32.2 & 0.0 & 1.1 & 59.2 & 54.7 & 1.2 & 65.1 & 6.1 & 78.0 & $-$1.60 & 92.3 & \\
    J1629$+$43 & 0.1812 & 7.3 & 0.0 & 0.2 & 11.4 & 13.1 & 1.0 & 13.2 & $\ldots$ & $\ldots$ & $\ldots$ & $\ldots$ & 1, 3, 6 \\
    J1645$+$1012 & 0.4109 & 36.2 & 0.0 & 1.6 & 49.1 & 28.1 & 1.5 & 42.9 & 2.3 & 69.0 & $-$3.00 & 94.7 & \\
    J1647$+$66 & 1.5998 & 22.5 & 0.0 & 0.8 & 25.5 & 7.0 & 1.1 & 7.8 & $\ldots$ & $\ldots$ & $\ldots$ & $\ldots$ & 1, 3, 6 \\
    J1649$+$2533 & 1.0153 & 34.5 & 0.1 & 0.7 & 22.8 & 6.6 & 1.5 & 9.6 & 7.4 & 12.0 & $-$1.10 & 13.5 & \\
    J1652$+$2651 & 0.9158 & 40.7 & 0.0 & 0.2 & 18.8 & 5.6 & 1.0 & 5.8 & 11.3 & 20.0 & $-$1.00 & 22.2 & \\
    J1706$+$59 & 1.4766 & 30.6 & 0.0 & 0.4 & 69.2 & 15.3 & 1.0 & 15.7 & $\ldots$ & $\ldots$ & $\ldots$ & $\ldots$ & 1, 3, 6 \\
    J1736$+$05 & 0.9992 & 38.7 & 0.0 & 2.0 & 15.0 & 10.3 & 1.9 & 19.4 & $\ldots$ & $\ldots$ & $\ldots$ & $\ldots$ & 3, 5, 6 \\
    J1741$+$2758 & 1.3607 & 29.1 & 0.0 & 1.9 & 49.4 & 12.4 & 1.7 & 21.6 & 3.0 & 30.0 & $-$1.10 & 33.7 & \\
    J1746$+$2540 & 1.0582 & 51.3 & 0.0 & 2.8 & 8.8 & 2.4 & 3.7 & 8.9 & 1.2 & 3.7 & $-$0.60 & 3.9 & \\
    J1752$+$2359 & 0.4091 & 36.2 & 0.0 & 1.0 & 21.9 & 6.5 & 1.2 & 7.6 & 3.5 & 5.3 & $-$1.30 & 6.1 & \\
    J1758$+$3030 & 0.9473 & 35.1 & 0.0 & 2.7 & 69.9 & 18.4 & 3.6 & 66.1 & 8.9 & 56.0 & $-$1.60 & 66.3 & \\
    J1800$+$50 & 0.5784 & 22.7 & 0.0 & 2.0 & 17.3 & 3.5 & 1.9 & 6.6 & $\ldots$ & $\ldots$ & $\ldots$ & $\ldots$ & 1, 3, 6 \\
    J1806$+$28 & 0.0151 & 18.7 & 0.1 & 0.4 & 100.2 & 44.2 & 1.1 & 47.8 & $\ldots$ & $\ldots$ & $\ldots$ & $\ldots$ & 1, 3 \\
    J1811$+$0702 & 0.4617 & 58.3 & 0.0 & 1.4 & 7.9 & 9.1 & 1.3 & 12.1 & 2.2 & $\ldots$ & $\ldots$ & 10.1 & 3 \\
    J1814$+$1130 & 0.7513 & 64.6 & 0.0 & 1.3 & 14.9 & 6.8 & 1.3 & 8.9 & 0.7 & $\ldots$ & $\ldots$ & 3.3 & 3 \\
    J1815$+$55 & 0.4268 & 59.0 & 0.0 & 0.8 & 15.1 & 3.6 & 1.1 & 3.9 & $\ldots$ & $\ldots$ & $\ldots$ & $\ldots$ & 1, 3, 6 \\
    J1821$+$1715 & 1.3667 & 60.2 & 0.0 & 2.0 & 17.0 & 8.2 & 1.9 & 15.5 & 3.7 & 11.0 & $-$0.50 & 11.6 & \\
    J1821$+$41 & 1.2619 & 40.7 & 0.0 & 0.5 & 36.2 & 6.6 & 1.0 & 6.9 & $\ldots$ & $\ldots$ & $\ldots$ & $\ldots$ & 1, 3, 6 \\
    J1822$+$0705 & 1.3628 & 62.4 & 0.1 & 0.8 & 19.9 & 14.2 & 1.5 & 21.9 & 3.8 & $\ldots$ & $\ldots$ & 17.4 & 3 \\
    J1832$+$27 & 0.6317 & 47.4 & 0.0 & 2.0 & 7.7 & 3.5 & 1.9 & 6.7 & $\ldots$ & $\ldots$ & $\ldots$ & $\ldots$ & 2, 3, 4, 6 \\
    J1836$+$51 & 0.6919 & 43.8 & 0.0 & 0.4 & 16.4 & 4.7 & 1.0 & 4.8 & $\ldots$ & $\ldots$ & $\ldots$ & $\ldots$ & 1, 3, 4, 6 \\
    J1838$+$1650 & 1.9020 & 33.1 & 0.0 & 1.6 & 13.9 & 7.1 & 1.5 & 10.5 & $\ldots$ & $\ldots$ & $-$1.60 & $\ldots$ & 6 \\
    J1844$+$21 & 0.5946 & 28.8 & 0.0 & 1.0 & 6.3 & 10.0 & 1.2 & 11.7 & $\ldots$ & $\ldots$ & $\ldots$ & $\ldots$ & 3, 4, 6 \\
    J1848$+$0647 & 0.5060 & 24.7 & 0.1 & 0.9 & 13.6 & 16.8 & 1.7 & 29.2 & 2.3 & $\ldots$ & $\ldots$ & 10.5 & 3, 5 \\
    J1849$+$2423 & 0.2756 & 62.3 & 0.1 & 0.4 & 11.4 & 7.0 & 1.4 & 9.8 & 2.3 & 8.1 & $-$0.70 & 8.7 & \\
    J1851$-$0053 & 1.4091 & 24.5 & 0.1 & 0.9 & 12.6 & 24.2 & 1.2 & 29.0 & $\ldots$ & $\ldots$ & $\ldots$ & $\ldots$ & 3 \\
    J1900$+$30 & 0.6022 & 71.8 & 0.0 & 1.1 & 16.3 & 5.5 & 1.2 & 6.7 & $\ldots$ & $\ldots$ & $-$1.50 & $\ldots$ & 5, 6 \\
    J1906$+$1854 & 1.0191 & 156.8 & 0.0 & 1.0 & 10.1 & 7.1 & 1.2 & 8.4 & 4.6 & 18.0 & $-$1.20 & 20.4 & \\
    J1907$+$57 & 0.4237 & 54.5 & 0.0 & 0.5 & 13.2 & 3.7 & 1.0 & 3.8 & $\ldots$ & $\ldots$ & $\ldots$ & $\ldots$ & 1, 3, 4, 6 \\
    J1912$+$2525 & 0.6220 & 37.8 & 0.0 & 2.1 & 50.0 & 12.1 & 2.0 & 24.2 & 1.6 & 18.0 & $-$1.90 & 22.0 & \\
    J1921$+$42 & 0.5952 & 53.2 & 0.0 & 0.7 & 30.9 & 8.4 & 1.1 & 9.1 & $\ldots$ & $\ldots$ & $\ldots$ & $\ldots$ & 1, 3, 6 \\
    J1922$+$58 & 0.5296 & 53.7 & 0.0 & 1.1 & 30.7 & 7.2 & 1.2 & 8.8 & $\ldots$ & $\ldots$ & $\ldots$ & $\ldots$ & 1, 3, 6 \\
    J1929$+$00 & 1.1669 & 42.8 & 0.0 & 0.5 & 35.4 & 32.0 & 1.0 & 33.1 & $\ldots$ & $\ldots$ & $\ldots$ & $\ldots$ & 3, 5, 6 \\
    J1929$+$62 & 1.4561 & 67.8 & 0.0 & 1.0 & 13.8 & 2.3 & 1.2 & 2.6 & $\ldots$ & $\ldots$ & $\ldots$ & $\ldots$ & 1, 3, 6 \\
    J1930$-$01 & 0.5937 & 36.5 & 0.0 & 1.0 & 28.4 & 26.3 & 1.2 & 30.9 & $\ldots$ & $\ldots$ & $\ldots$ & $\ldots$ & 3, 4, 6 \\
    J1935$+$52 & 0.5684 & 71.3 & 0.0 & 0.9 & 13.6 & 4.1 & 1.1 & 4.7 & $\ldots$ & $\ldots$ & $\ldots$ & $\ldots$ & 1, 3, 6 \\
    J1938$+$0650 & 1.1216 & 75.1 & 0.0 & 1.2 & 9.7 & 7.1 & 1.2 & 8.8 & 3.2 & $\ldots$ & $\ldots$ & 14.6 & 3, 5 \\
    J1938$+$14 & 2.9025 & 74.0 & 0.0 & 0.7 & 28.3 & 8.9 & 1.1 & 9.5 & $\ldots$ & $\ldots$ & $\ldots$ & $\ldots$ & 2, 3, 6 \\
    J1941$+$0121 & 0.2173 & 52.3 & 0.0 & 0.6 & 17.0 & 23.7 & 1.1 & 25.4 & $\ldots$ & $\ldots$ & $\ldots$ & $\ldots$ & 3 \\
    J1941$+$43 & 0.8409 & 79.3 & 0.0 & 0.9 & 24.9 & 6.8 & 1.1 & 7.6 & $\ldots$ & $\ldots$ & $\ldots$ & $\ldots$ & 1, 3, 6 \\
    \hline
  \end{tabular}
\end{table*}

\setcounter{table}{0}
\begin{table*}
  \footnotesize
  \caption{Continued.}
  \begin{tabular}{lrrrrrrrrrrrrl}
    \hline
    PSR & \multicolumn{1}{c}{$P$} & \multicolumn{1}{c}{$\mathrm{DM}$} & \multicolumn{1}{c}{$\theta_\mathrm{TAB}$} & \multicolumn{1}{c}{$\theta_\mathrm{SAP}$} & \multicolumn{1}{c}{$\sigma$} & \multicolumn{1}{c}{$S_{135}^\mathrm{uncor}$} & \multicolumn{1}{c}{$S_{135}^{\times}$} & \multicolumn{1}{c}{$S_{135}^\mathrm{cor}$} & \multicolumn{1}{c}{$S_{400}$} & \multicolumn{1}{c}{$S_{150}$} & \multicolumn{1}{c}{$\alpha$} & \multicolumn{1}{c}{$S_{135}^\mathrm{exp}$} & \multicolumn{1}{c}{Notes} \\
    & \multicolumn{1}{c}{(s)} & \multicolumn{1}{c}{(pc\,cm$^{-3}$)} & \multicolumn{1}{c}{(\degr)} & \multicolumn{1}{c}{(\degr)} & & \multicolumn{1}{c}{(mJy)} & & \multicolumn{1}{c}{(mJy)} & \multicolumn{1}{c}{(mJy)} & \multicolumn{1}{c}{(mJy)} & & \multicolumn{1}{c}{(mJy)} & \\
    \hline
    J1942$+$81 & 0.2036 & 40.3 & 0.0 & 0.9 & 22.1 & 4.4 & 1.1 & 4.9 & $\ldots$ & $\ldots$ & $\ldots$ & $\ldots$ & 1, 3, 6 \\
    J1943$+$0609 & 0.4462 & 70.7 & 0.0 & 1.5 & 7.4 & 8.1 & 1.4 & 11.4 & $\ldots$ & $\ldots$ & $\ldots$ & $\ldots$ & 3 \\
    J1945$+$07 & 1.0740 & 62.3 & 0.0 & 0.7 & 8.9 & 4.6 & 1.1 & 5.0 & $\ldots$ & $\ldots$ & $\ldots$ & $\ldots$ & 2, 3, 6 \\
    J1952$+$30 & 1.6657 & 189.8 & 0.0 & 0.5 & 7.9 & 4.4 & 1.0 & 4.6 & $\ldots$ & $\ldots$ & $\ldots$ & $\ldots$ & 3, 4, 6 \\
    J1954$+$43 & 1.3870 & 130.2 & 0.0 & 0.8 & 6.7 & 5.4 & 1.1 & 6.0 & $\ldots$ & $\ldots$ & $\ldots$ & $\ldots$ & 1, 3, 6 \\
    J2000$+$29 & 3.0738 & 132.3 & 0.0 & 0.9 & 12.2 & 3.8 & 1.1 & 4.3 & $\ldots$ & $\ldots$ & $\ldots$ & $\ldots$ & 1, 3, 4, 6 \\
    J2001$+$42 & 0.7192 & 54.9 & 0.0 & 1.0 & 26.3 & 12.2 & 1.2 & 14.1 & $\ldots$ & $\ldots$ & $\ldots$ & $\ldots$ & 1, 3, 6 \\
    J2005$-$0020 & 2.2797 & 36.2 & 0.2 & 0.9 & 19.1 & 10.1 & 2.0 & 20.2 & 8.0 & $\ldots$ & $\ldots$ & 36.6 & 3 \\
    J2007$+$0809 & 0.3257 & 53.4 & 0.0 & 2.6 & 14.3 & 16.3 & 3.3 & 53.8 & $\ldots$ & 32.0 & $\ldots$ & 37.1 & 3 \\
    J2007$+$0910 & 0.4587 & 48.7 & 0.0 & 1.5 & 28.4 & 12.4 & 1.5 & 18.1 & 1.5 & 26.0 & $-$2.70 & 34.6 & \\
    J2008$+$2513 & 0.5892 & 60.6 & 0.2 & 0.7 & 19.6 & 7.5 & 1.7 & 13.0 & 2.7 & 3.8 & $-$1.20 & 4.3 & \\
    J2010$+$2845 & 0.5654 & 112.4 & 0.5 & 0.5 & 12.9 & 11.6 & $\ldots$ & $\ldots$ & $\ldots$ & $\ldots$ & $\ldots$ & $\ldots$ & 3 \\
    J2016$+$1948 & 0.0649 & 33.8 & 0.1 & 0.7 & 31.5 & 16.2 & 1.4 & 22.1 & 3.3 & $\ldots$ & $\ldots$ & 15.1 & 3 \\
    J2017$+$2043 & 0.5371 & 60.5 & 0.1 & 0.7 & 20.6 & 8.2 & 1.2 & 10.2 & 1.5 & 14.0 & $-$1.50 & 16.4 & \\
    J2017$+$59 & 0.4035 & 60.3 & 0.0 & 0.2 & 17.7 & 8.3 & 1.0 & 8.4 & $\ldots$ & $\ldots$ & $\ldots$ & $\ldots$ & 1, 3, 6 \\
    J2027$+$74 & 0.5152 & 11.6 & 0.0 & 0.4 & 15.9 & 9.2 & 1.0 & 9.5 & $\ldots$ & $\ldots$ & $\ldots$ & $\ldots$ & 1, 3, 6 \\
    J2030$+$55 & 0.5789 & 59.6 & 0.0 & 0.5 & 19.5 & 8.6 & 1.0 & 8.9 & $\ldots$ & $\ldots$ & $\ldots$ & $\ldots$ & 3, 6 \\
    J2033$+$0042 & 5.0134 & 37.8 & 0.1 & 0.8 & 12.6 & 6.2 & 1.4 & 8.9 & $\ldots$ & $\ldots$ & $\ldots$ & $\ldots$ & 3 \\
    J2036$+$2835 & 1.3587 & 84.3 & 0.1 & 0.5 & 19.9 & 5.6 & 1.4 & 8.1 & $\ldots$ & 9.6 & $-$1.90 & 11.7 & 5 \\
    J2038$+$35 & 0.1602 & 57.9 & 0.0 & 0.4 & 30.1 & 16.8 & 1.0 & 17.2 & $\ldots$ & $\ldots$ & $\ldots$ & $\ldots$ & 3, 6 \\
    J2040$+$1657 & 0.8656 & 50.7 & 0.0 & 1.1 & 8.2 & 4.4 & 1.2 & 5.4 & 0.6 & 9.4 & $-$2.60 & 12.4 & \\
    J2043$+$2740 & 0.0961 & 21.0 & 0.0 & 2.2 & 165.5 & 75.5 & 2.2 & 165.6 & 15.0 & 140.0 & $-$1.30 & 160.6 & \\
    J2043$+$7045 & 0.5896 & 57.5 & 0.0 & 0.7 & 12.2 & 5.8 & 1.1 & 6.2 & $\ldots$ & $\ldots$ & $\ldots$ & $\ldots$ & 3, 6 \\
    J2045$+$0912 & 0.3956 & 31.4 & 0.0 & 1.6 & 13.1 & 4.0 & 1.5 & 5.8 & 3.5 & 9.6 & $-$1.00 & 10.7 & \\
    J2102$+$38 & 1.1899 & 86.2 & 0.0 & 0.4 & 11.5 & 8.2 & 1.0 & 8.4 & $\ldots$ & $\ldots$ & $\ldots$ & $\ldots$ & 3, 6 \\
    J2105$+$28 & 0.4057 & 62.3 & 0.0 & 0.5 & 16.2 & 5.8 & 1.0 & 6.0 & $\ldots$ & $\ldots$ & $\ldots$ & $\ldots$ & 1, 3, 6 \\
    J2113$+$67 & 0.5521 & 54.7 & 0.0 & 0.2 & 10.4 & 4.3 & 1.0 & 4.3 & $\ldots$ & $\ldots$ & $\ldots$ & $\ldots$ & 1, 3, 6 \\
    J2137$+$64 & 1.7510 & 105.9 & 0.0 & 0.9 & 7.8 & 2.6 & 1.1 & 3.0 & $\ldots$ & $\ldots$ & $\ldots$ & $\ldots$ & 1, 3, 6 \\
    J2139$+$2242 & 1.0835 & 44.2 & 0.0 & 1.1 & 75.1 & 33.0 & 1.2 & 40.0 & $\ldots$ & 46.0 & $-$0.20 & 47.0 & \\
    J2145$-$0750 & 0.0161 & 9.0 & 6.1 & 4.3 & 21.8 & 33.4 & $\ldots$ & $\ldots$ & 100.0 & 162.0 & $-$1.80 & 195.8 & \\
    J2155$+$2813 & 1.6090 & 77.2 & 0.0 & 1.1 & 22.4 & 4.0 & 1.2 & 4.8 & 2.1 & 9.6 & $-$1.40 & 11.1 & \\
    J2156$+$2618 & 0.4982 & 48.5 & 0.2 & 0.5 & 8.6 & 3.1 & 2.4 & 7.2 & 2.7 & 2.9 & $-$1.00 & 3.2 & \\
    J2202$+$21 & 1.3583 & 17.8 & 0.0 & 1.0 & 15.3 & 3.3 & 1.2 & 3.9 & $\ldots$ & $\ldots$ & $\ldots$ & $\ldots$ & 2, 3, 4, 6 \\
    J2203$+$50 & 0.7454 & 76.3 & 0.0 & 0.9 & 5.6 & 4.5 & 1.1 & 5.0 & $\ldots$ & $\ldots$ & $\ldots$ & $\ldots$ & 3, 5, 6 \\
    J2205$+$1444 & 0.9380 & 36.8 & 0.1 & 0.8 & 14.0 & 4.2 & 1.4 & 5.7 & 1.5 & 4.6 & $-$2.10 & 5.7 & \\
    J2207$+$40 & 0.6370 & 11.8 & 0.0 & 1.0 & 66.7 & 31.8 & 1.1 & 36.5 & $\ldots$ & $\ldots$ & $\ldots$ & $\ldots$ & 1, 3, 6 \\
    J2208$+$5500 & 0.9332 & 104.7 & 0.1 & 0.6 & 15.7 & 8.3 & 1.2 & 10.0 & $\ldots$ & $\ldots$ & $\ldots$ & $\ldots$ & 3, 5 \\
    J2215$+$1538 & 0.3742 & 29.2 & 0.0 & 1.6 & 27.6 & 11.4 & 1.5 & 16.7 & 3.7 & 6.8 & $-$0.20 & 6.9 & \\
    J2222$+$2923 & 0.2814 & 49.4 & 0.0 & 1.2 & 13.2 & 3.0 & 1.2 & 3.8 & $\ldots$ & 4.3 & $-$1.10 & 4.8 & \\
    J2222$-$0137 & 0.0328 & 3.3 & 0.1 & 1.8 & 19.7 & 5.5 & 2.1 & 11.5 & $\ldots$ & $\ldots$ & $\ldots$ & $\ldots$ & 3 \\
    J2227$+$30 & 0.8424 & 20.0 & 0.0 & 1.1 & 48.7 & 11.9 & 1.2 & 14.5 & $\ldots$ & $\ldots$ & $\ldots$ & $\ldots$ & 3, 5, 6 \\
    J2228$+$40 & 0.2727 & 74.2 & 0.0 & 0.6 & 13.9 & 5.7 & 1.1 & 6.1 & $\ldots$ & $\ldots$ & $\ldots$ & $\ldots$ & 3, 4, 6 \\
    J2234$+$2114 & 1.3587 & 35.3 & 0.0 & 1.1 & 45.2 & 10.2 & 1.2 & 12.3 & 2.6 & 13.0 & $-$1.30 & 14.9 & \\
    J2235$+$1506 & 0.0598 & 18.1 & 0.1 & 0.2 & 17.3 & 6.8 & 1.1 & 7.6 & 3.0 & 6.0 & $-$2.90 & 8.1 & \\
    J2243$+$69 & 0.8554 & 67.8 & 0.0 & 0.9 & 13.0 & 4.3 & 1.1 & 4.8 & $\ldots$ & $\ldots$ & $\ldots$ & $\ldots$ & 1, 3, 6 \\
    J2248$-$0101 & 0.4772 & 29.0 & 0.0 & 1.2 & 20.8 & 10.5 & 1.2 & 13.0 & 11.0 & $\ldots$ & $\ldots$ & 50.3 & \\
    J2253$+$1516 & 0.7922 & 29.2 & 0.0 & 1.7 & 26.0 & 9.0 & 1.6 & 14.4 & 2.4 & 6.0 & $-$1.00 & 6.7 & \\
    J2301$+$48 & 0.7420 & 72.8 & 0.0 & 0.2 & 14.8 & 4.8 & 1.0 & 4.8 & $\ldots$ & $\ldots$ & $\ldots$ & $\ldots$ & 3, 4, 6 \\
    J2302$+$6028 & 1.2064 & 156.7 & 0.0 & 1.8 & 42.3 & 40.1 & 1.7 & 68.5 & 12.0 & $\ldots$ & $\ldots$ & 54.9 & 3 \\
    J2315$+$58 & 1.0616 & 73.2 & 0.0 & 2.0 & 20.3 & 15.3 & 1.9 & 29.0 & $\ldots$ & $\ldots$ & $\ldots$ & $\ldots$ & 3, 6 \\
    J2316$+$69 & 0.8134 & 71.4 & 0.0 & 0.7 & 35.1 & 15.9 & 1.1 & 17.3 & $\ldots$ & $\ldots$ & $\ldots$ & $\ldots$ & 1, 3, 6 \\
    J2326$+$6141 & 0.7900 & 34.0 & 0.0 & 0.4 & 15.5 & 8.3 & 1.0 & 8.5 & $\ldots$ & $\ldots$ & $\ldots$ & $\ldots$ & 3, 6 \\
    J2340$+$08 & 0.3033 & 23.7 & 0.1 & 0.9 & 20.2 & 7.6 & 1.2 & 9.3 & $\ldots$ & $\ldots$ & $\ldots$ & $\ldots$ & 2, 3 \\
    J2347$+$02 & 1.3861 & 16.2 & 0.0 & 1.0 & 19.5 & 5.2 & 1.2 & 6.1 & $\ldots$ & $\ldots$ & $\ldots$ & $\ldots$ & 2, 3, 4, 6 \\
    J2352$+$65 & 1.1649 & 154.3 & 0.0 & 0.6 & 6.4 & 4.0 & 1.1 & 4.3 & $\ldots$ & $\ldots$ & $\ldots$ & $\ldots$ & 3, 5, 6 \\
    J2353$+$85 & 1.0117 & 38.2 & 0.0 & 0.4 & 14.2 & 2.4 & 1.0 & 2.4 & $\ldots$ & $\ldots$ & $\ldots$ & $\ldots$ & 1, 3, 6 \\
    J2356$+$22 & 1.8410 & 21.9 & 0.0 & 0.9 & 20.6 & 3.2 & 1.1 & 3.7 & $\ldots$ & $\ldots$ & $\ldots$ & $\ldots$ & 1, 3, 4, 6 \\
    \hline
  \end{tabular}
  \tablefoot{[1] Pulsars discovered by the GBNCC survey
    \citep{slr+14}, [2] Pulsars discovered by the AO327 survey
    \citep{dsm+13}, [3] Pulsars previously not reported by other LOFAR
    censuses~\citep{phs+16,bkk+16,kvh+16}, [4] Pulsars not in v1.59 of
    the ATNF pulsar catalogue \citep{mht+05}, [5] Pulsars with
    $|\delta\mathrm{DM}|>2.0$\,pc\,cm$^{-3}$ compared to v1.54 of the
    ATNF pulsar catalogue. [6] This pulsar has no precise position
    measurement. The pulsar is assumed to be located at the centre of
    the detected TAB and only the offset from the SAP is considered
    for the flux density value correction. }
\end{table*}

\end{appendix}

\end{document}